\documentclass[a4paper,11pt]{article}
\usepackage{jheppub}

\usepackage{xspace}
\usepackage{placeins}
\usepackage{wasysym}
\usepackage{slashed}
\usepackage{braket}
\usepackage{footmisc}
\usepackage{enumitem}
\usepackage{comment}
\usepackage{empheq}
\usepackage[font={bf, footnotesize}]{subfig}
\usepackage[usenames,dvipsnames]{xcolor}
\usepackage[cal=cm,scr=euler]{mathalpha}

\usepackage{axodraw4j}
\usepackage{pstricks}
\usepackage{color}

\usepackage{amsmath}

\allowdisplaybreaks

\DeclareMathOperator{\arctanh}{arctanh}
\DeclareMathOperator{\Li}{Li}

\newcommand{\cev}[1]{\reflectbox{\ensuremath{\vec{\reflectbox{\ensuremath{#1}}}}}}

\usepackage{marginnote}
\usepackage[normalem]{ulem}

\newcommand{\pv}{ {\bf p} \cdot {\bf v} }
\newcommand{\nn}{ \nonumber }

\title{No Einstein Relation, No Problem: How Relativistic Weakly Coupled Heavy Quarks Equilibrate Beyond Leading Logarithm Anyway}

\preprint{MIT-CTP/6123}

\author[1,2]{Jean F. Du Plessis,}

\author[2]{and Bruno Scheihing-Hitschfeld}

\affiliation[1]{
Center for Theoretical Physics --- A Leinweber Institute,\\ Massachusetts Institute of Technology, Cambridge, Massachusetts 02139, USA
}

\affiliation[2]{
Kavli Institute for Theoretical Physics,\\ University of California, Santa Barbara, California 93106, USA
}

\emailAdd{jeandp@mit.edu}
\emailAdd{bscheihi@kitp.ucsb.edu}

\abstract{Relativistic heavy quarks equilibrate even when their microscopic drag and momentum diffusion violate the Einstein relation. We investigate how equilibration proceeds as a function of the coupling and compared to simplified Fokker-Planck models. To do this, we present a detailed calculation of the momentum-transfer kernel in weakly coupled non-Abelian plasmas through strict $\mathcal{O}(g^4)$, and study the resulting dynamics at leading order in the inverse heavy-quark mass. Beyond leading logarithm, its non-Gaussian structure plays a crucial role in equilibration. We examine closely the analytic structure of the evolution kernel, determine the large-order asymptotics of the momentum-transfer cumulants, and show how the kernel's bounded analytic domain fixes asymmetric exponential tails at the level of the momentum transfer probability. Strongly coupled $\mathcal{N}=4$ SYM also features an approximately Gaussian core and exponential tails, but differs in the nature of the kernel's singularities and the velocity dependence of the transport coefficients and tail exponents. These differences have distinct dynamical consequences: for the parameters studied, the persistence of the relativistic population in steeply falling spectra is largely explained by drag alone at very weak coupling, whereas fluctuation-induced survivor bias produces a large relative enhancement at strong coupling. Nevertheless, the shared qualitative structure between weak and strong coupling points the way towards a practical phenomenological description of heavy quarks anchored in first-principles field theory.}

\begin{document}
\maketitle
\flushbottom

\section{Introduction}

Understanding how heavy quarks get stopped and equilibrate as they propagate through quark-gluon plasma (QGP) provides a unique window to characterize the non-perturbative and out of equilibrium dynamics of QCD. QGP, the state that QCD matter forms when it is heated up to temperatures above $T_c \approx 150$ MeV is one of the most striking examples of the emergence of fluid dynamics from strongly coupled quantum matter — the same quantum matter that gives rise to atomic nuclei. This extreme state of matter that filled the universe a microsecond after the Big Bang can be studied by colliding ultrarelativistic heavy nuclei. However, the very short lifetime of the droplet of QGP formed in these collisions makes it impossible to study with genuinely external probes. Instead, we have to rely only on particles produced in the same collisions.

In addition to all of the matter that soon after the collision starts behaving hydrodynamically, a comparatively small number of highly energetic particles are created, which then act as probes whose characteristic energies are well separated from those of the thermal environment that will soon form around it. 
A heavy quark in this context means a charm or bottom quark, i.e., those with a mass larger than the temperature of the plasma and that are stable on the timescales of the QGP medium evolution. Due to their large mass, their thermal production is exponentially suppressed, making heavy flavor approximately conserved during the evolution of the plasma. As such, having these ``external’’ probes of QGP that can be tagged and thus isolated in experiment are invaluable resources to increase our understanding not only of the fluid dynamics of QGP, but also of QCD as a whole.

Heavy Ion Collision (HIC) experiments have provided a wealth of data in the form of extensive measurements of heavy-flavor flow and nuclear modification factors, both at RHIC~\cite{PHENIX:2006iih,
PHENIX:2010xji,
STAR:2014wif,
STAR:2017kkh,
STAR:2018zdy,
Shi:2024eyk} and at the LHC
\cite{ALICE:2013olq,
ALICE:2014qvj,
ALICE:2017pbx,
CMS:2017qjw,
CMS:2017uuv,
ATLAS:2018hqe,
ALICE:2018gif,
ATLAS:2020yxw,
ALICE:2020iug,
CMS:2020bnz,
ATLAS:2021xtw,
ALICE:2021rxa,
ALICE:2021kfc,
CMS:2022sxl,
ALICE:2022tji,
CMS:2022vfn,
ALICE:2023gjj,
ALICE:2023hou,CMS:2024vip}, to answer how heavy quarks equilibrate with QGP.  From a theory perspective, variants of this question have been asked many times before~\cite{Svetitsky:1987gq,Braaten:1991jj,Braaten:1991we,vanHees:2004gq,vanHees:2005wb,Mustafa:2004dr,Peshier:2008bg}, perhaps most famously in the seminal work by Moore and Teaney~\cite{Moore:2004tg}, and steady progress has followed since then. The frameworks that are used in this endeavor go under the name of Heavy Quark Transport (for reviews see~\cite{Prino:2016cni,Xu:2018gux,Rapp:2018qla,Cao:2018ews,Dong:2019unq,Dong:2019byy,He:2022ywp}), which include Boltzmann transport models, Fokker-Planck/Langevin dynamics, and, more recently, Kolmogorov dynamics~\cite{Rajagopal:2026urq}. Each of these frameworks provides a concrete way to evolve the phase space distribution of heavy quarks $\mathscr{P}({\bf p},t)$, and thus of characterizing the microscopic momentum transfer probability distribution $P({\bf k};{\bf v}, \Delta t)$ that effects time evolution at a fundamental level
\begin{equation}
	\mathscr{P}({\bf p},t+\Delta t) = \int d^3{\bf k} \, P({\bf k}; {\bf v} ({\bf p} + {\bf k}), \Delta t) \, \mathscr{P}({\bf p} + {\bf k},t) \, , \label{eq:momentum-transfer}
\end{equation}
where ${\bf v}({\bf p})$ is the heavy quark velocity. All of these frameworks use a Markovian assumption in their derivation, justified by the separation of scales between the heavy quark mass $M$ and the QGP temperature $T$.

Both the Boltzmann transport and Kolmogorov dynamics frameworks provide a quite flexible description of $P({\bf k};{\bf v}, \Delta t)$. Relative to the quantum-mechanical starting point, the problem is simplified to studying a classical stochastic process. Fokker-Planck/Langevin dynamics goes one step further: In the heavy quark mass limit $M \to \infty$, any change in the quark's momentum will not change its velocity ${\bf v}$, even after a very long propagation time. Then, in the long time limit, the central limit theorem (CLT) takes over as the total momentum transfer becomes a sum of identically distributed random variables, and a Gaussian stochastic process emerges, characterized by i) its mean, which is a mean velocity-dependent momentum transfer rate (usually described in terms of a drag coefficient $\eta_D$ or, equivalently, a drag force $F_D = \eta_D p$), and ii) two variances, one for longitudinal momentum broadening and the other for transverse momentum broadening (characterized by momentum broadening coefficients $\kappa_L$ and $\kappa_T$, respectively). Crucially, this mean and these variances are determined by those of the microscopic momentum transfer distribution $P({\bf k};{\bf v}, \Delta t)$. Therefore, if all the approximations hold, this Langevin description is both predictive and directly informative of fundamental properties of QCD. That being said, whether the mass of $b$ or $c$ quarks is large enough for the CLT to completely take over the dynamics is an approximation whose validity has to be quantified --- for comparisons of Langevin/Fokker-Planck to Boltzmann transport simulations see~\cite{Das:2013kea}; for a comparison between Langevin/Fokker-Planck and Kolmogorov see~\cite{Rajagopal:2026urq}.

With the goal of establishing a sharp connection between experimental data and first-principles theory, in the past couple of decades a great deal of effort has gone into the theoretical characterization of the heavy quark diffusion coefficient~\cite{Moore:2004tg,Casalderrey-Solana:2006fio}, where perturbative QCD~\cite{Caron-Huot:2007rwy,Caron-Huot:2008dyw,Burnier:2010rp,Eller:2019spw,Scheihing-Hitschfeld:2022xqx,Scheihing-Hitschfeld:2023tuz,delaCruz:2024cix} and lattice QCD~\cite{Petreczky:2005nh,Caron-Huot:2009ncn,Banerjee:2011ra,Brambilla:2020siz,Altenkort:2020fgs,Brambilla:2022xbd,Altenkort:2023oms,Altenkort:2023eav} have worked in tandem all the way to having obtained a quite satisfying picture at all temperatures above the crossover temperature $T_c$~\cite{HotQCD:2025fbd}. Somewhat surprisingly, at low temperatures above the crossover temperature, lattice data is well described by the result for the diffusion coefficient in a different gauge theory: strongly coupled $\mathcal{N}=4$ supersymmetric Yang-Mills theory (SYM).

There is a limitation, however: due to the Euclidean nature of lattice calculations, the heavy quark coefficient so determined may then be used to describe the dynamics of \textit{non-relativistic} heavy quarks, but it does not determine the behaviour of a moving heavy quark beyond the non-relativistic regime (i.e., with relative velocity $v>0$ to the QGP). Moreover, for general $v > 0$, the momentum diffusion coefficients $\kappa_L$ and $\kappa_T$ are velocity-dependent, and explicit calculations both at weak~\cite{Moore:2004tg} and strong coupling~\cite{Herzog:2006gh,Gubser:2006bz,Gubser:2006nz} show that the Einstein relation $v \kappa_L = 2 T F_D$ --- the necessary condition for kinetic equilibration to take place in the Langevin approach to heavy particle dynamics --- does not hold.

This means that, when confronted with experimental data --- where most features of interest lie in the partially equilibrated, relativistic heavy quark regime --- the Langevin approach runs into a problem: it cannot self-consistently describe kinetic equilibration while using the microscopic QFT input provided by $\eta_D$, $\kappa_L$ and $\kappa_T$. This indicates, as illustrated later in Fig~\ref{fig:cltplot}, that the approximation where the CLT takes over the dynamics is insufficient to describe relativistic heavy quark equilibration.

Therefore, in order to capitalize on the tremendous progress made in determining the heavy quark diffusion coefficient from first principles and use this information to describe experimental data, these results must be embedded in a more general framework. The key ingredient that needs to be added is the inclusion of non-Gaussian features in the fluctuations of the momentum transfer~\cite{Rajagopal:2025rxr}. Both the Boltzmann transport formalism and Kolmogorov dynamics provide such a framework. However, as we will argue later on in this work, the connection between the transport coefficients and the ensuing dynamics is most transparent when a Kolmogorov equation is used --- even if, as we will also show, it may be equivalently cast as a Boltzmann equation (which is a natural way of rewriting the former when the medium is made up of weakly coupled quasiparticles).

The program of developing Kolmogorov dynamics to a point where it may be applicable to the dynamics of heavy quarks in realistic descriptions of heavy-ion collisions started with the calculation of the complete non-Gaussian momentum transfer probability distribution in strongly coupled $\mathcal{N}=4$ SYM~\cite{Rajagopal:2025ukd}. 
In this work we compute the complete non-Gaussian momentum transfer distribution for a heavy quark of arbitrary velocity in weakly coupled non-Abelian plasmas --- some aspects of which were already presented in~\cite{DuPlessis:2026pyr} --- detailing the calculation and conceptual steps needed to arrive at the result. This brings the state of the art of heavy quark transport in perturbative QCD (and related gauge theories) to the same level of completeness that is now available at strong coupling. Going beyond the drag and momentum broadening coefficients that have been the focus of previous work in perturbative gauge theories~\cite{Moore:2004tg,Chesler:2006gr}, we obtain the entire tower of cumulants of the distribution, show that the generating function has a bounded domain of analyticity, and identify the exponential tails and the velocity-dependent tail exponent that this bound implies.  
These tails are not incidental: they are essential for equilibration at finite velocity, and they are precisely what a Gaussian truncation discards.

The rest of our paper is organized as follows: in section~\ref{sec:setup} we discuss the conceptual framework in which our results are derived, starting from Heavy Quark Effective Theory (HQET), discussing how Kolmogorov dynamics naturally follows from it, and the structural constraints on how thermalization takes place. 
In section~\ref{sec:calculation} we describe the weakly coupled calculation of the evolution kernel that enters the Kolmogorov equation, discuss various cross-checks with results  from the literature, show that its domain of analyticity is bounded --- which fixes the exponential tails of the momentum transfer distribution --- and compare the results in different gauge theories of theoretical interest as well as with the strongly coupled results of~\cite{Rajagopal:2025ukd}.
In section~\ref{sec:Kolmogorov} we discuss the dynamics of heavy quarks in these weakly coupled gauge theories, and compare them with their strongly coupled counterpart in $\mathcal{N}=4$ SYM. We give our conclusions in section~\ref{sec:conclusions-outlook}, together with an outlook of the remaining steps ahead to arrive at a comprehensive understanding of heavy quark transport in heavy-ion collisions.

\section{Setup} \label{sec:setup}

In this section we give a self-contained exposition of all of the ingredients that go into the description of heavy quarks in a thermal environment. Parts of this discussion can be found in a more compact version in~\cite{Rajagopal:2025rxr,DuPlessis:2026pyr}. However, with our later calculation in mind, at various points we shall advance some of the qualitative and quantitative features that appear in the specific gauge theories of interest in this work, as well as aspects of their comparison to the strongly coupled $\mathcal{N}=4$ SYM theory results. We start in subsection~\ref{sec:HQET} by describing the kinematic aspects of our setup in the context of HQET. Then in subsection~\ref{sec:Kolmogorov-setup} we proceed to discuss how the HQET matrix elements enter the dynamics of individual heavy quarks and ensembles thereof in the presence of a thermal medium. We conclude this section by discussing the structural properties of the operator that generates the time evolution of an ensemble of heavy quarks in subsection~\ref{sec:equilibration-setup}.

\subsection{Kinematics: Heavy Quark Effective Theory} \label{sec:HQET}

Due to their large mass, the dynamics of heavy quarks may be studied using a simpler kinetic term than that of a 4-component Dirac spinor field. The HQET Lagrangian~\cite{Georgi:1990um}
\begin{equation}
    \mathcal{L}_{\rm HQET} = \bar{Q} (i v \cdot D) Q + \mathcal{L}_{\rm light} + \mathcal{O}(1/M) \, , \label{eq:HQET}
\end{equation}
captures all of the interactions between the heavy quark and the gauge field $A_\mu$ at leading order in $1/M$. In certain theories, such as $\mathcal{N}=4$ SYM, heavy quarks may also be naturally coupled to scalars. Here we only consider the case where the gauge field mediates all of the momentum transfer to and from the medium\footnote{This may seem at odds with the setup that~\cite{Casalderrey-Solana:2006fio,Casalderrey-Solana:2007ahi} used in $\mathcal{N}=4$ SYM theory to define and calculate the momentum diffusion coefficients at strong coupling, where the scalar fields of the theory enter explicitly. However, as was later clarified in~\cite{Alday:2007he,Polchinski:2011im} (see also~\cite{Casalderrey-Solana:2011dxg}), the difference between the relevant Wilson loop expectation values (which Eq.~\eqref{eq:Wloop-Wlines} later exemplifies) only enters at subleading order in the strong coupling expansion --- therefore the strongly coupled results~\cite{Herzog:2006gh,Gubser:2006bz,Casalderrey-Solana:2006fio,Gubser:2006nz,Casalderrey-Solana:2007ahi} apply whether the heavy quark is coupled to the scalars in $\mathcal{N}=4$ SYM or not. We note that the weakly coupled counterpart of the heavy quark diffusion coefficient with the scalars in $\mathcal{N}=4$ SYM does differ from the one without them~\cite{Chesler:2006gr,Caron-Huot:2008dyw}.\label{ft:scalarsN4}}. While formally $Q$ is a two-component spinor field, spin degrees of freedom only enter the dynamics at subleading order in the large mass expansion, and we shall thus omit them throughout. Due to the large mass, the 4-velocity $v^\mu =  (\gamma, \gamma {\bf v} )$ is treated as a constant in this Lagrangian [where we have introduced the Lorentz boost factor $\gamma = (1-v^2)^{-1/2}$], labeling the ``hard'' momentum of the heavy quark $p^\mu = Mv^\mu$.

It follows that the equation of motion for the heavy quark field $Q(x)$ is
\begin{equation}
    v^\mu \left( \partial_\mu - i g A_\mu \right) Q(x) = 0 \, ,
\end{equation}
and therefore the effect of the gauge field on heavy quark propagation is described by
\begin{equation}
    Q^l(x_f) = W_{[x_f,x_i]}^{lk} Q^k(x_i) \, ,
\end{equation}
where $l,k$ are color indices and the Wilson line $W$ is a path-ordered exponential
\begin{equation}
    W_{[x_f,x_i]}^{lk} =  \left[ {\rm P} \exp \left( i g \int_{x_i}^{x_f} dx^\mu A_\mu  \right) \right]^{lk} \, ,
\end{equation}
where the color indices stem from the fact that the gauge field $A_\mu = A^a_\mu T^a $ is matrix-valued. The matrices $T^a$ are the generators of the gauge group representation under which the heavy quarks transform. We take it to be the fundamental representation of SU($N_c$) throughout this work; generalizing our results to other representations is mostly straightforward. Furthermore, the initial and final positions $x_f$ and $x_i$ are not independent. Rather, they are the endpoints of a path determined by the velocity and the propagation time, parametrized by
\begin{equation}
    x^\mu(t) = x_i^\mu + (t - t_i) n^\mu \, ,
\end{equation}
where $n^\mu = (1,{\bf v})=v^\mu/\gamma$ is the tangent to the HQ trajectory with respect to coordinate time, $x_i^\mu = x^\mu(t_i) = (t_i, {\bf x}_i) $, and the final position is determined by $x_f^\mu = x^\mu(t_f) = (t_f, (t_f - t_i) {\bf v} + {\bf x}_i )$. 

It follows that the momentum change amplitude for the momentum eigenstates of the quark field is given by
\begin{equation}
    \langle {\bf p} -  {\bf k} |_l U_{\rm HQET}(t_f-t_i) | {\bf p} \rangle_k = \int \mathrm{d}^3{\bf x}_f \ e^{i  {\bf k} \cdot {\bf x}_f } W^{lk}_{[x_f,x_i]} \, , \label{eq:amplitude}
\end{equation}
where $U_{\rm HQET}$ is the time evolution operator that the Lagrangian in Eq.~\eqref{eq:HQET} defines, and $|{\bf p}\rangle_k$ is a heavy quark state with momentum ${\bf p}$ and color $k$. We stress that this momentum change amplitude has open color indices, and therefore is not a gauge-invariant quantity. To define a physical momentum change probability, initial and final states containing a heavy quark have to be specified.

These considerations lead to a momentum transfer probability for a heavy quark moving with velocity ${\bf v} = {\bf v}({\bf p}) = {\bf p}/\sqrt{{\bf p}^2 + M^2}$ through a thermal environment at temperature $T = 1/\beta$ during a time $\Delta t$ given by
\begin{align}
    P( {\bf k};{\bf v}) &=
    \frac{{\rm Tr} \! \left[ |{\bf p} - {\bf k}\rangle_{k} \tilde{\rho}^Q_{ki} \langle {\bf p}-{\bf k}|_i U_{\rm HQET}(\Delta t) |{\bf p}\rangle_j e^{-\beta H} \rho_{jl}^{Q} \langle {\bf p}|_l U_{\rm HQET}^\dagger(\Delta t) \right]}{{\rm Tr} \! \left[ U_{\rm HQET}(\Delta t) |{\bf p}\rangle_j e^{-\beta H} \rho_{jl}^{Q} \langle {\bf p}|_l U_{\rm HQET}^\dagger(\Delta t) \right]} \nonumber \\
    &= \int  \frac{\mathrm{d}^3{\bf L}}{(2\pi)^3} \, e^{-i  {\bf k} \cdot {\bf L}  } \langle W_{\bf v} \rangle ({\bf L};\Delta t) \, , \label{eq:P-from-W} 
\end{align}
where we have introduced the Wilson loop $ \langle W_{\bf v} \rangle ({\bf L};\Delta t)$, explicitly given by
\begin{align}
    \langle W_{\bf v} \rangle ({\bf L};\Delta t) = \frac{1}{Z} {\rm Tr}_{\mathcal{H}} \! \left[ W^{lk}_{[(0, {\bf L}),(\Delta t, {\bf v} \Delta t + {\bf L} )]} \tilde{\rho}^Q_{ki} W^{ij}_{[(\Delta t, {\bf v} \Delta t ),(0,{\bf 0})]} e^{-\beta H}  \rho_{jl}^Q \right] \, , \label{eq:Wloop-Wlines}
\end{align}
where $Z$ is the normalization factor in the denominator of the first line of  Eq.~\eqref{eq:P-from-W}, which enforces\footnote{Note that this prefactor is time-independent, because the evolution operators $U_{\rm HQET}$ can be moved around the trace and canceled out. In turn, also note that the final state projection $\tilde{\rho}^Q$ must be equal to the identity matrix when ${\bf L} = 0$ for these expressions to be consistent.} $\langle W_{\bf v} \rangle ({\bf L} = {\bf 0}) = 1$. The color matrices $\rho^Q$ and $\tilde{\rho}^Q$ encode the initial state preparation and final state projections, which are functionals of the gauge field and are also ${\bf L}$-dependent, but will turn out to be trivial at the order we are considering.
In this way the expectation value of the Wilson loop gives the \textit{characteristic function} of the momentum transfer distribution.


It proves useful to note that in the large time limit $\Delta t \gg \beta, |{\bf L}|$ the time dependence of the Wilson loop simplifies
\begin{equation}
    \langle W_{\bf v} \rangle ({\bf L};\Delta t) = \exp \left[ - (\Delta t) T S({\bf L}; {\bf v}) + \ldots \right] \, , \label{eq:W-S}
\end{equation}
due to time-translation invariance of the dynamics of the gauge theory. The separation between $\Delta t$ and all other scales characterizing the environment is what ultimately allows for an overall shift symmetry to emerge, with which the leading dependence on $\Delta t$ becomes extensive in the exponent. The emergence of this shift symmetry is equivalent to the emergence of a Markovian regime, as the time derivative of the Wilson loop~\eqref{eq:W-S} depends only on its value at the same time (see Eq.~\eqref{eq:W-time-derivative} later). 

The function $S({\bf L};{\bf v})$ is a central object in our subsequent calculation. It has a natural interpretation in terms of moments of the momentum change probability $P({\bf k};{\bf v})$, because
being the rescaled logarithm of the characteristic function of the distribution means that its analytic continuation $-S(\bf{L}=i\bf{x})$ is the (scaled) cumulant generating function of the momentum change distribution.
Furthermore, spacetime translations plus parity and time reversal imply that~\cite{Rajagopal:2025rxr}
\begin{equation}
    S({\bf L};{\bf v}) = S(-{\bf L}+i {\bf v}/T;{\bf v}) \, . \label{eq:S-analytic-property}
\end{equation}
We will refer to $S$ as the \textit{kernel}, due to its direct relation with the evolution kernel of the Kolmogorov equation~\eqref{eq:Kolmogorov-final}, which is the focus of the next subsection.

Through the Fourier transform in Eq.~\eqref{eq:Wloop-Wlines} it is easy to show that~\eqref{eq:S-analytic-property} is equivalent to the detailed balance requirement
\begin{equation}
    P({\bf{k}};{\bf{v}})e^{-{\bf{v}}\cdot{\bf{k}}/(2T)}=P(-{\bf{k}};{\bf{v}})e^{{\bf{v}}\cdot{\bf{k}}/(2T)},\label{eq:detailed-balance}
\end{equation}
demanding a balance between momentum gain and loss. 

\begin{figure}
    \centering
    \includegraphics[width=0.65\linewidth]{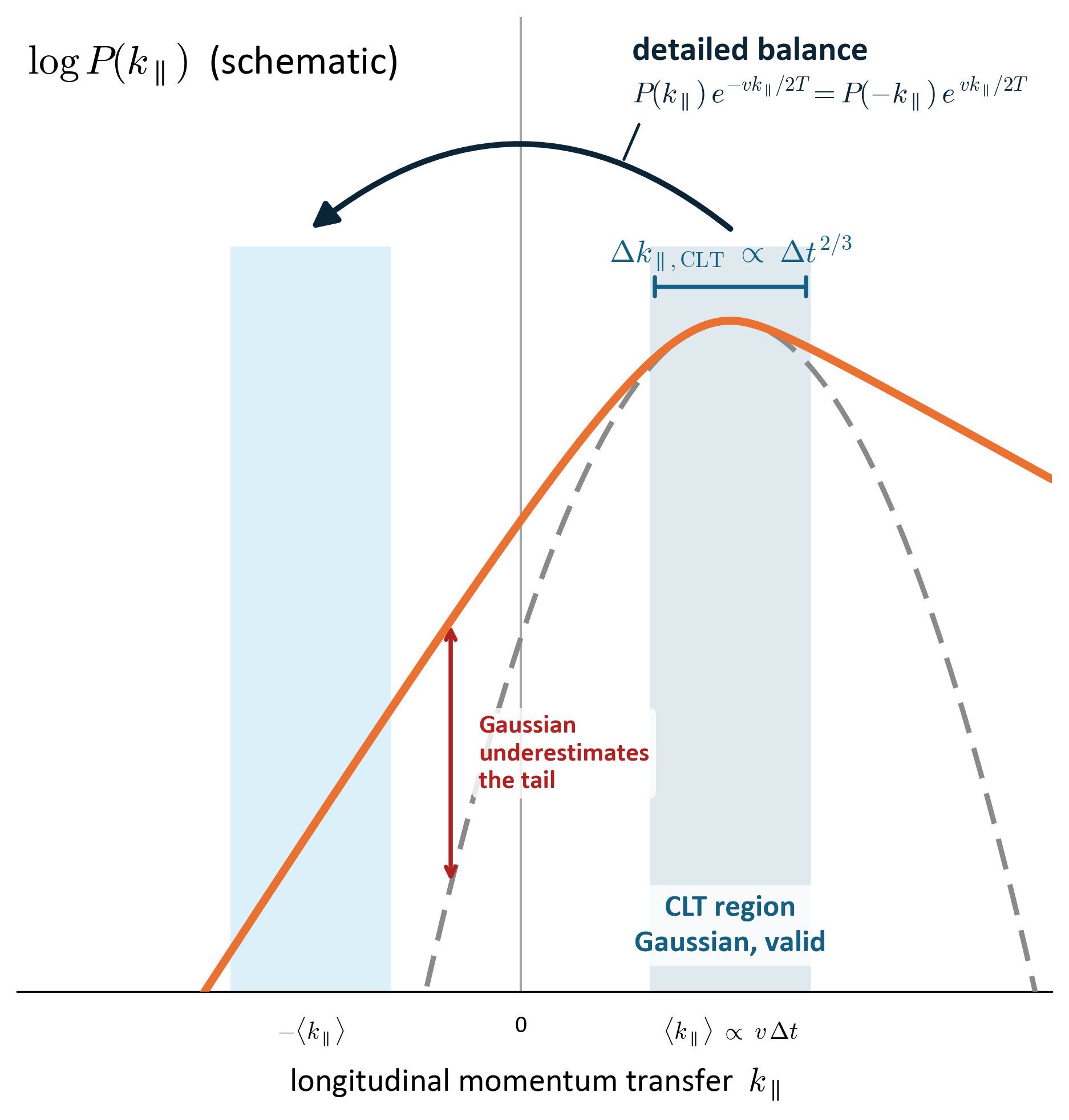}
    \caption{Schematic illustrating why the central limit theorem does not, by itself, imply the Einstein relation at nonzero velocity. Positive $k_\parallel$ denotes momentum transferred from the heavy quark to the medium. At late times the Gaussian/CLT regime is localized around the mean momentum loss $\langle k_\parallel\rangle\propto\Delta t$ and grows only sublinearly\protect\footref{ft:dt23} with $\Delta t$, whereas detailed balance relates this region to momenta near $-\langle k_\parallel\rangle$, parametrically outside the CLT regime. Equilibration therefore depends on the non-Gaussian tails of the momentum transfer distribution even when its central region is accurately Gaussian.}
    \label{fig:cltplot}
\end{figure}

\paragraph{A comment on the Central Limit Theorem} It is tempting to consider simplifying the description even further. For example, if one additionally assumes the exact Gaussian form
\begin{equation}
    -\ln P({\bf{k}},{\bf{v}})=\frac{(k_\parallel-\Delta t\, \eta_D(v) \, M \gamma v )^2}{2\Delta t\,\kappa_L(v)}+\frac{k_\perp^2}{2\Delta t\,\kappa_T(v)},
\end{equation}
then this detailed balance is equivalent to the Einstein relation $\kappa_L=2TE\eta_D$. In fact, one might be tempted to invoke the CLT to argue for the generic application of this relation at late times, regardless of the underlying dynamics. The CLT, however, controls the distribution near its mean. The window over which the Gaussian approximation remains accurate grows only sublinearly\footnote{For a cumulant generating function analytic near the origin, the first non-Gaussian correction to the exponent is $\mathfrak M_3(\delta k_\parallel)^3/[6(\Delta t T)^2\mathfrak M_2^3]$, where $\delta k_\parallel=k_\parallel-\langle k_\parallel\rangle$. Relative Gaussian accuracy therefore requires $|\delta k_\parallel|/T=o((\Delta t T)^{2/3})$ at fixed velocity and coupling. See~\cite[Corollary~4.12 and Section~4.4]{Fray2016}.\label{ft:dt23}} with $\Delta t$. At nonzero velocity, the mean momentum transfer grows linearly with $\Delta t$, so detailed balance relates this Gaussian region to momenta parametrically outside it, as illustrated in Fig.~\ref{fig:cltplot}. This shows that equilibration generally depends on details of the non-Gaussian tails of the momentum transfer distribution.

One can include higher order in $1/M$ corrections to Eq.~\eqref{eq:Wloop-Wlines} by adding such corrections up to the desired order to the time evolution operator $U_{\rm HQET}$
--- in principle, obtaining an exact result if all terms are restored. By comparing the leading-order term to the subleading ones one can estimate the regime of validity of calculations in this setup. More specifically, starting from a given initial state, corrections to the leading-order result can be quantified in time-dependent perturbation theory by calculating the matrix elements of the subleading terms in the HQET Lagrangian. These are parametrically controlled by the 4-momentum change in the transition amplitude through the magnitude of $\Delta t (k^2 - (k\cdot v)^2)/M$ and by the magnitude of gauge field fluctuations in the thermal state through $\Delta t T^2/M$. In what follows, we will always assume that $M$ is sufficiently large that these corrections are negligible as we set up the evolution equations. However, while we can always set up our calculation such that $T/M \ll 1$, the question of whether the support of the momentum change probability as a function of ${\bf k}$ satisfies this assumption can only be assessed after a calculation of the aforementioned subleading terms have been carried out, which is beyond the scope of this work.

\subsection{Dynamics: Evolution of a Heavy Quark Distribution} \label{sec:Kolmogorov-setup}

Equipped with a heavy quark momentum change probability $P({\bf k};{\bf v})$, given in terms of a Wilson loop by Eq.~\eqref{eq:P-from-W}, the evolution of the heavy quark phase (momentum) space distribution $\mathscr{P}({\bf p})$ follows from the convolution
\begin{align}
    \mathscr{P}({\bf p},t+\Delta t) &= \int d^3 {\bf k} P({\bf k};{\bf v}({\bf p} + {\bf k}),\Delta t) \mathscr{P}({\bf p} + {\bf k},t) \nonumber \\
    &= \int \frac{d^3 {\bf k} \, d^3 {\bf L}}{(2\pi)^3} \, e^{-i {\bf k} \cdot {\bf L}} \langle W_{{\bf v}({\bf p} + {\bf k})} \rangle({\bf L}; \Delta t)  \mathscr{P}({\bf p} + {\bf k},t) \, . \label{eq:convolution}
\end{align}

Taking a time step $\Delta t$ sufficiently long such that the Wilson loop behaves exponentially as in Eq.~\eqref{eq:W-S}, one has that
\begin{equation}
    \partial_{\Delta t} \langle W_{{\bf v}} \rangle({\bf L}; \Delta t) = - T S({\bf L};{\bf v}) \langle W_{{\bf v}} \rangle({\bf L}; \Delta t) \, . \label{eq:W-time-derivative}
\end{equation}

Then, taking a derivative of Eq.~\eqref{eq:convolution} with respect to $\Delta t$, one finds
\begin{align}
    \partial_{\Delta t} \mathscr{P}({\bf p},t+\Delta t) &= - T \int \frac{d^3 {\bf k} \, d^3 {\bf L}}{(2\pi)^3} \, e^{-i {\bf k} \cdot {\bf L}} S({\bf L};{\bf v}({\bf p} + {\bf k})) \langle W_{{\bf v}({\bf p} + {\bf k})} \rangle({\bf L}; \Delta t)  \mathscr{P}({\bf p} + {\bf k},t) \nonumber \\
    &= - T \int \frac{d^3 {\bf k} \, d^3 {\bf L}}{(2\pi)^3} \, e^{-i {\bf k} \cdot {\bf L}} S(i \cev{\partial}_{\bf k};{\bf v}({\bf p} + {\bf k})) \langle W_{{\bf v}({\bf p} + {\bf k})} \rangle({\bf L}; \Delta t)  \mathscr{P}({\bf p} + {\bf k},t) \nonumber \\
    &= - T \int \frac{d^3 {\bf k} \, d^3 {\bf L}}{(2\pi)^3} \, e^{-i {\bf k} \cdot {\bf L}} S(-i \vec{\partial}_{\bf k};{\bf v}({\bf p} + {\bf k})) \langle W_{{\bf v}({\bf p} + {\bf k})} \rangle({\bf L}; \Delta t)  \mathscr{P}({\bf p} + {\bf k},t) \nonumber \\
    &= - T \int \frac{d^3 {\bf k} \, d^3 {\bf L}}{(2\pi)^3} \, e^{-i {\bf k} \cdot {\bf L}} S(-i \vec{\partial}_{\bf p};{\bf v}({\bf p} + {\bf k})) \langle W_{{\bf v}({\bf p} + {\bf k})} \rangle({\bf L}; \Delta t)  \mathscr{P}({\bf p} + {\bf k},t)\, , \label{eq:Kolmogorov-derivation}
\end{align}
where from the third to the fourth line we have used the fact that the rest of the integrand only depends on the combination ${\bf p} + {\bf k}$, and therefore a derivative with respect to ${\bf p}$ is equivalent to a derivative with respect to ${\bf k}$.

At this point, the separation of scales provided by the large mass kicks in, which will allow us to arrive at our result for the evolution equation, which we give in Eq.~\eqref{eq:Kolmogorov-final}. There are (at least) two ways of arriving at the same conclusion:
\begin{enumerate}
    \item Keeping $\Delta t > 0$, provided the effected momentum change is sufficiently small compared to the mass of the heavy quark, we may approximate ${\bf v}({\bf p} + {\bf k})$ in the second argument of $S$ as ${\bf v}({\bf p})$. It then follows that
\begin{align}
    \partial_t \mathscr{P}({\bf p},t+\Delta t) &= - T S(-i \vec{\partial}_{\bf p};{\bf v}({\bf p})) \int \frac{d^3 {\bf k} \, d^3 {\bf L}}{(2\pi)^3} \, e^{-i {\bf k} \cdot {\bf L}}  \langle W_{{\bf v}({\bf p} + {\bf k})} \rangle({\bf L}; \Delta t)  \mathscr{P}({\bf p} + {\bf k},t) \nonumber \\
    &= - T S(-i \vec{\partial}_{\bf p};{\bf v}({\bf p})) \mathscr{P}({\bf p},t+\Delta t) \, , \label{eq:Kolmogorov-derivation-2}
\end{align}
where the derivatives $\partial_{\bf p}$ act on all of the momentum dependence of the right hand side of this expression, including ${\bf v}({\bf p})$. Relabeling $t + \Delta t \to t$, we obtain the evolution equation we are after.

It is worth taking a moment to emphasize why we did not set ${\bf k} \to 0$ in the velocity dependence of $\langle W_{\bf v} \rangle$ in Eq.~\eqref{eq:Kolmogorov-derivation-2}, and why we only did it for $S$ after the integration by parts had been carried out. The reason is, ultimately, that in this line of derivation we wish to preserve the cumulative effect of $\Delta t$ without approximations. One can also appreciate this by repeating the first line in Eq.~\eqref{eq:Kolmogorov-derivation} and considering the case where ${\bf v}$ does not change at all, in which case it follows without any approximations that
\begin{align}
     \partial_t \mathscr{P}_{{\rm const.}\, {\bf v}}({\bf p},t) &= - T \int \frac{d^3 {\bf k} \, d^3 {\bf L}}{(2\pi)^3} \, e^{-i {\bf k} \cdot {\bf L}} S({\bf L};{\bf v})   \mathscr{P}_{{\rm const.}\, {\bf v}}({\bf p} + {\bf k},t) \, .
\end{align}
Since this equation describes the action of a linear operator on the phase space distribution, the natural identification to make in the case where the velocity is a function of momentum is to use ${\bf v} = {\bf v}({\bf p} + {\bf k})$ under the integral sign, which is in fact equivalent to Eq.~\eqref{eq:Kolmogorov-derivation-2}.

\item We can instead assume that Eq.~\eqref{eq:W-S} holds for arbitrary $\Delta t$, keeping in mind that for the relevant time scales in which the distribution function evolves significantly (which, as we will see below, is a factor $\sim M/T$ longer than the time scale in which the environment evolves) this time should be seen as a small interval, we may simply set $\Delta t = 0$ in the last line of Eq.~\eqref{eq:Kolmogorov-derivation}. This amounts to neglecting all memory effects from Eq.~\eqref{eq:W-S}, keeping only the Markovian term that dominates the long-time dynamics. It then follows that
\begin{align}
    \partial_t \mathscr{P}({\bf p},t) &= - T \int \frac{d^3 {\bf k} \, d^3 {\bf L}}{(2\pi)^3} \, e^{-i {\bf k} \cdot {\bf L}} S(-i \vec{\partial}_{\bf p};{\bf v}({\bf p} + {\bf k}))  \mathscr{P}({\bf p} + {\bf k},t) \nonumber \\ 
     &= - T \int d^3 {\bf k} \, \delta^{(3)}({\bf k}) S(-i \vec{\partial}_{\bf p};{\bf v}({\bf p} + {\bf k}))  \mathscr{P}({\bf p} + {\bf k},t) \, , \label{eq:Kolmogorov-derivation-3}
\end{align}
where in the last line we have carried out the integral over ${\bf L}$.

\end{enumerate}

In summary, we have obtained, using two different methods, the result that the dynamics of the heavy quark phase space distribution is governed by
\begin{align}
    \partial_t \mathscr{P}({\bf p},t) &= - T S(-i \vec{\partial}_{\bf p};{\bf v}({\bf p}) ) \mathscr{P}({\bf p},t) \nonumber \\
    &= - T \int \frac{d^3 {\bf k} \, d^3 {\bf L}}{(2\pi)^3} \, e^{-i {\bf k} \cdot {\bf L}} S({\bf L};{\bf v}({\bf p} + {\bf k})) \mathscr{P}({\bf p} + {\bf k},t) \, . \label{eq:Kolmogorov-final}
\end{align}
Within the approximations we have made, the time evolution encoded in these equations is equivalent to that determined by the convolution in Eq.~\eqref{eq:convolution}. We emphasize that these equations apply for relativistic and non-relativistic heavy quarks at the same time, as their complete dispersion relation is being used to determine their velocity ${\bf v}({\bf p})$. This is possible because we have applied the HQET Lagrangian~\eqref{eq:HQET} sector by sector --- i.e., one velocity $v^\mu$ at a time, which can be \textit{any} timelike velocity --- and used the resulting amplitudes at a fixed ${\bf v}$ to define the dynamics of a statistical ensemble of heavy quarks. Only at this point different velocity ``bins'' get in contact with each other. 

Another appealing property of Eq.~\eqref{eq:Kolmogorov-final} is that heavy quark number conservation is explicit. This follows from the fact that the derivatives in the first argument of $S$ in the first line act on all of the remaining momentum dependence of this line, and that $S({\bf 0},{\bf v}) = 0$, which together imply that the right hand side of the equation is a total derivative. Then, integrating over ${\bf p}$ on both sides of this equation to obtain the variation of the total number of heavy quarks, one obtains that this number is conserved.

Note that the second line of Eq.~\eqref{eq:Kolmogorov-final} also closely resembles the structure of a collision kernel. In fact, if the thermal environment in which the heavy quark propagates is made up of quasiparticles, it is natural that such a structure would appear. More specifically, we will see in our calculation in section~\ref{sec:calculation} that the way in which $S({\bf L};{\bf v})$ appears will be naturally describable in terms of a collision rate $\Gamma({\bf q};{\bf v})$, appearing as
\begin{equation}
    - T S({\bf L};{\bf v}) = \int d^3 {\bf q} \left[ e^{i {\bf q} \cdot {\bf L} } -1 \right] \Gamma({\bf q};{\bf v}) \, , \label{eq:Gamma-def}
\end{equation}
in which case the result of evaluating the ${\bf L}$ argument of $S$ with the momentum derivative in the evolution equation~\eqref{eq:Kolmogorov-final} automatically becomes
\begin{equation}
    \partial_t \mathscr{P}({\bf p},t) = \int d^3 {\bf q} \Big[ \Gamma({\bf q}; {\bf v}({\bf p} + {\bf q}) ) \mathscr{P}({\bf p}+ {\bf q},t) - \Gamma({\bf q}; {\bf v}({\bf p}) ) \mathscr{P}({\bf p},t) \Big] \label{eq:Boltzmann}
\end{equation}
a form which is familiar from Boltzmann transport studies~\cite{Svetitsky:1987gq,Peshier:2008bg,Das:2013kea,Prino:2016cni,Rapp:2018qla}. (There is also extensive discussion in the literature regarding kinetic descriptions of hard massless partons that are weakly coupled to the QGP, see for example~\cite{Blaizot:2012fh,Blaizot:2013vha,Barata:2020rdn,Boguslavski:2024jwr,Soudi:2026fls}.)

There is one further simplification that appears in the dynamics at leading order in $1/M$. Concretely, because we expect the late-time equilibrium state should be
\begin{equation}
    \mathscr{P}_{\rm eq} = \exp \left( - \sqrt{{\bf p}^2 + M^2}/T \right) \, , \label{eq:P-equilibrium}
\end{equation}
it proves convenient to reorganize the dynamics in terms of an expansion in $T/M$ as
\begin{equation}
    \mathscr{P}({\bf u},\tau) = \exp \left[  \sum_{a=-1}^\infty \left( \frac{T}{M} \right)^a f_a({\bf u},\tau) \right] \, ,
\end{equation}
where ${\bf u} = {\bf p}/M$ and $\tau = t T^2/M$. At zeroth order in $T/M$, the dynamics then simplifies to~\cite{Rajagopal:2025ukd,Rajagopal:2025rxr}
\begin{align}
    \partial_\tau f_{-1}({\bf u},\tau) &= - S(-i T^{-1} \partial_{\bf u}f_{-1} ; {\bf v}({\bf u}) ) \nonumber \\
    &\equiv - K( T^{-1} \partial_{\bf u}f_{-1} ; {\bf v}({\bf u})) \, . \label{eq:K-def}
\end{align}

Therefore, at leading order in the $1/M$ expansion, the dynamics is completely characterized by the function $K({\bf x};{\bf v})$, which is given by the logarithm of the Wilson loop in its Markovian regime~\eqref{eq:W-S} analytically continued to imaginary spatial separations.

It is worth noting that Eq.~\eqref{eq:K-def} no longer looks like a kinetic theory. It is, in fact, highly nonlinear. However, it makes no reference to the mass of the heavy quark, and as such, provides a clean way to examine the dynamics at leading order in $1/M$ with no contamination from higher order terms. This comes at the cost that information contained at subleading order in $1/M$, such as the total number of heavy quarks, is not automatically captured by the dynamics. As such, reconstructing the full phase space distribution from $f_{-1}({\bf u})$ in a way that preserves the total particle number requires an additional prescription. In practice, the simplest procedure is to reconstruct
\begin{equation}
    \mathscr{P}_{\rm LO}({\bf u},\tau) = \frac{\int d^3 {\bf w} e^{M f_{-1}({\bf w},0)/T} }{\int d^3 {\bf w} e^{M f_{-1}({\bf w},\tau)/T}  } \exp \left( \frac{M}{T} f_{-1}({\bf u},\tau) \right)
\end{equation}
such that the total number of particles is automatically preserved. Note that this normalization prefactor only enters the dynamics at subleading order in $T/M$, because the value of $f_{-1}$ at its local maxima (or minima) is preserved by the dynamics in Eq.~\eqref{eq:K-def} --- and as such the exponential enhancement in the numerator and denominator of this prefactor are exactly balanced for all $\tau$.

\subsection{Symmetries: Equilibrium Condition and Structural Properties}
\label{sec:equilibration-setup}

Having established that the dynamics is completely determined by the function $K({\bf x},{\bf v})$ (which we shall henceforth refer to as the ``evolution kernel''), it is desirable to characterize its structural properties --- and as such, inform our expectations for what properties resulting from the calculation of this object will be generic, and which ones will be theory-specific.

\paragraph{Equilibrium Condition.}

Ref.~\cite{Rajagopal:2025rxr} showed that regardless of the details of the theory, this function satisfies
\begin{equation}
    K({\bf x};{\bf v}) = K(-{\bf x} - {\bf v}/T;{\bf v}) \, , \label{eq:K-analytic-property}
\end{equation}
as a consequence of a generalized KMS relation between line operators, relying only on time-reversal, parity and spacetime translation symmetries of the gauge theory that describes the environment, which we gave before in terms of $S$ in Eq.~\eqref{eq:S-analytic-property}. In other words, the evolution kernel is an even function around ${\bf x} = - {\bf v}/(2T)$ as a function of ${\bf x}$. Furthermore, unitarity implies that $K({\bf 0};{\bf v}) = 0$ --- because when the Wilson lines in Eq.~\eqref{eq:Wloop-Wlines} are at the same spatial position they define unitary operators that are the exact inverses of each other.
That is to say, regardless of the gauge theory to which heavy quarks are coupled, $K({\bf x};{\bf v})$ is an even function of ${\bf y} = {\bf x} + {\bf v}/(2T)$ and vanishes at ${\bf x} = 0$ and ${\bf x} = - {\bf v}/T$. The latter implies that $f_{-1} = -E({\bf p})/M$, corresponding to a Boltzmann distribution as in Eq.~\eqref{eq:P-equilibrium}, is indeed an equilibrium solution. 

\paragraph{Relating $K({\bf x};{\bf v})$ and the momentum transfer probability.}

More insight into the shape of $K$ can be gained by examining how it is related to the momentum change probability $P({\bf k};{\bf v})$. In the late-time limit defined by the emergence of~\eqref{eq:W-S}, we have that\footnote{ Note that we are using the convention where ${\bf k}$ is the momentum transferred from the heavy quark to the medium. Positive $k_\parallel$ therefore corresponds to momentum lost by the heavy quark.}
\begin{equation}
    P({\bf k};{\bf v}) = \frac{1}{(2\pi)^3} \int d^3 {\bf L} e^{-i {\bf k} \cdot {\bf L} } e^{- t T S({\bf L};{\bf v}) } \, . \label{eq:P-Markovian}
\end{equation}
For a heavy quark moving with an approximately constant velocity through the thermal environment, the total momentum change is naturally described in terms of a momentum change rate ${\bf C} 
= {\bf k}/(t T)$, where we have added a factor of $T$ as a reference time scale. In terms of this rate, the probability reads
\begin{equation}
    P({\bf C};{\bf v}) = \frac{1}{(2\pi)^3} \int d^3 {\bf L} \exp\left[  - t T \left( i {\bf C} \cdot {\bf L} + S({\bf L};{\bf v}) \right)\right] \, . \label{eq:P-of-S-Fourier}
\end{equation}

In the long time limit $t T \to \infty$, the dominant contribution to this integral comes from the extrema of the function in the exponent --- i.e., from the values of ${\bf L}$ such that
\begin{equation}
    i {\bf C} + \frac{\partial S}{\partial {\bf L}} = 0 \, , \label{eq:L-of-C}
\end{equation}
leading to a momentum change probability given by
\begin{equation}
    P({\bf C};{\bf v}) = \exp \left( - t T \tilde{S}({\bf C};{\bf v}) + \mathcal{O}(t^0) \right) \, ,\label{eq:legendresaddledistribution}
\end{equation}
where
\begin{equation}
    \tilde{S}({\bf C};{\bf v}) = \left[ S({\bf L};{\bf v}) - {\bf L} \cdot \frac{\partial S}{\partial {\bf L}}  \right]_{{\bf L}={\bf L}({\bf C};{\bf v})} \, , \label{eq:tildeS-from-S}
\end{equation}
where ${\bf L}({\bf C};{\bf v})$ is determined by the solution to Eq.~\eqref{eq:L-of-C}. Furthermore, if one takes Eq.~\eqref{eq:P-Markovian} as an exact result --- i.e., if one assumes that the dynamics is exactly Markovian, then the subleading contributions in the expansion in inverse powers of $t$ are also calculable. For example, the factor controlled by the $t^0$ piece can be obtained by calculating the Hessian matrix of $S$ and using the well-known saddle-point approximation formulas. However, since the precise form of the right hand side in Eq.~\eqref{eq:P-Markovian} is itself a consequence of the large time limit, here we restrict ourselves to the leading behavior, which is the piece that is extensive in $t$.

Conversely, from the inverse relation
\begin{equation}
     \langle W_{\bf v} \rangle({\bf L}) = \int d^3 {\bf k} \, e^{i {\bf k}\cdot {\bf L}} P({\bf k};{\bf v}) = \int d^3 {\bf C} \exp\left[  - t T \left( -i {\bf C} \cdot {\bf L} + \tilde{S}({\bf C};{\bf v}) \right)\right]   \, , \label{eq:W-from-P}
\end{equation}
one concludes that the late-time exponential behavior of the Wilson loop $S$ is determined by that of the momentum change probability $\tilde{S}$ as
\begin{equation}
    S({\bf L};{\bf v}) =  \left[ \tilde{S}({\bf C};{\bf v}) - {\bf C} \cdot \frac{\partial \tilde{S}}{\partial {\bf C}}  \right]_{{\bf C}={\bf C}({\bf L};{\bf v})}  \, ,  \label{eq:S-from-tildeS}
\end{equation}
with $C({\bf L})$ determined by
\begin{equation}
    -i {\bf L} + \frac{\partial \tilde{S}}{\partial {\bf C}} = 0 \, .  \label{eq:C-of-L}
\end{equation}

It is worth noting that from Eq.~\eqref{eq:W-from-P} it follows that the cumulants of the momentum transfer distribution can be directly computed by taking derivatives of $S({\bf L})$:
\begin{equation}
    \langle k_{i_1} \cdots k_{i_n} \rangle_c = - t T \left. \left(-i \frac{\partial}{\partial L_{i_1}}  \right) \cdots \left(-i \frac{\partial}{\partial L_{i_n}}  \right) S({\bf L};{\bf v}) \right|_{{\bf L} = {\bf 0}} \, . \label{eq:cumulants}
\end{equation}

Noting that $K({\bf x};{\bf v}) = S(-i {\bf x}; {\bf v})$, Eq.~\eqref{eq:S-from-tildeS} yields a powerful conclusion: because $\tilde{S}$ is a real-valued function for real momentum transfer ${\bf C}$, it follows that solutions to Eq.~\eqref{eq:L-of-C} can be found for real ${\bf x} = i {\bf L}$ as
\begin{equation}
    {\bf x} = \left. \frac{\partial \tilde{S}}{\partial {\bf C}} \right|_{{\bf C} = {\bf C}({\bf x};{\bf v}) } \, , \label{eq:C-of-x}
\end{equation}
which in turn determine
\begin{equation}
    K({\bf x};{\bf v}) = \left[ \tilde{S}({\bf C};{\bf v}) - {\bf C} \cdot \frac{\partial \tilde{S}}{\partial {\bf C}}  \right]_{{\bf C}={\bf C}({\bf x};{\bf v})} \, . \label{eq:K-from-tildeS}
\end{equation}

From these seemingly innocuous manipulations, it follows that $K({\bf x};{\bf v})$ and $\tilde{S}({\bf C};{\bf v})$ are Legendre transforms of each other. Then, because $P$ is a probability distribution,
\begin{enumerate}
    \item $K$ is a real function for real ${\bf x}$ that admit a solution for ${\bf C}$ in Eq.~\eqref{eq:C-of-x},
    \item The minimum of $\tilde{S}$, i.e., the most likely momentum transfer rate $\langle {\bf C}({\bf v})\rangle$ (denoted as an expectation value because in the large time limit the mean and the most likely value coincide), corresponds to ${\bf x} = 0$. That is to say, ${\bf C}({\bf x} = {\bf 0};{\bf v}) = \langle {\bf C}({\bf v})\rangle$.
    \item By virtue of the symmetry property~\eqref{eq:K-analytic-property} together with the reciprocal Legendre transform 
    \begin{equation}
        \tilde{S}({\bf C};{\bf v}) = \left[ K({\bf x};{\bf v}) - {\bf x} \cdot \frac{\partial K}{\partial {\bf x}} \right]_{{\bf x} = {\bf x}({\bf C})} \, ,
    \end{equation}
    where ${\bf x}({\bf C})$ is determined by ${\bf C} + \frac{\partial K}{\partial {\bf x}} = 0$, ${\bf x} = - {\bf v}/T$ corresponds to the additive inverse of the most likely momentum transfer rate, ${\bf C}({\bf x} = -{\bf v}/T;{\bf v}) = -\langle {\bf C}({\bf v})\rangle$. 
    \item The additive inverse of the evolution kernel, $-K$, is the \textit{scaled cumulant generating function} of $P$.
\end{enumerate}

The last of these can also be interpreted as a manifestation of the fact that $\tilde{S}$ may always be decomposed as~\cite{Rajagopal:2025rxr}
\begin{equation}
    \tilde{S}({\bf C};{\bf v}) = \tilde{S}_e({\bf C};{\bf v}) - \frac{{\bf v} \cdot {\bf C}}{2T} \, .
\end{equation}
where $S_e$ is an even function of ${\bf C}$.

It is interesting to note that many of the above statements can be made in the mathematical framework of large deviation theory~\cite{TOUCHETTE20091}. We discuss some of these aspects in Appendix~\ref{app:large-deviation}.

\paragraph{Concavity of $K({\bf x};{\bf v})$.}

Note that the map ${\bf C}({\bf x})$ defined by Eq.~\eqref{eq:C-of-x} requires $\tilde{S}$ to be a strictly convex function of ${\bf C}$ --- otherwise, there can be more than one solution to this equation. Note that $\tilde{S}$ being a convex function is equivalent to $K$ being a concave function, by virtue of the properties of the Legendre transform. 

This actually also follows from unitarity, or rather, more precisely, from the fact that $P({\bf k};{\bf v})$ is a probability distribution. To begin, note that $P$ is real, and therefore the complex conjugate of the Wilson loop is equal to itself with ${\bf L} \to -{\bf L}$. Therefore,
\begin{equation}
    S^*({\bf L};{\bf v}) = S(-{\bf L};{\bf v}) \, , \label{eq:S-conj}
\end{equation}
where the star denotes complex conjugation. Consequently, we see that ${\rm Re}\{S\}$ is symmetric under ${\bf L} \to -{\bf L}$, and therefore it has a maximum over the real ${\bf L}$ domain at ${\bf L} = 0$.
Indeed, we know that ${\rm Re}\{S\} \geq 0$ for real ${\bf L}$ by unitarity, because the correlation function of two unitary operators cannot be bigger in magnitude than its value in the configuration in which these unitary operators are the exact inverses of each other --- as is the case when ${\bf L} = 0$. 

Viewed as a function of complex separation between the lines --- for definiteness ${\bf L} \to {\bf L} - i{\bf x}$, consistent with the rest of our notation --- there is in fact a stronger statement one can make. As we show in Appendix~\ref{app:proof-maximum}, the Hessian matrix with respect to real separation variations
\begin{equation}
    \frac{\partial^2 {\rm Re}\{S\} }{\partial L_i \partial L_j}({\bf L}={\bf 0},{\bf x}) \succeq 0
\end{equation}
is a positive definite matrix for all imaginary separations ${\bf x}$.

Then, because $S$ is an analytic function, it follows that ${\rm Re}\{S\}$ is a harmonic function as a consequence of the Cauchy-Riemann equations for the real and imaginary parts of $S$, and therefore the Hessian matrix with respect to imaginary separation variations
\begin{equation}
    \frac{\partial^2 {\rm Re}\{S\} }{\partial x_i \partial x_j}({\bf L}={\bf 0},{\bf x}) = \frac{\partial^2 K }{\partial x_i \partial x_j}({\bf x};{\bf v})\preceq 0
\end{equation}
is negative definite at all imaginary separations ${\bf x}$. It follows that $K({\bf x};{\bf v})$ is a concave function of ${\bf x}$.

This property is verified by the strongly coupled calculation in~\cite{Rajagopal:2025ukd}.
And, indeed, we shall see in our calculation in Section~\ref{sec:calculation} that $K$ is a concave function 
throughout the (real) convergence domain of the fixed-order integral representation. 

In fact, in weakly coupled setups it is useful to study $S$ in terms of its Fourier transform --- closely related to $\Gamma({\bf q};{\bf v})$ discussed earlier, introduced in Eq.~\eqref{eq:Gamma-def} --- which we will introduce in terms of a function that we denote by $G({\bf q};{\bf v})$ as
\begin{equation}
    -T S({\bf L};{\bf v}) = \int d^3{\bf q} \, e^{i{\bf q} \cdot {\bf L}} G({\bf q};{\bf v}) \, ,
\end{equation}
which, in light of Eq.~\eqref{eq:S-conj}, is a real function. Furthermore, since we know that $S({\bf 0};{\bf v})=0$, there is no change if we instead write
\begin{equation}
    -T S({\bf L};{\bf v}) = \int d^3{\bf q} \left[e^{i{\bf q} \cdot {\bf L}} -1 \right] G({\bf q};{\bf v}) \, ,
\end{equation}
meaning that $G$ and the more often quoted quantity in the perturbative literature $\Gamma$ are equal up to a delta function at zero momentum $G({\bf q};{\bf v}) - \Gamma({\bf q};{\bf v}) \propto \delta({\bf q})$.
Whenever $G$ can be identified with a positive interaction rate $\Gamma$, concavity of $K$ follows from its relation with $S$ and Eq.~\eqref{eq:Gamma-def}
\begin{equation}
    -T \frac{\partial^2 K}{\partial x_i \partial x_j} = \int d^3 {\bf q} \, q_i q_j \, e^{{\bf q} \cdot {\bf x}} \, \Gamma({\bf q};{\bf v}) \succeq 0 \, ,
\end{equation}
because $\Gamma({\bf q};{\bf v}) \geq 0$.

\paragraph{Domain of $K({\bf x};{\bf v})$ and Large-Order Cumulants.}

\begin{figure}[t]
    \centering
    \includegraphics[width=\linewidth]{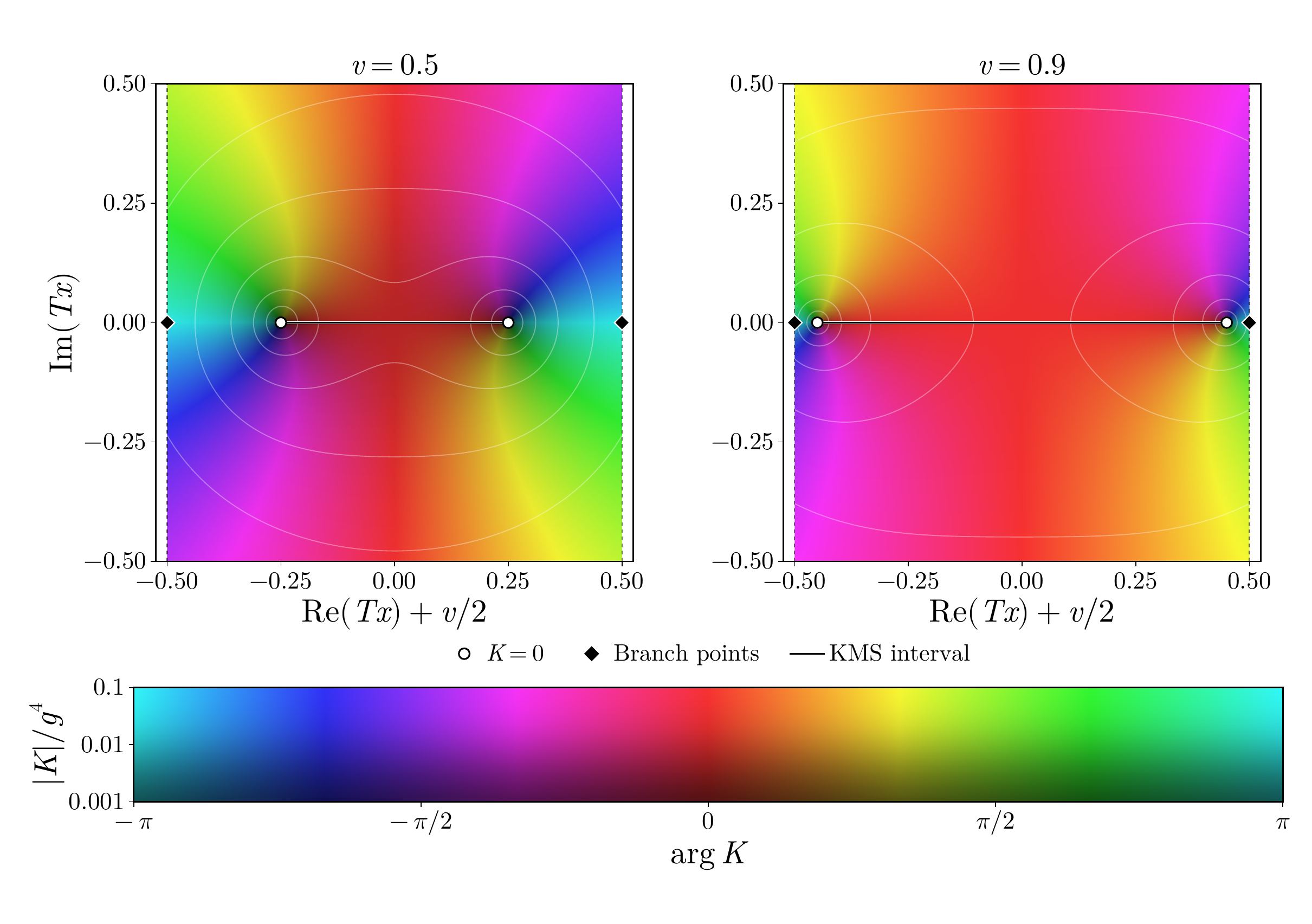}
    \caption{Longitudinal evolution kernel $K(x;v)$ in QCD at $g=0.1$, for $v=0.5$ (left) and $v=0.9$ (right). Hue denotes $\arg K$ and brightness denotes $|K|/g^4$. The plot is restricted to the convergence strip of the fixed-order integral, bounded by $\operatorname{Re}(Tx)+v/2=\pm1/2$ (dashed lines). The analytic continuation beyond this strip is discussed in App.~\ref{app:moment-resummation}. White circles mark the zeros at $x=0$ and $x=-v/T$, joined by the KMS interval; black diamonds mark the first branch points. Thin white contours are spaced by factors of two in $|K|/g^4$, starting at $0.0025$.}
    \label{fig:complexKplot}
\end{figure}

Recently, it was shown that~\cite{Rajagopal:2026urq} the domain of the evolution kernel $K({\bf x};{\bf v})$ at strong coupling is bounded, but strictly greater than the strip $-v/T < {\bf x} \cdot \hat{\bf v} < 0$. This can be understood by examining the radius of convergence of the series that characterizes $K({\bf x};{\bf v})$ in terms of the longitudinal cumulants of the momentum change distribution,
\begin{equation}
    K(x;v) = - \sum_{n=1}^\infty \frac{x^n}{n!} \mathfrak{M}_n(v) \, ,
\end{equation}
where we have introduced $x \equiv {\bf x} \cdot \hat{\bf v}$.

Figure~\ref{fig:complexKplot} illustrates these properties for the weakly coupled longitudinal evolution kernel that we shall calculate in Sec.~\ref{sec:calculation}, evaluated at complex $x$. The zeros at $x=0$ and $x=-v/T$ and the two branch points are paired by the equilibration condition $K(x) = K(-x-v/T)$. The figure is restricted to the convergence strip, stopping at the first branch points on the real axis. In Sec.~\ref{sec:cumulants} we determine their locations and show how they fix the asymmetric exponential tails of the momentum-transfer distribution.

In strongly coupled $\mathcal{N}=4$ SYM, the coefficients $\mathfrak{M}_n(v)$ can be calculated analytically for any given $n$ (although we have not found a closed form expression for generic $n$). At large $n$, they scale as
\begin{align}
    \mathfrak{M}_n(v=0) \sim 
    \left( \frac{16 \sqrt{2} \, \Gamma(7/4)^2 }{9\sqrt{\pi}} \right)^n T^{n+1} n! \, , & & \mathfrak{M}_n(v \to 1) \sim 
    \left( \frac{8 \sqrt{2} \Gamma(7/4)^2}{3\sqrt{\pi}} \right)^n T^{n+1} n! \gamma^{3n/2-1/2} \, ,
\end{align}
defining a convergence radius $ R(v=0) = 9\sqrt{\pi/2}/(16\Gamma(7/4)^2 T)$ at zero velocity, and $R(v\to 1) = 2R(v=0)/(3\gamma^{3/2})$ as $\gamma \to \infty$. The exact domain of the $K(x;v)$ function is known at all $v$ and one can verify that these convergence radii indeed match the extent of the domain around $x=0$, given by
\begin{equation}
    R(v) = \frac{9 \sqrt{\pi} \left[ 4 (1-v^2)^{3/4} - v\, {\rm B} \left( 1 - v^2, \frac34, \frac12 \right) \right] }{64 \sqrt{2} \, \Gamma(7/4)^2 \, T } \, .
\end{equation}

\begin{figure}
    \centering
    \includegraphics[width=0.9\linewidth]{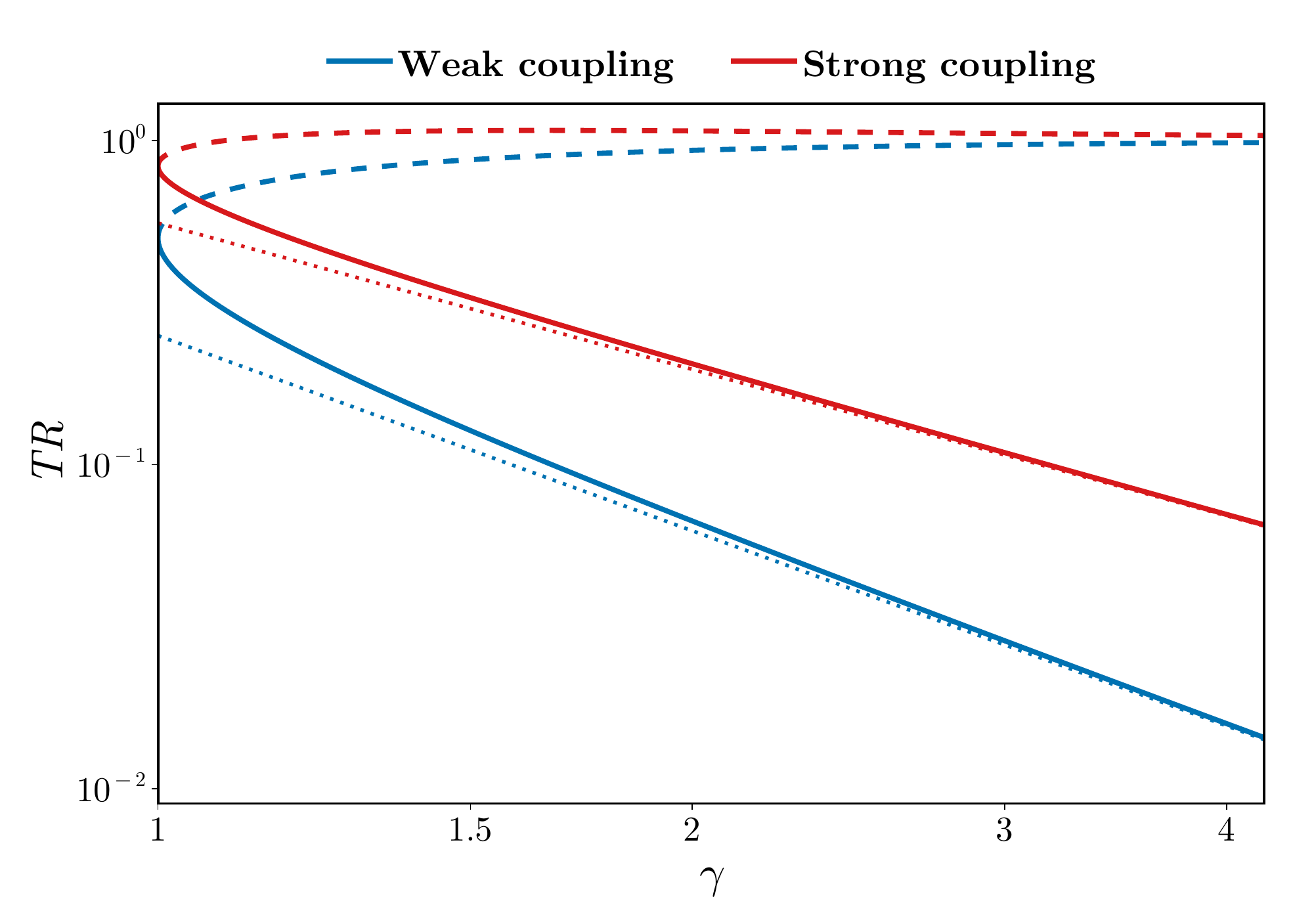}
    \caption{Longitudinal tail exponents as a function of boost factor $\gamma$ at weak and strong coupling. Solid curves show the positive-$k_L$ (momentum-loss) exponent $TR_+$, while dashed curves show the negative-$k_L$ (momentum-gain) exponent $TR_-=TR_++v$, as required by detailed balance. At weak coupling (blue) $TR_+=(1-v)/2\to1/(4\gamma^2)$ (dotted line), whereas the strong-coupling (red) positive-tail exponent decreases more slowly as $TR_+\to1/(r_0\gamma^\frac32)$. Consequently the weak-coupling momentum-loss tail is parametrically broader at large boost, while in both theories the momentum-gain exponent remains of order unity. The exponents are independent of the coupling at the leading order in each regime.}
    \label{fig:radconv}
\end{figure}

Even more interestingly, we shall find that the coefficients $\mathfrak{M}_n$ in weakly coupled gauge theories have a large $n$ scaling given by
\begin{align}
    \mathfrak{M}_n \sim \frac{n!}{n^2}
 \left(\frac{2T}{1-v}\right)^n
\end{align}
which signals that the finite domain of the evolution kernel $K$ is not an accidental feature of the strongly coupled calculation --- it might in fact be generic. We compare these in Figure~\ref{fig:radconv}. We derive different asymptotics of the weakly coupled moments in App.~\ref{app:large-order-arbitrary-v}. This radius of convergence $R(v)$ will end up being the \textit{tail exponent}~\cite{DuPlessis:2026pyr}, setting the exponential decay in the tails of the non-Gaussian momentum transfer distribution tails.

These structural similarities between the weakly and strongly coupled limits of the same theory ($\mathcal{N}=4$ SYM) give us confidence that the applicability of the perturbative results in QCD may be broader than one might have guessed. In the rest of this work, we shall see via explicit calculation how these and other quantitative features of the momentum transfer dynamics emerge, while at the same time keeping an eye on the strongly coupled result --- with which the comparison of our results may reveal new, yet to be established, universal features of the momentum transfer mechanism in gauge theories at finite temperature.

\section{Evolution Kernel in Weakly Coupled Plasmas} \label{sec:calculation}

We now turn to the task of calculating the expectation value of the Wilson loop~\eqref{eq:Wloop-Wlines} in weakly coupled gauge theories, and more specifically, of its behavior at late times~\eqref{eq:W-S}.
Having a small coupling $g \ll 1$ allows the use of perturbative methods. Our sign conventions and notation for the various quantities associated with the SU$(N_c)$ algebra follows those of~\cite{Schwartz:2014sze}.

We start this section by discussing the structure of the calculation in subsection~\ref{sec:calculation-structure} --- specifically, how the contributing Wick contractions make their appearances and how these should be grouped together to extract the late-time behavior of the Wilson loop. In a nutshell, following the analysis of~\cite{DEramo:2012uzl}, the expectation value of the Wilson loop can be reconstructed by exponentiating the fully connected Feynman diagrams that appear from the Wick contractions, of which the $t$-extensive contribution can be mapped directly to the exponent of the Wilson loop in the late-time limit. Keeping in mind that the definition of the Wilson loop is manifestly gauge-invariant (we will discuss this in the calculation as well), whenever we give explicit expressions for individual diagram contributions to the Wilson loop, we shall do so in Feynman gauge.

The logic of the rest of this section runs as follows. Section~\ref{sec:matching} addresses the one place where the perturbative expansion of Sec.~\ref{sec:calculation-structure} is not sufficient on its own: the soft sector, which requires HTL resummation and a matching prescription to avoid double counting. The matched kernel that results, given in Sec.~\ref{sec:result-analytic}, is correct at $\mathcal{O}(g^4)$, but it is contaminated by higher-order terms which the HTL resummation drags along and which render every cumulant beyond the second ultraviolet divergent. In Sec.~\ref{sec:fixed-order} we show that removing this unphysical contamination renders the entire tower of cumulants finite. 
The resulting fixed-order integral converges beyond the KMS strip (the interval $-v/T < x < 0$) and admits a Taylor expansion in its finite cumulants. Section~\ref{sec:cumulants} uses this expansion twice: at low order, to test whether the Einstein relation holds, and at large order, to determine the domain of analyticity of $K$ and hence the shape of the tails of the momentum transfer distribution. Section~\ref{sec:result-numerics} then evaluates the resulting distributions in pure Yang-Mills theory, massless QCD and $\mathcal{N}=4$ SYM, and compares them against the strongly coupled $\mathcal{N}=4$ SYM result of~\cite{Rajagopal:2025ukd}. Technical details of the analytic calculation are collected in Appendix~\ref{app:calculation-evol}, while the representation used for the numerical evaluation of the strict fixed-order kernel is given in Appendix~\ref{app:coord_transforms}. Some limiting forms of the cumulants and a closer look at the domain of analyticity of the kernel are presented in Appendix~\ref{app:momentasymptotics}.

Throughout this section we present results for three weakly coupled theories:
\begin{enumerate}
    \item $SU(3)$ Yang--Mills: $N_c=3$, $N_X=N_f=0$;
    \item massless QCD: $N_c=3$, $N_f=3$ with $T_R=\tfrac12$, $N_X=0$;
    \item $\mathcal{N}=4$ SYM: $N_c$ is free (see below), $N_X=6$ real adjoint scalars, and $N_f = 2$ Dirac adjoint fermions\footnote{Strictly speaking, $4$ adjoint Weyl fermions, but for our present purposes it makes no difference to describe them in terms of $2$ Dirac fermions.} which have $T_R=N_c$.
\end{enumerate}
Unless stated otherwise, we use $g=0.1,\lambda=0.03$ for plotting numerical values of the quantities we calculate. Note that, up to an overall normalization, the $\mathcal{N}=4$ SYM result depends on the 't~Hooft coupling $\lambda = g^2N_c$ rather than on $g$ and $N_c$ separately, so it can be straightforwardly compared against the large $N_c$ AdS/CFT result, which only features an explicit dependence on $\lambda$.

\subsection{Perturbative Structure of the Wilson Loop} \label{sec:calculation-structure}

The central object in this work is the Wilson loop introduced in Eq.~\eqref{eq:Wloop-Wlines}, which we repeat here
\begin{equation}
    \langle W_{\bf v} \rangle ({\bf L};t) = \frac{1}{Z} {\rm Tr}_{\mathcal{H}} \! \left[ W^{lk}_{[(0, {\bf L}),( t, {\bf v} t + {\bf L} )]} \tilde{\rho}^Q_{ki} W^{ij}_{[( t, {\bf v} t ),(0,{\bf 0})]} e^{-\beta H}  \rho_{jl}^Q \right] \, . \label{eq:Wloop-Wlines-2}
\end{equation}
To calculate it at weak coupling, we shall first expand all of the terms appearing in the trace in a power series in $g$ up to $\mathcal{O}(g^4)$. The $g^4$ terms provide the leading contribution to the result, because, as we shall see, the $\mathcal{O}(g^2)$ contribution vanishes.

All of the operators in Eq.~\eqref{eq:Wloop-Wlines-2} depend on the coupling $g$. We can classify them in three groups:
\begin{enumerate}
    \item The Wilson lines $W$. Because they are of the form\footnote{Here we are using the canonical normalization for free fields, so that the quadratic part of the kinetic term of the gauge field in the Yang-Mills Lagrangian is all that survives the $g\to 0$ limit. There is an alternative way of defining $A_\mu$, in which one lets $gA_\mu \to A_\mu$. If one does so, then the expansion of the Wilson lines is simply an expansion in powers of $A_\mu$.} ${\rm P} \exp \left( ig \int dx^\mu A_\mu \right)$, these have an explicit dependence on the coupling that we will track by expanding the path ordered exponential in the Wilson lines up to fourth order.
    \item The initial and final state colored matrices $\rho$ matrices. These are coupling dependent because they are $A_\mu$- and gauge-dependent ---  transforming in the opposite way as the endpoints of the Wilson lines under gauge transformations, and as such may in principle be expanded in $g$. At the lowest order, we have $\rho^Q_{ij} = \tilde{\rho}^Q_{ij} = \delta_{ij} + \mathcal{O}(g)$, because they implement the sum over initial and final color states in the probability derived from the heavy quark amplitude~\eqref{eq:amplitude}. And, in fact, this is all we need for our present calculation, because any Wick contraction\footnote{Except for the axial gauge $n^\mu A_\mu = 0$, which is known to be singular. See~\cite{Scheihing-Hitschfeld:2022xqx} for a discussion on the subtleties that appear when one tries to impose the corresponding gauge-fixing condition in a path integral that contains Wilson lines of infinite extent along $n^\mu$.} of a gauge field coming from $\rho$ or $\tilde{\rho}$ with a gauge field from the timelike Wilson lines will not give a $t$-extensive contribution to the final result. Therefore, as far as the contributions to $S({\bf L};{\bf v})$ go in perturbation theory, we may drop the contributions from these matrices altogether and simply set them to identity color matrices $\delta_{ij}$.
    \item The gauge theory Hamiltonian $H$. Note that in addition to its explicit appearance in the thermal density matrix $e^{-\beta H}$, it also appears implicitly in the time evolution of the gauge field $A_\mu$ in the Wilson lines. The perturbative expansion of this object is equivalent to the perturbative expansion of correlation functions in terms of Feynman diagrams. However, for our present purposes, writing out the diagrams is not more efficient as a computational method than simply doing all of the Wick contractions --- which is why we only discuss some notions usually associated to diagrams and their Feynman rules in Appendix~\ref{app:calculation-evol} as they become convenient.
\end{enumerate}

Then, setting the $\rho$ matrices to unity, expanding the Wilson lines, and grouping terms according to the number of gauge fields in each term (equivalently, powers of the coupling), one obtains
\begin{align} \label{eq:W-expansion}
    \langle W_{\bf v} \rangle({\bf L};t) = 1 + g W_1({\bf L};{\bf v},t) + g^2 W_2({\bf L};{\bf v},t) + g^3 W_3({\bf L};{\bf v},t) + g^4 W_4({\bf L};{\bf v},t) + \ldots 
\end{align}
where $W_1 = 0$ because it is proportional to the trace of a traceless matrix. Also, $W_3$ vanishes in Feynman gauge up to $\mathcal{O}(g^4)$ (inclusive) because the only contribution to the gauge field $3$-point function $\langle A_\mu^a A_\nu^b A_\rho^c \rangle$ is antisymmetric in the spacetime indices --- and in this particular expectation value all of these indices are contracted with the same vector $n^\mu$. We show this in a more detailed way in Appendix~\ref{app:calculation-structure-3}.

We are left with two nonzero contributions. On the one hand, we have the contribution from the $2$-point function
\begin{align}
    W_2({\bf L};{\bf v},t) = \frac{T_F}{N_c}  n^\mu n^\nu \int_0^t dt' \int_0^t dt'' \Big[  \langle A^a_\mu(t'-t'',(t'-t''){\bf v} + {\bf L} ) A^a_\nu(0,{\bf 0} ) \rangle_T &\nonumber \\ - \langle A^a_\mu(t'-t'',(t'-t''){\bf v}  ) A^a_\nu(0,{\bf 0} ) \rangle_T &\Big] \, , \label{eq:W2}
\end{align}
where we defined the thermal average of an operator as
\begin{equation}
    \langle O \rangle_T \equiv \frac{{\rm Tr}_{\mathcal{H}} \! \left[ O e^{-\beta H} \right]}{{\rm Tr}_{\mathcal{H}} \! \left[ e^{-\beta H} \right]}  \, ,
\end{equation}
and $T_F$ sets the normalization of the $SU(N_c)$ generators of the representation in which the heavy quark transforms ($T_F = 1/2$ for the fundamental representation). Furthermore, a gauge-fixing condition must be used in order to evaluate the $2$-point correlators in Eq.~\eqref{eq:W2}\footnote{This is an intrinsic requirement of perturbative calculations, where the $2$-point function of the gauge field --- the natural field variable in the problem, but at the same time a gauge-dependent quantity --- appears explicitly. However, if the thermal ensemble of the same theory could be written in terms of different variables, a gauge-fixing condition may not be necessary. This is what happens, for example, in the holographic calculation of this Wilson loop in $\mathcal{N}=4$ SYM~\cite{Rajagopal:2025ukd}.}.  

Similarly, $W_4$ is given by
\begin{align} \label{eq:W4}
    W_4({\bf L};{\bf v},t) &= \frac{1}{N_c} {\rm Tr}_c \! \left[ T^a T^b T^c T^d \right] n^\mu n^\nu n^\rho n^\sigma  \\
    & \quad \times \Bigg[ \int_0^t \!\! dt_4 \! \int_{t_4}^t \!\! dt_3 \! \int_{t_3}^t \!\! dt_2 \! \int_{t_2}^t \!\! dt_1  \langle A_\mu^a(h_{{\bf v}}^{{\bf L}}(t_4)) A_\nu^b(h_{{\bf v}}^{{\bf L}}(t_3)) A_\rho^c(h_{{\bf v}}^{{\bf L}}(t_2)) A_\sigma^d(h_{{\bf v}}^{{\bf L}}(t_1)) \rangle_T \nonumber \\
    & \quad \,\,\, -\int_0^t \!\! dt_4 \! \int_{t_4}^t \!\! dt_3 \! \int_{t_3}^t \!\! dt_2 \! \int_{0}^t \!\! dt_1 \langle A_\mu^a(h_{{\bf v}}^{{\bf L}}(t_4)) A_\nu^b(h_{{\bf v}}^{{\bf L}}(t_3)) A_\rho^c(h_{{\bf v}}^{{\bf L}}(t_2)) A_\sigma^d(h_{{\bf v}}^{{\bf 0}}(t_1)) \rangle_T \nonumber \\
    & \quad \,\,\, + \int_0^t \!\! dt_4 \! \int_{t_4}^t \!\! dt_3 \! \int_{0}^t \!\! dt_2 \! \int_{0}^{t_2} \!\! dt_1 \langle A_\mu^a(h_{{\bf v}}^{{\bf L}}(t_4)) A_\nu^b(h_{{\bf v}}^{{\bf L}}(t_3)) A_\rho^c(h_{{\bf v}}^{{\bf 0}}(t_2)) A_\sigma^d(h_{{\bf v}}^{{\bf 0}}(t_1)) \rangle_T \nonumber \\
    & \quad \,\,\, - \int_0^t \!\! dt_4 \! \int_{0}^t \!\! dt_3 \! \int_{0}^{t_3} \!\! dt_2 \! \int_{0}^{t_2} \!\! dt_1 \langle A_\mu^a(h_{{\bf v}}^{{\bf L}}(t_4)) A_\nu^b(h_{{\bf v}}^{{\bf 0}}(t_3)) A_\rho^c(h_{{\bf v}}^{{\bf 0}}(t_2)) A_\sigma^d(h_{{\bf v}}^{{\bf 0}}(t_1)) \rangle_T \nonumber \\
    & \quad \,\,\, + \int_0^t \!\! dt_4 \! \int_{0}^{t_4} \!\! dt_3 \! \int_{0}^{t_3} \!\! dt_2 \! \int_{0}^{t_2} \!\! dt_1 \langle A_\mu^a(h_{{\bf v}}^{{\bf 0}}(t_4)) A_\nu^b(h_{{\bf v}}^{{\bf 0}}(t_3)) A_\rho^c(h_{{\bf v}}^{{\bf 0}}(t_2)) A_\sigma^d(h_{{\bf v}}^{{\bf 0}}(t_1)) \rangle_T  \nonumber
    \Bigg] \, ,
\end{align}
where, to save space, we have introduced the coordinate vector
\begin{equation}
    h_{{\bf v}}^{{\bf L}}(t') = (t', t'{\bf v} + {\bf L} ) \, .
\end{equation}

One can see that the normalization condition $\langle W_{\bf v} \rangle({\bf L} ={\bf 0};t) = 1$ is trivially satisfied in this expansion because
\begin{equation}
    W_2({\bf L } = {\bf 0};{\bf v},t) = W_4({\bf L } = {\bf 0};{\bf v},t) = 0 \, ,
\end{equation}
as there is an exact cancellation in Eq.~\eqref{eq:W2}, and the integration regions in Eq.~\eqref{eq:W4} with opposite overall signs add up to exactly the same integration region, and thus cancel.

We remark that we did not include any contribution from the time-independent normalization prefactor $Z$ in this calculation (other than a constant factor of $N_c$). However, in principle, this prefactor could be coupling-dependent. The reason why it will not appear in our calculation is because we are interested in calculating the leading $t$-dependence at leading nontrivial order in powers of $g$, which starts at $\mathcal{O}(g^4)$, and hence any coupling dependence in $Z$ will only enter at subleading order in the large time expansion. 

That being said, to show that our object of interest $S({\bf L};{\bf v})$ starts at $\mathcal{O}(g^4)$, some further explanation is needed, given that $W_2$ seemingly gives an $\mathcal{O}(g^2)$ contribution. The reason is that the $t$-extensive piece in $W_2$ itself starts at $\mathcal{O}(g^2)$. To see this, first note that at large $t$ we have\footnote{Note that this expression (at lowest order in perturbation theory) is explicitly gauge invariant as any transformation in a transverse direction will be projected out by $n^\mu n^\nu$ and any transformation of $A_\mu$ parallel to $n^\mu$ will be a total derivative and integrate to zero along the path of the heavy quark (otherwise the gauge transformation would be modifying the boundary conditions of the gauge field at infinity). At higher order in perturbation theory, gauge invariance is in general only verified after summing over all diagrams.}
\begin{align}
    W_2({\bf L};{\bf v},t) = t \frac{T_F}{N_c} n^\mu n^\nu \!\! \int_{-\infty}^\infty \!\! dt' \Big[  \langle A^a_\mu(t',t'{\bf v} + {\bf L} ) A^a_\nu(0,{\bf 0} ) \rangle_T- \langle A^a_\mu(t',t'{\bf v}  ) A^a_\nu(0,{\bf 0} ) \rangle_T \Big] + \mathcal{O}(t^0) \, , \label{eq:W2-t}
\end{align}
and that the two-point function for a free gauge field can be written as
\begin{equation}
    \langle A^a_\mu(t',t'{\bf v} + {\bf L} ) A^b_\nu(0,{\bf 0} ) \rangle_T = \delta^{ab} \int_{\omega,{\bf k}} e^{-i\omega t'} e^{i {\bf k} \cdot (t' {\bf v} + {\bf L}) } P_{\mu \nu}(\omega, {\bf k} ) D^>(\omega, {\bf k} ) \, ,
\end{equation}
where we introduced the shorthand notation $\int_{\omega,{\bf k}} = \int \frac{d\omega}{(2\pi)} \frac{d^3 {\bf k} }{(2\pi)^3}$. Here, $P_{\mu \nu}$ is the polarization tensor, whose precise expression is different in different gauges, and $D^>(\omega, {\bf k}) = [1 - e^{-\beta \omega}]^{-1} \varrho(\omega, {\bf k} ) $ is the free massless scalar Wightman function, with
\begin{equation}
    \varrho(\omega, {\bf k}) = {\rm sgn}(\omega) (2\pi) \delta(\omega^2 - {\bf k}^2) \label{eq:rho-free} 
\end{equation}
the free massless scalar spectral function\footnote{Note that some authors define the relation between $\varrho$ and $D^>$ with an additional factor of $2$ relative to our convention.}. In our calculations, we will use the Feynman gauge expression $P_{\mu \nu} = - g_{\mu \nu}$.

Second, note that the integral over $t'$ in~\eqref{eq:W2-t} imposes a constraint over the momentum integrals
\begin{equation}
    \int_{-\infty}^{\infty} e^{-i\omega t'} e^{i t' {\bf k} \cdot  {\bf v}  } = (2\pi) \delta(\omega - {\bf v} \cdot {\bf k} ) \, ,
\end{equation}
which, together with the on-shell delta function $\delta(\omega^2 - {\bf k}^2 )$ coming from the free gauge field propagator, constrains the contribution to $W_2$ to be entirely localized\footnote{This relies on the strict inequality $v = |{\bf v}| < 1$. If $v = 1$, the constraint would amount to having ${\bf k}$ collinear with ${\bf v}$.} at ${\bf k} = 0$. However, the difference between the two $2$-point functions in Eq.~\eqref{eq:W2-t} in Fourier space amounts exactly to a prefactor of $e^{i {\bf k} \cdot {\bf L}}$. Hence, the contributions coming from the infinitesimal vicinity of ${\bf k} = 0$ is equal in both terms, and therefore they cancel out.\footnote{We note that in the (for our present purposes, unrealistic) scenario where the medium features an effective speed of gluon propagation $c_{\rm eff}$ that is less than one, the $\mathcal{O}(g^2)$ terms describe Cherenkov radiation when $v > c_{\rm eff}$~\cite{Lin:2026nlr}.} We provide a more extensive discussion in Appendix~\ref{app:calculation-structure-2}.

Therefore, the leading contribution to~\eqref{eq:W2-t} comes from the weak coupling expansion of the gauge field self-energy, which brings with it an additional power of $g^2$, and therefore it enters at the same order as~\eqref{eq:W4} in the calculation of $\langle W_{\bf v} \rangle({\bf L};t)$. Note that $W_2$ automatically obeys the same property under parity as Eq.~\eqref{eq:S-analytic-property} because the Wightman correlator of any bosonic field satisfies $D^>(t,{\bf x}) = D^>(-t-i\beta,-{\bf x})$ in thermal equilibrium --- regardless of whether the theory is interacting or non-interacting. $W_4$ also obeys it, albeit verifying it is less straightforward.

Having observed this property of the free gauge field $2$-point function, an additional simplification emerges in Eq.~\eqref{eq:W4}. Note that this expression for $W_4$ is to be evaluated using non-interacting Wick contractions because any interacting contribution comes with additional powers of the coupling
$g$. After doing the Wick contractions, the resulting $2$-point functions can be classified by whether the spatial separation between the operators depends on ${\bf L}$ or not. In fact, for all except one of the expectation values under integral signs in Eq.~\eqref{eq:W4}, the Wick contractions yield at most one Wightman function where ${\bf L}$ appears explicitly, which, as per our previous discussion of the $2$-point function, are actually ${\bf L}$-independent contributions once one carries out the time integrals in the $t\to \infty$ limit. We show this in detail in Appendix~\ref{app:calculation-structure-4}. Therefore, the one line that yields an ${\bf L}$-dependent contribution is the third line in the square brackets, and the rest of them may be interpreted as simply providing the appropriate quantity to be subtracted such that $W_4({\bf L} = {\bf 0}) = 0$. 

As such, the analysis of $W_4$ is simplified to
\begin{align}
    W_4({\bf L};{\bf v},t) &= \frac{1}{N_c} n^\mu n^\nu n^\rho n^\sigma \int_0^t \!\! dt_4 \! \int_{t_4}^t \!\! dt_3 \! \int_{0}^t \!\! dt_2 \! \int_{0}^{t_2} \!\! dt_1 \int_{\omega,{\bf k}} \int_{\Omega,{\bf q}} e^{i ( {\bf k} + {\bf q}) \cdot {\bf L}  } D^>(\omega,{\bf k}) D^>(\Omega,{\bf q}) \nonumber \\
    & \quad \times \Bigg[ {\rm Tr}_c \! \left[ T^a T^b T^a T^b \right]  e^{-i(\omega - {\bf k} \cdot {\bf v} ) (t_4-t_2) } e^{-i(\Omega - {\bf q}\cdot {\bf v}) (t_3-t_1)} P_{\mu \rho}(\omega,{\bf k}) P_{\nu \sigma}(\Omega,{\bf q}) \nonumber \\
    & \quad \quad +  {\rm Tr}_c \! \left[ T^a T^b T^b T^a \right] e^{-i(\omega - {\bf k} \cdot {\bf v} ) (t_4-t_1) } e^{-i(\Omega - {\bf q}\cdot {\bf v}) (t_3-t_2)} P_{\mu \sigma}(\omega,{\bf k}) P_{\nu \rho}(\Omega,{\bf q}) 
    \Bigg] \nonumber \\ 
    & \quad - \Big\{ ({\bf L} = {\bf 0})  \Big\} \, ,    \label{eq:W4-simp}
\end{align}
where the subtraction in the last line stands for subtracting the value of the rest of the expression on the right hand side evaluated at ${\bf L} = {\bf 0}$.

In Appendix~\ref{app:calculation-structure-4} we work through the time integrals and find that the Feynman gauge expression for $W_4$ is
\begin{align}
    W_4({\bf L};{\bf v},t) &=  t (n^2)^2 C_F \frac{N_c}2 \int_{\omega,{\bf k}} \int_{p_0,{\bf p}}  e^{i {\bf p} \cdot {\bf L}  } D^>(\omega,{\bf k}) D^>(p_0 - \omega,{\bf p}- {\bf k} ) \frac{(2\pi) \delta(p_0 - {\bf p} \cdot {\bf v} ) }{( \omega - {\bf k} \cdot {\bf v} - i0^+ )^2}  \nonumber \\
    & \quad + \frac12 \left( t C_F n^2 \int_{\omega,{\bf k}} e^{i {\bf k} \cdot {\bf L} } D^>(\omega,{\bf k}) (2\pi) \delta(\omega - {\bf k} \cdot {\bf v}) \right)^2 - \Big\{ ({\bf L} = {\bf 0})  \Big\} + \mathcal{O}(t^0,g^5)
    \nonumber \\
    &= \big( \tilde{W}_4 \big)  +  \frac{1}{2} \big( \tilde{W}_2 \big)^2 - \Big\{ ({\bf L} = {\bf 0})  \Big\} + \mathcal{O}(t^0,g^5) \label{eq:W4-result}
\end{align}
which, consistent with the fact that the leading contribution to the exponent of the Wilson loop is linear in time, contains a linear piece and a quadratic piece corresponding to the square of the (seemingly) ${\bf L}$-dependent contribution at $\mathcal{O}(g^2)$. We have used the $\tilde{W}_n$ symbol to denote the fixed-order contribution to each of $W_2$ and $W_4$ that appears in this expression, which fully contain the ${\bf L}$-dependence of the result (even though $\tilde{W}_2$ is actually ${\bf L}$-independent).

Collecting all of our results, we have
\begin{align}
    S({\bf L};{\bf v}) &= - \frac{g^4 C_F}{T} \Bigg[ \frac{(1-v^2)^2 N_c}{2} \int_{\omega,{\bf k}} \int_{p_0,{\bf p}}  e^{i {\bf p} \cdot {\bf L}  } D^>(\omega,{\bf k}) D^>(p_0 - \omega,{\bf p}- {\bf k} ) \frac{(2\pi) \delta(p_0 - {\bf p} \cdot {\bf v} ) }{( \omega - {\bf k} \cdot {\bf v} - i0^+ )^2} \nonumber \\
    & \quad\quad\quad\quad\quad  + n^\mu n^\nu \int_{-\infty}^\infty dt' \left[ D^>_{\mu \nu}(t', t'{\bf v} + {\bf L} ) \right]_{\mathcal{O}(g^2)} - \Big\{ ({\bf L} = {\bf 0})  \Big\} \Bigg] + \mathcal{O}(g^5) \, , \label{eq:S-setup-result}
\end{align}
where $\left[ D^>_{\mu \nu}(t', t'{\bf v} + {\bf L} ) \right]_{\mathcal{O}(g^2)}$ represents the term of the color-averaged Wightman function of the gauge field whose coefficient in a weak coupling expansion is $g^2$.

\subsection{Matching the Hard and Soft Sectors}\label{sec:matching}

There is one final subtlety that needs to be addressed, which has to do with the time integral in the last line of Eq.~\eqref{eq:S-setup-result}. Because this integral may receive contributions from arbitrarily long wavelength modes, it is necessary to employ the Hard Thermal Loop (HTL) resummation~\cite{Braaten:1989mz,Braaten:1990az,Braaten:1991gm,Caron-Huot:2007cma} to accurately describe the soft physics in the integral. However, while this resummation guarantees a correct description of the IR contributions to the integral at leading order in perturbation theory, it does not correctly describe the UV contributions (in fact, it gives rise to unphysical divergences), which can in principle be calculated directly in perturbation theory without need of any resummation. From the point of view of the latter, the need to use the HTL resummation arises from the fact that the ``naive'' perturbative contributions (i.e., the ones that may be directly calculated in terms of Feynman diagrams without any resummation) are IR divergent. This raises a problem, because in order to accurately calculate the result in perturbation theory one needs to include both kinds of contributions --- and also avoid double-counting degrees of freedom with the momentum integrals. This problem is well-known to specialists, and various ways of implementing the subtraction to avoid double-counting exist, dating back at least to the early works by Braaten and Thoma on the energy loss of a heavy quark~\cite{Braaten:1991jj,Braaten:1991we}.  

The approach we follow is most closely inspired by the work of Burnier, Laine, Langelage and Mether~\cite{Burnier:2010rp}.
Much like in our present situation, in that work its authors calculate the spectral function of two chromoelectric fields inserted along a Wilson line that wraps around the SK contour --- which may in fact be used to extract the heavy quark diffusion coefficient. Their solution to avoid double-counting, adapted to our present calculation, is to consider the sum of the ``naive'' perturbative contributions and the HTL-resummed Wightman function, and subtract from it the HTL Wightman function \textit{without} doing the resummation (in other words, the HTL Wightman function expanded up to $\mathcal{O}(g^2)$ inclusive and truncated at that order). Explicitly,
\begin{equation}
    D^>_{\mu \nu}(t', t'{\bf v} + {\bf L} ) = \left[ D^>_{\mu \nu}(t', t'{\bf v} + {\bf L} ) \right]_{\rm pert.} + \left[ D^>_{\mu \nu}(t', t'{\bf v} + {\bf L} ) \right]_{\rm HTL} - \left[ D^>_{\mu \nu}(t', t'{\bf v} + {\bf L} ) \right]_{\rm unres. \, HTL} \label{eq:Wightman-decomp}
\end{equation}
takes care of the UV and IR contributions in a way that is guaranteed to be accurate at lowest order in perturbation theory. Divergences are canceled
\begin{enumerate}
    \item in the IR between the perturbative and the unresummed HTL contributions, and
    \item in the UV between the two HTL contributions,
\end{enumerate}
making Eq.~\eqref{eq:Wightman-decomp} manifestly finite. The cancellation in the IR means that in that regime the HTL-resummed contribution dominates, and gives rise to terms that depend on the coupling as $g^4\ln(1/g)$ in addition to the $g^4$ terms.

That being said, Eq.~\eqref{eq:Wightman-decomp} also contains terms that are higher order in the coupling than $\mathcal{O}(g^4)$ in the HTL-resummed Wightman function. To give an unambiguous answer at $\mathcal{O}(g^4)$ (including both $g^4$ and $g^4\ln (1/g)$), we need to remove these higher order contributions from the calculation.

To do this, it helps to make use of the additional structure provided by the spectral decomposition of the Wightman functions. Explicitly, one has
\begin{equation}
    \int_{-\infty}^\infty dt' D^>_{\mu \nu}(t',t'{\bf v} + {\bf L}) = \int \frac{d^3 {\bf p} }{(2\pi)^3} \frac{e^{i {\bf p} \cdot {\bf L} } }{1 - e^{- {\bf p} \cdot {\bf v}/T } } \varrho_{\mu\nu}(p_0 = {\bf p} \cdot {\bf v},{\bf p}) \, , \label{eq:Wightman-spectral}
\end{equation}
in terms of the spectral function $\varrho_{\mu\nu}$ of the gauge field --- which, as we mentioned earlier, we shall evaluate in Feynman gauge. Here we used that the integral over $t'$ enforces $p_0 = {\bf p} \cdot {\bf v}$, effectively selecting the ``zero mode'' of the spectral function along the heavy quark propagation direction. One can check that this term satisfies the KMS relation~\eqref{eq:S-analytic-property} by letting ${\bf p} \to -{\bf p}$ and using that the spectral function is odd under a spacetime inversion.

With Eq.~\eqref{eq:Wightman-spectral} in hand, it pays off to note that, from the point of view of a Taylor series expansion of $S({\bf L};{\bf v})$, the aforementioned cancellation between the perturbative and HTL-unresummed contributions that prevents IR divergences from appearing is only relevant for derivatives with respect to ${\bf L}$ (or, equivalently, moments with respect to ${\bf p}$) up to a finite order. As our calculation later in this section will show, this finite order is $2$. Because of this, we may trivially rearrange Eq.~\eqref{eq:Wightman-spectral} into
\begin{align}
    \int_{-\infty}^\infty dt' D^>_{\mu \nu}(t',t'{\bf v} + {\bf L}) &= \int \frac{d^3 {\bf p} }{(2\pi)^3} \frac{e^{i {\bf p} \cdot {\bf L} } - \left[1 + i{\bf p} \cdot {\bf L} - \frac12 ({\bf p} \cdot {\bf L})^2 \right] }{1 - e^{- {\bf p} \cdot {\bf v}/T } } \varrho_{\mu\nu}(p_0 = {\bf p} \cdot {\bf v},{\bf p}) \nonumber \\
    &  + \int \frac{d^3 {\bf p} }{(2\pi)^3} \frac{ 1 + i{\bf p} \cdot {\bf L} - \frac12 ({\bf p} \cdot {\bf L})^2  }{1 - e^{- {\bf p} \cdot {\bf v}/T } } \varrho_{\mu\nu}(p_0 = {\bf p} \cdot {\bf v},{\bf p}) \, , \label{eq:Wightman-spectral-add-sub}
\end{align}
where the first line is explicitly insensitive to IR features of $\varrho_{\mu\nu}$, and the second line compensates for the subtracted contributions.

Crucially, both lines on the right hand side of Eq.~\eqref{eq:Wightman-spectral-add-sub} are separately finite. Furthermore, the first line is separately IR finite for each contribution in the decomposition~\eqref{eq:Wightman-decomp}. This implies that the difference between the HTL resummed and unresummed pieces in the first line contributes to the result only at $\mathcal{O}(g^6)$, and may thus be dropped. As such, up to $\mathcal{O}(g^4)$, the contributions from the spectral function in the first line of Eq.~\eqref{eq:Wightman-spectral-add-sub} are exactly described by the ``naive'' perturbative contribution alone --- and no factors of $\ln(1/g)$ appear in the result from this contribution.

In fact, all of the $g^4 \ln(1/g)$ contributions to $S({\bf L};{\bf v})$, which can be traced back to the HTL resummation, are contained in the second line of Eq.~\eqref{eq:Wightman-spectral-add-sub}. Note that this second line is qualitatively simpler than the first, because all of the ${\bf L}$-dependence of this expression can be factored out (term by term). As such, we only need to evaluate it once for every ${\bf v}$, without needing to scan over ${\bf L}$ --- which is a substantial simplification. On the other hand, care needs to be taken in order to isolate the $\mathcal{O}(g^4)$ contributions, as there are higher order pieces that need to be subtracted by hand. We shall discuss this in detail in subsection~\ref{sec:fixed-order}.

With all the pieces now in place, we will proceed to evaluate Eq.~\eqref{eq:S-setup-result} in the remainder of this section. Note that, in light of our discussion in the last few paragraphs, we can unambiguously decompose our result at $\mathcal{O}(g^4)$ between the $\ln(1/g)$-enhanced pieces and the rest as
\begin{align}\label{eq:Sdecomp}
    S({\bf L};{\bf v}) = g^4 \ln(1/g) \left[ -i \bar{F}_D^{LL} ({\bf L} \cdot \hat{\bf v} ) + \frac{\bar{\kappa}_L^{\rm LL}}{2}  ({\bf L} \cdot \hat{\bf v})^2 + \frac{\bar{\kappa}_T^{\rm LL}}{2} \Big( {\bf L}^2 - ({\bf L} \cdot \hat{\bf v})^2 \Big) \right] + g^4 \bar{S}_{\rm SLL}({\bf L};{\bf v}) \, .
\end{align}
We shall verify that, writing $m_D^2 = \frac13g^2 T^2 \mathcal{N}$ with
$\mathcal{N}\equiv N_c+\frac{N_c N_X}{2}+N_fT_R$, the reduced leading-log
coefficients in~\eqref{eq:Sdecomp} are
\begin{align}
    \bar{F}_D^{\rm LL}(v) &= \frac{C_F \mathcal{N} T}{24\pi}\,
        \frac{v-(1-v^2)\arctanh v}{v^2} \, , \\
    \bar{\kappa}_L^{\rm LL}(v) &= \frac{C_F \mathcal{N} T^2}{12\pi}\,
        \frac{v-(1-v^2)\arctanh v}{v^3} \,=\frac{2}{v}\bar{F}_D^{\rm LL}(v)\, , \\
    \bar{\kappa}_T^{\rm LL}(v) &= \frac{C_F \mathcal{N} T^2}{12\pi}\,
        \frac{3v^3+(1-v^2)^2\arctanh v - v}{2v^3} \, ,
\end{align}
normalized such that $F_D = g^4\ln(1/g)\,T\,\bar{F}_D^{\rm LL}$ and
$\kappa_{L,T} = g^4\ln(1/g)\,T\,\bar{\kappa}_{L,T}^{\rm LL}$ at leading log, with
$\kappa_T$ referring to a single Cartesian component transverse to
${\bf v}$. These reproduce the leading-log results of~\cite{Moore:2004tg},
satisfy the Einstein relation $v\bar{\kappa}_L^{\rm LL} = 2T\bar{F}_D^{\rm LL}$
exactly, and are isotropic as $v\to0$:
$\bar{\kappa}_L^{\rm LL},\bar{\kappa}_T^{\rm LL} \to C_F \mathcal{N} T^2/(18\pi)$.

The leading log results define an exactly Gaussian stochastic process for heavy quarks, because a quadratic $S$ maps to a Gaussian $P({\bf k};{\bf v})$. All of the non-Gaussian structure of the momentum transfer distribution therefore resides in $\bar{S}_{\rm SLL}({\bf L};{\bf v})$. It too satisfies the analytic property~\eqref{eq:S-analytic-property}, and hence the equilibration condition~\eqref{eq:K-analytic-property} --- but in a more non-trivial way. In the quadratic case that condition collapses to the Einstein relation between two coefficients; for $\bar{S}_{\rm SLL}$ it holds only through a balance among all of the cumulants at once. This is the precise sense in which equilibration at finite velocity is a statement about the entire momentum transfer distribution rather than about its first two moments, as anticipated in Sec.~\ref{sec:HQET}.

\subsection{The Matched Kernel} \label{sec:result-analytic}

Without further ado, leaving the intermediate details to Appendix~\ref{app:calculation-evol}, we start this section by giving the explicit expression for $S({\bf L};{\bf v})$ that follows from Eqs.~\eqref{eq:S-setup-result} and~\eqref{eq:Wightman-decomp}. This is an instrumental step towards our goal of deriving the fixed order kernel later in the section. It also already allows us to do multiple consistency checks. It is simply given by
\begin{align} \label{eq:bigintegral}
S({\bf L};{\bf v}) &= - \frac{g^2 C_F}{T} \int_{\bf p} \frac{e^{i {\bf p} \cdot {\bf L} } - 1}{1 - e^{- {\bf p} \cdot {\bf v}/T } } \\
& \times \Bigg\{  \frac{g^2 N_c}{2} \int_{k_0}  \frac{e^{{\bf p} \cdot {\bf v}/T} - 1 }{|e^{k_0/T}-1| |e^{({\bf p} \cdot {\bf v}-k_0)/T} - 1| }  
\frac{1}{ 4 |{\bf p}|}  \Theta \! \left( 1 - u^2 \right)  \nonumber \\
& \quad \times \Bigg(  \frac{ (1- {\bf v}^2 )^2  \left| {\rm sgn}(k_0) - u \frac{{\bf v} \cdot {\bf p}}{|{\bf p}|}  \right| }{  k_0^2  \left[ \left({\rm sgn}(k_0) - u \frac{{\bf v} \cdot {\bf p}}{|{\bf p}|} \right)^2 - (1-u^2) \left( {\bf v}^2 - \frac{({\bf v} \cdot {\bf p})^2}{{\bf p}^2} \right)   \right]^{3/2} } - \frac{4(1-{\bf v}^2) }{  {\bf p}^2 -  ( {\bf p} \cdot {\bf v} )^2   }  \nonumber \\ 
& \quad \quad \quad  + \frac{8 + 4N_X }{ [{\bf p}^2 -  ( {\bf p} \cdot {\bf v} )^2 ]^2 } \left[ \left( k_0 - |k_0| \frac{{\bf p} \cdot {\bf v}}{|{\bf p}|} u \right)^2  + \frac{k_0^2}{2} \left( {\bf v}^2 - \frac{ ( {\bf p} \cdot {\bf v} )^2 }{{\bf p}^2} \right) \left( 1 - u^2 \right) \right] \Bigg) \nonumber \\
& \quad + 2 g^2 N_f T_R  \int_{k_0} \frac{ e^{k_0/T} (1 - e^{- {\bf p} \cdot {\bf v}/T }) }{(e^{k_0/T}+1) (e^{(k_0-{\bf p} \cdot {\bf v})/T} + 1) }   
\frac{1}{ 4 |{\bf p}| }  \Theta \! \left( 1 - u^2 \right) \Bigg( - \frac{1-{\bf v}^2}{ {\bf p}^2 - ({\bf p}\cdot {\bf v})^2 }  \nonumber \\
& \quad \quad \quad 
+ \frac{4}{[{\bf p}^2 - ({\bf p}\cdot {\bf v})^2 ]^2} \Bigg[ \left( k_0 -  |k_0| \frac{{\bf p} \cdot {\bf v}}{|{\bf p}|} u \right)^2  + \frac{k_0^2}2  \left( {\bf v}^2 - \frac{ ( {\bf p} \cdot {\bf v} )^2 }{{\bf p}^2} \right) \left( 1 - u^2 \right) \Bigg] \Bigg) \nonumber \\
& \quad +  \frac{ 2 \Gamma_E( ({\bf p} \cdot {\bf v})/|{\bf p}|  )   }{ \Sigma_E^2( p_0 = {\bf p} \cdot {\bf v} , p ) + \Gamma_E^2( ({\bf p} \cdot {\bf v})/|{\bf p}|  ) } -  \frac{ 2 \Gamma_E( ({\bf p} \cdot {\bf v})/|{\bf p}|  )   }{ |{\bf p}|^4 } \nonumber \\
& \quad   + \frac{2 {\bf v}^2 \left( 1 - \frac{({\bf p} \cdot {\bf v})^2}{{\bf p}^2 {\bf v}^2} \right) \, \Gamma_T( ({\bf p} \cdot {\bf v})/|{\bf p}| )  }{ \Sigma_T^2( p_0 = {\bf p} \cdot {\bf v}, p ) + \Gamma_T^2( ({\bf p} \cdot {\bf v})/|{\bf p}| ) } 
- \frac{2 {\bf v}^2 \left( 1 - \frac{({\bf p} \cdot {\bf v})^2}{{\bf p}^2 {\bf v}^2 } \right) \, \Gamma_T( ({\bf p} \cdot {\bf v})/|{\bf p}|  )  }{ |{\bf p}|^4 \left(1 -  \frac{({\bf p} \cdot {\bf v})^2}{{\bf p}^2  }  \right)^2 } 
  \Bigg\} \nonumber 
\, , 
\end{align}
where we have introduced $u$ as
\begin{equation}\label{eq:u-def}
	u = \frac{ {\bf p}^2 + 2 ({\bf p} \cdot {\bf v}) k_0 - ({\bf p} \cdot {\bf v})^2 }{2|{\bf p}| |k_0|} \, ,
\end{equation}
describing the cosine of the angle between the ${\bf k}$ vector in the ``loop'' integral and ${\bf p}$.
As before, here $m_D^2 = \tfrac13 g^2T^2 \mathcal{N}$ is the Debye mass, with the HTL self-energies given by textbook results~\cite{Laine:2016hma}:
\begin{align}
    \Sigma_T(k_0=k\eta,k) &= k^2(1-\eta^2)
    + \frac{m_D^2}{2}\left[\eta^2 + \frac{\eta(1-\eta^2)}{2}
    \ln\left|\frac{1+\eta}{1-\eta}\right|\right] , \\
    \Sigma_E(k_0=k\eta,k) &= k^2 + m_D^2\left[1 - \frac{\eta}{2}
    \ln\left|\frac{1+\eta}{1-\eta}\right|\right] , \\
    \Gamma_T(\eta) &= \frac{\pi m_D^2}{4}\eta(1-\eta^2) \, , \qquad
    \Gamma_E(\eta) = \frac{\pi m_D^2}{2}\eta \, .
\end{align}

Having in mind that we will evaluate this expression in three different theories, we have written this expression allowing for $N_X$ scalars in the adjoint representation and $N_f$ Dirac fermions in representation $R$ as constituents of the QGP that provides the environment for heavy quark propagation.

The first three lines inside the curly brackets $\{ \cdot \}$ in Eq.~\eqref{eq:bigintegral} describe the bosonic contribution to the spectral function, containing the contributions from gluons and adjoint scalars; the next two lines describe the fermionic contribution; the last two lines contain the HTL resummed expressions for the gauge field spectral function minus their unresummed versions. At weak coupling, the ``hard'' momentum $\sim T$ physics in the integral is well described by the first five lines in the curly brackets, whereas the ``soft'' momentum physics $\sim gT$ is well described by the first terms in the last two lines --- with the second terms in the last two lines having the role of subtracting the unphysical divergences in each term and avoid double-counting in the momentum integrals. Note that we have included the ${\bf L} = {\bf 0}$ subtraction in the first line by explicitly subtracting $1$ from $e^{i {\bf p} \cdot {\bf L} }$. This is the structure we anticipated in Secs.~\ref{sec:Kolmogorov-setup} and~\ref{sec:equilibration-setup} for weakly coupled theories, where the integrand multiplying $[e^{i{\bf p}\cdot{\bf L}}-1]$ can be identified with the collision rate of a heavy quark (compare with Eq.~\eqref{eq:Gamma-def}).

It is convenient to abbreviate Eq.~\eqref{eq:bigintegral} as
\begin{align} 
    S({\bf L};{\bf v}) = -\frac{g^2 C_F}{T}\int_{\bf p}
    \frac{e^{i{\bf p}\cdot{\bf L}}-1}{1-e^{-{\bf p}\cdot{\bf v}/T}}
    \Big\{ & g^2\,\mathcal{I}_{\rm Hard}({\bf p},{\bf v};T,N_c,N_X,N_f) \nonumber \\
    &+ \mathcal{I}_{\rm Soft}({\bf p},{\bf v};m_D^2) 
    - \mathcal{I}_{\rm Match}({\bf p},{\bf v};m_D^2) \Big\} , \label{eq:kernel-compact}
\end{align}
where $g^2\mathcal{I}_{\rm Hard}$ denotes the first five lines inside the curly
brackets of Eq.~\eqref{eq:bigintegral}, $\mathcal{I}_{\rm Soft}$ the two
HTL-resummed terms, and $\mathcal{I}_{\rm Match}$ their unresummed counterparts.
This is the matched kernel at $\mathcal{O}(g^4)$. It is not yet the object we shall use:
Sec.~\ref{sec:fixed-order} extracts from it the strict fixed-order kernel, which enters
everything that follows.

Two properties of the decomposition~\eqref{eq:kernel-compact} will be central in what follows. First, the step function in
$\mathcal{I}_{\rm Hard}$ restricts the loop energy to either
\begin{equation} \label{eq:k0-threshold}
    2k_0 \geq |{\bf p}| + {\bf p}\cdot{\bf v} \quad (k_0 > 0) \, ,
    \qquad {\rm or} \qquad
    2k_0 \leq |{\bf p}| - {\bf p}\cdot{\bf v} \quad (k_0 < 0) \, ,
\end{equation}
and hence in either case to
$2|k_0| \geq |{\bf p}| - |{\bf p}\cdot{\bf v}| \geq (1-v)|{\bf p}|$. This is the
usual $2\to2$ on shell elastic scattering kinematic bound (e.g.~\cite{Moore:2004tg}): a massless medium parton must carry at least this energy in order to deliver three-momentum ${\bf p}$ at energy transfer $p_0 = {\bf p}\cdot{\bf v}$ with both legs on shell. Together with the statistical factors it makes $\mathcal{I}_{\rm Hard}$ exponentially suppressed as $e^{-(1-v)|{\bf p}|/2T}$ at large $|{\bf p}|$. Second, $\mathcal{I}_{\rm Match}$ is by construction the leading term in the expansion of $\mathcal{I}_{\rm Soft}$ in powers of $m_D^2/|{\bf p}|^2$, so that
\begin{equation} \label{eq:high-p-HTL-cancel}
    \mathcal{I}_{\rm Soft} - \mathcal{I}_{\rm Match}
    = \mathcal{O}\!\left(m_D^4/|{\bf p}|^6\right) .
\end{equation}

As with any lengthy expression, it is worth checking whether Eq.~\eqref{eq:bigintegral} reproduces known results from the literature --- first as a cross-check, but also as a way of getting familiar with subsequent calculations that follow from it.
 
\paragraph{Cross-check: The Heavy Quark Diffusion Coefficient at $v=0$.}
At $v=0$ there is no preferred direction, and the second cumulant defines
the heavy quark momentum diffusion coefficient for a single Cartesian
component, $\kappa \equiv T \mathfrak{M}_2(v=0)$. Evaluating it from
Eq.~\eqref{eq:bigintegral} (see Appendix~\ref{app:v0kappa}) gives
\begin{equation} \label{eq:kappa-v0}
    \kappa = \frac{g^2 C_F m_D^2 T}{6\pi}
    \left[ \frac12 + \frac{N_c N_X}{4\mathcal{N}} - \gamma_{\rm E}
    + \ln\!\left(\frac{2T}{m_D}\right)
    + \frac{T_R N_f}{\mathcal{N}}\ln 2 + \frac{\zeta'(2)}{\zeta(2)} \right] \, .
\end{equation}
Setting $N_X=0$ this reproduces the known leading-order result~\cite{Moore:2004tg,Burnier:2010rp}. To our knowledge the $N_X\neq0$ generalization where the heavy quarks are not directly coupled to scalars has not appeared previously. The difference with the well-known calculations in perturbative $\mathcal{N}=4$ SYM~\cite{Chesler:2006gr,Caron-Huot:2008dyw} is that those calculations feature Wilson lines that include scalars in their definition, thus accounting for a direct coupling of the heavy quark to the scalar. See also our discussion in footnote\footref{ft:scalarsN4}.

\paragraph{Cross-check: The Leading Logarithm at Arbitrary Velocity.}
The logarithmically enhanced contribution comes entirely from the soft
sector, and is obtained by evaluating the HTL contribution with the
self-energies dropped. As we show in \cite{DuPlessis:2026pyr},
this yields
\begin{align} \label{eq:S-LL}
    S_{\rm LL}({\bf L};{\bf v}) = \frac{g^2 C_F m_D^2}{4\pi}
    \ln\!\left(\frac{m_D}{T}\right) \Bigg[ &
    \frac{(v^2-1)\arctanh(v)+v}{2v^3}\, iL_\parallel\!\left(iL_\parallel+\frac{v}{T}\right)
    \nonumber \\ & +
    \frac{3v^3+(1-v^2)^2\arctanh(v)-v}{4v^3}\,(iL_\perp)^2 \Bigg] \, ,
\end{align}
which is precisely the quadratic form defined by
$\bar{F}_D^{\rm LL},\bar{\kappa}_L^{\rm LL},\bar{\kappa}_T^{\rm LL}$ of
Section~\ref{sec:matching}, and hence agrees with the leading-log
transport coefficients of~\cite{Moore:2004tg}. Being quadratic in ${\bf L}$,
it describes an exactly Gaussian stochastic process; the combination
$iL_\parallel(iL_\parallel + v/T)$ makes the equilibration
condition~\eqref{eq:S-analytic-property} --- equivalently the Einstein
relation --- manifest.

These are two valuable checks on the derived expression against known results.

\subsection{The Fixed-Order Kernel} \label{sec:fixed-order}

Equation~\eqref{eq:bigintegral} is correct through $\mathcal{O}(g^4)$, but the
HTL-resummed propagators also carry contributions of higher order in $g$. These
are harmless for the first two cumulants, where the soft physics is physical and
required, but they render every cumulant with $n\geq3$ ultraviolet divergent. By inspecting 
Eq.~\eqref{eq:high-p-HTL-cancel} together with the overall $g^2$ prefactor,
it is clear that these apparent UV divergences are of strictly higher order than the $g^4$ terms we are after, and yet they contribute infinitely if we take Eq.~\eqref{eq:bigintegral} literally. Physically, they are also in the wrong part of momentum space: the soft sector describes many small kicks and should not be controlling cumulants that probe hard momentum transfer.

The strict fixed-order kernel was constructed by us in Ref.~\cite{DuPlessis:2026pyr} by
separating the two regimes $p<\Lambda,p>\Lambda$ with Heaviside functions:
the HTL pieces may be dropped
altogether for $|{\bf p}|>\Lambda$ with $\Lambda\sim g^0T$, while in the soft
region the radial integral must be performed exactly \emph{before} expanding in
$m_D/T$, since the two operations do not commute. This is clear, because a power series expansion in $m_D/T$ cannot account for the $g^4\ln(1/g)$ terms that are manifest in the exact calculation. 
Setting $\Lambda=T$, the result is
\begin{align}
    &S_{\rm FO}({\bf L};{\bf v}) = \nonumber \\
    &\frac{g^2 C_F m_D^2}{4\pi}\ln\!\left(\frac{m_D}{T}\right)
    \Bigg[
    \frac{(v^2-1)\arctanh(v)+v}{2v^3}\,
    iL_\parallel\!\left(iL_\parallel+\frac{v}{T}\right)+\frac{3v^3+(1-v^2)^2\arctanh(v)-v}{4v^3}(iL_\perp)^2
    \Bigg]
    \nonumber\\
    &+\frac{g^2 C_F m_D^2}{4\pi}
    \int_0^1 d\mu
    \Bigg[
    \mu^2\,iL_\parallel\!\left(iL_\parallel+\frac{v}{T}\right)
    +\frac{1-\mu^2}{2}(iL_\perp)^2
    \Bigg]\Bigg[
    \Phi\!\left(a_E,b_E\right)
    +\frac{v^2(1-\mu^2)}{2(1-v^2\mu^2)}
    \Phi\!\left(a_T,b_T\right)
    \Bigg] \nonumber \\
    &-\frac{g^2 C_F}{T} \int_{\bf p}
    \left\{ g^2
    \frac{e^{i{\bf p}\cdot{\bf L}}-1}{1-e^{-{\bf p}\cdot{\bf v}/T}}\, \mathcal I_{\rm Hard}
    -\theta(T-|{\bf p}|)\, ({\bf p} \cdot i{\bf L})  \left[{\bf p} \cdot\left(i{\bf L} +\frac{{\bf v}}{T} \right)\right] \frac{T  }{2\,{\bf p}\! \cdot\!{\bf v}} \, \mathcal{I}_{\rm Match}
    \right\},
    \label{eq:SFO-main}
\end{align}
where\footnote{For $b>0$, $\arctan(b/a)$ denotes the angle in $(0,\pi)$, equivalently $\operatorname{atan2}(b,a)$.}
\begin{equation}
    \Phi(a,b)\equiv \tfrac14\ln(a^2+b^2)+\tfrac{a}{2b}\arctan(b/a) \, ,
\end{equation}
and 
\begin{align}
    a_E &\equiv 1-\tfrac{\mu v}{2}\ln\!\left(\tfrac{1+\mu v}{1-\mu v}\right), &
    b_E &\equiv \tfrac{\pi}{2}\mu v \, , \\
    a_T &\equiv \tfrac12\left[\tfrac{(\mu v)^2}{1-(\mu v)^2}
        +\tfrac{\mu v}{2}\ln\!\left(\tfrac{1+\mu v}{1-\mu v}\right)\right], &
    b_T &\equiv \tfrac{\pi}{4}\mu v \, .
\end{align}
Eq.~\eqref{eq:SFO-main} is the strict $\mathcal{O}(g^4)$ fixed order kernel, which we use throughout the remainder of this work.
The first line of Eq.~\eqref{eq:SFO-main} is the logarithmically enhanced
contribution~\eqref{eq:S-LL}; the last two lines are the remaining fixed-order
piece. Reference~\cite{DuPlessis:2026pyr} verifies that $S_{\rm FO}$ is
independent of the separation scale $\Lambda$, and that its second cumulant at
${\bf v}={\bf 0}$ reproduces Eq.~\eqref{eq:kappa-v0}.

Upon analytic continuation to complex separation ${\bf L} = - i{\bf x}$, the fixed-order integral in Eq.~\eqref{eq:SFO-main} converges throughout an open interval $-R_-<\operatorname{Re}x<R_+$. We discuss the values of $R_\pm$ and their physical meaning in detail in the next subsection. Near the boundaries of this interval, the disappearing large momentum suppression of the integrand makes direct numerical integration more demanding. In practice, we use our implementation of the integration (discussed in Appendix~\ref{app:coord_transforms}) to evaluate the integral directly in the KMS strip $-v/T < x < 0$ and use the Taylor expansions about $x=0$ and $x=-v/T$, together with the resummation described in Appendix~\ref{app:moment-resummation}, elsewhere in the domain. These are alternative numerical representations of the same kernel, whose agreement in overlapping regions provides a consistency check. Figure~\ref{fig:stripschematic} compares their convergence intervals. Continuation beyond the first branch points is discussed separately in Appendix~\ref{app:moment-resummation} and illustrated in Fig.~\ref{fig:kernel-continuation}.

It is also important to note that the fixed order kernel inherits all of the consistency checks we performed on the matched kernel.

\subsection{Cumulants, Analyticity and Tails}\label{sec:cumulants}
To begin examining the physical implications of the fixed-order kernel, we proceed along the lines of~\cite{Rajagopal:2025rxr} to calculate the momentum transfer probability in the late-time limit. For the equilibration dynamics of an isotropic distribution function, only the longitudinal projection contributes in the heavy quark limit. We therefore start by introducing the longitudinal evolution kernel
\begin{equation}
    \!\!\!K\!\equiv\!
    S_{\rm FO}({\bf L}\!=\!-ix\hat{\bf v};{\bf v}\!=\!v\hat{\bf v})
    \!=
    \!-\!\sum_{n=1}^\infty \frac{\mathfrak{M}_n(v)}{n!}x^n,
    \label{eq:Kxv}
\end{equation}
where $\mathfrak{M}_n(v) \equiv \hat{\bf v}_{i_1}\! \cdots \hat{\bf v}_{i_n} M^{\rm FO}_{i_1 \ldots i_n}$ are the projected (fixed order) cumulants per unit time of the momentum change probability distribution obtained from the Taylor series of $S_{FO}$ as in Eq.~\eqref{eq:cumulants}. The remainder of this subsection examines the first two
cumulants, which carry the Gaussian characteristics, and then the large-$n$ behavior of
the tower, which controls the domain of $K$ and hence the shape of the tails.

\paragraph{Einstein Relation.}

\begin{figure}[ht]
    \centering
    \includegraphics[width=0.495\linewidth]{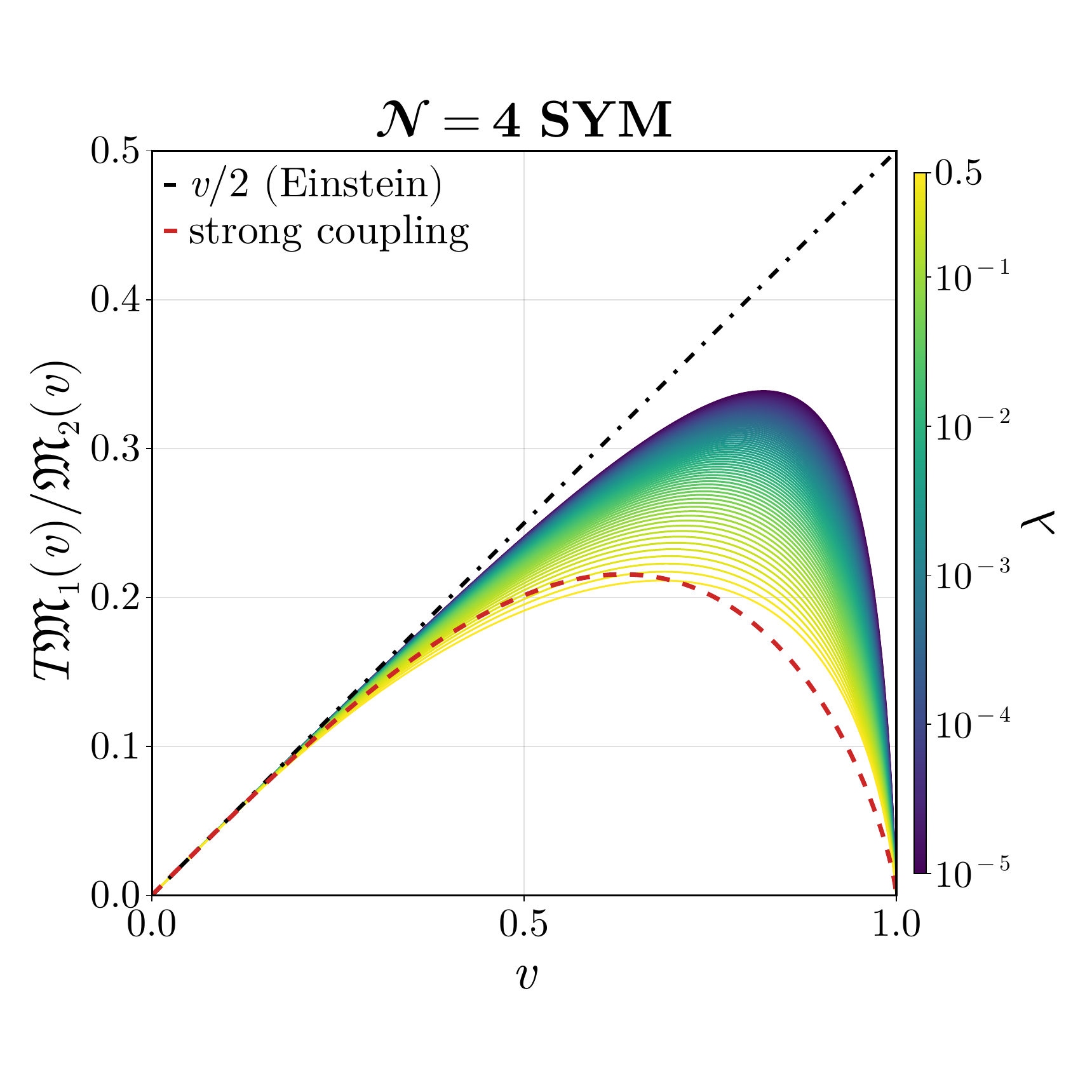}
    \includegraphics[width=0.495\linewidth]{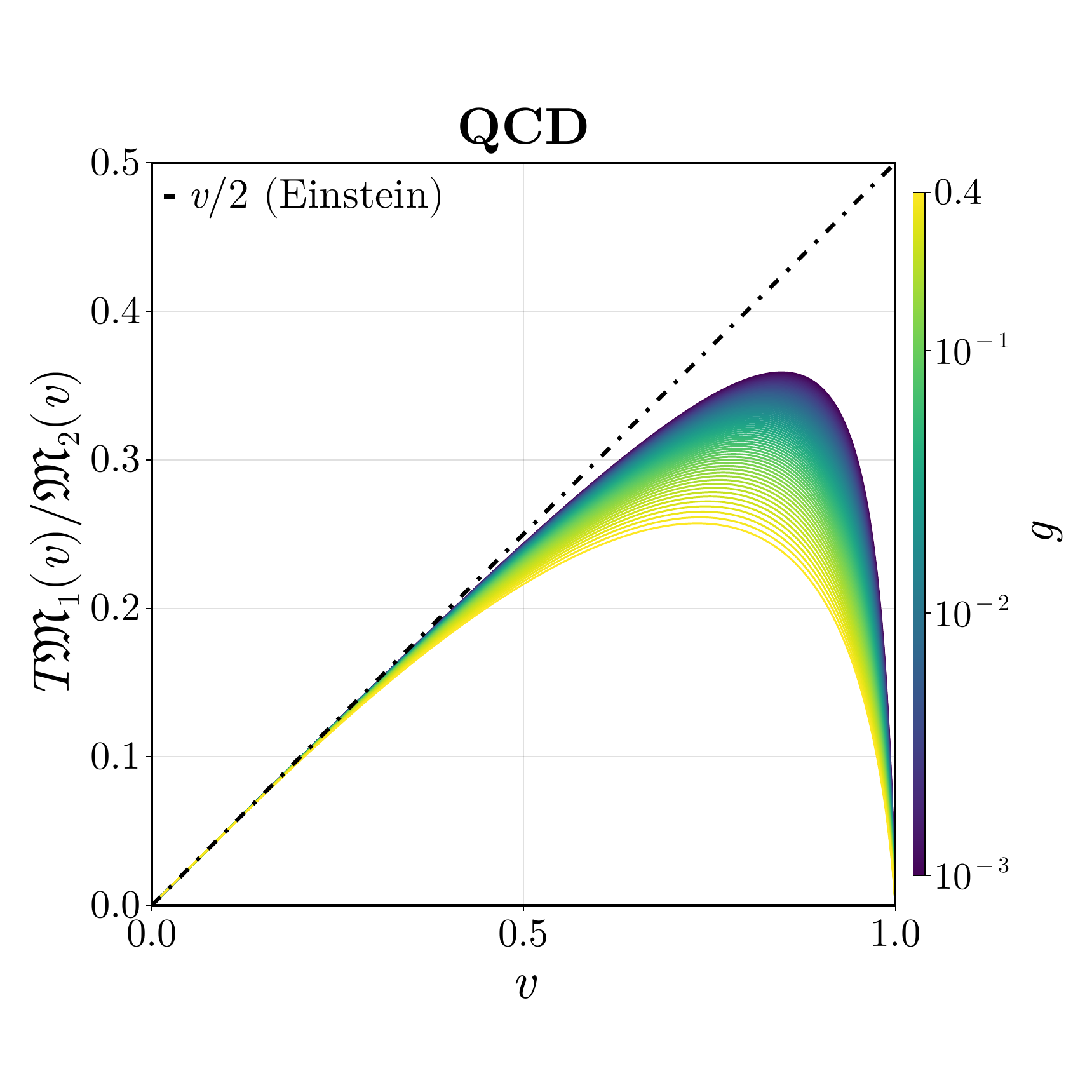}
    \caption{Ratio $T\mathfrak{M}_1(v)/\mathfrak{M}_2(v)$ of the first two longitudinal cumulants compared with the Einstein-relation value $v/2$ (black dot-dashed line). Left: weakly coupled $\mathcal{N}=4$ SYM for the indicated 't Hooft couplings $\lambda$, together with the strong-coupling result (red dashed). Right: QCD for the indicated gauge couplings $g$. For fixed $v<1$ the weak-coupling curves approach $v/2$ as the coupling decreases, and all results recover the Einstein relation as $v\to0$. At fixed nonzero coupling, however, the ratio bends away from $v/2$ at sufficiently large velocity, with the turnover moving toward $v=1$ as the coupling decreases, exhibiting the noncommutativity of the weak-coupling and ultrarelativistic limits.}
    \label{fig:einstein-ratio}
\end{figure}

Since $-K$ is the scaled cumulant generating function of the momentum
transfer distribution, Eq.~\eqref{eq:cumulants} gives
$\langle k_\parallel^n\rangle_c = t\,T\,\mathfrak{M}_n$, and therefore
\begin{align}
    \mathfrak{M}_1(v) = \frac{\eta_D(v)\, |{\bf p}|}{T} \, ,
    & &
    \mathfrak{M}_2(v) = \frac{\kappa_L(v)}{T} \, ,
\end{align}
both positive --- the latter guaranteed by the concavity of $K$ established
in Section~\ref{sec:equilibration-setup}. Consequently the Einstein
relation $\kappa_L = 2TE\eta_D$ is the statement
\begin{equation}
    \frac{T\,\mathfrak{M}_1(v)}{\mathfrak{M}_2(v)} = \frac{|{\bf p}|}{2E} = \frac{v}{2} \, ,
\end{equation}
which we plot as a function of velocity at several couplings in
Fig.~\ref{fig:einstein-ratio}. As expected, it is restored both in the
non-relativistic limit and as $g\to0$. In the left panel, weak and strongly coupled $\mathcal{N}=4$ SYM shows much the same qualitative behavior of this ratio as a function of velocity. It is also clear from this figure that the small coupling and large velocity limits do not commute.

\paragraph{Large-Order Cumulants and the Domain of \texorpdfstring{$K$}{K}.}

Because the soft and matching terms in Eq.~\eqref{eq:SFO-main} are at most
quadratic in ${\bf L}$, the cumulants with $n\geq3$ are determined entirely
by the hard sector. This makes their large-$n$ behavior --- and hence the domain of
analyticity of $K$ --- accessible analytically. At $v=0$ the cumulants can
be evaluated in closed form (see Appendix~\ref{app:moments}):
\begin{align} 
    \mathfrak{M}_n(0) = &\frac{3 g^2 C_F m_D^2\, 2^{n-1} T^{n-2} (n-1)!\, \zeta(n)}
    {\pi^3 (n+1)\, \mathcal{N}} \nonumber \\
    & \times \left[ \frac{N_c\big(N_X n(n+2) + 2(n^2-2n+8)\big)}{8(n-2)(n+2)}
    + \frac{(1-2^{1-n}) T_R N_f}{n-2} \right] \, , \label{eq:Mn-v0-closed}
\end{align}
for even $n\geq4$, with $\mathfrak{M}_n(0)=0$ for odd $n$. The explicit $(n-2)^{-1}$
pole reflects the infrared sensitivity of the diffusion coefficient, and is
precisely why $n=2$ had to be calculated separately including HTL-resummed physics supplemented by an appropriate matching calculation.

At arbitrary velocity, the large-$n$ asymptotics are controlled by a
codimension-two corner of phase space --- the threshold of the $k_0$
integration together with the collinear angular endpoint --- and we find
(see Appendix~\ref{app:large-order-arbitrary-v})
\begin{equation} \label{eq:Mn-large-n}
    \mathfrak{M}_n(v) \sim
    \frac{g^4 C_F N_c (N_X+2)}{32\pi^3}\, \frac{n!}{n^2}
    \left[ (1-v)\left(\frac{2T}{1-v}\right)^{\!n}
    + (-1)^n (1+v)\left(\frac{2T}{1+v}\right)^{\!n} \right] ,
\end{equation}
with the fermionic
contribution being suppressed by one further power of $1/n$.

This expression defines two radii
\begin{align}
    R_+(v) = \frac{1-v}{2T} \, , & & R_-(v) = \frac{1+v}{2T} \, ,
\end{align}
of which the smallest defines the radius of convergence of the series~\eqref{eq:Kxv} about $x=0$ as
\begin{equation}
R(v)^{-1}=\limsup_{n\to\infty}\left(\frac{|\mathfrak M_n(v)|}{n!}\right)^{1/n}=R_+(v)^{-1} \, .\footnote{At this point it becomes clear that, in the ultra-relativistic limit we have $R(v) = 1/[2\gamma^2(1+v)T] \to 1/(4\gamma^2 T)$ as $v\to1$.} \label{eq:R-in-terms-of-Rplus}
\end{equation}
 The presence of the second radius $R_-(v)$ is in fact a necessary consequence of the equilibration condition. In terms of the moment series, or equivalently in terms of a condition on the moments, the equilibration condition is
\begin{align}
    \sum_{n=0}^\infty \frac{x^n}{n!} \mathfrak{M}_n = \sum_{n=0}^\infty \frac{(-x-v/T)^n}{n!} \mathfrak{M}_n & & \iff & & \mathfrak{M}_n = \sum_{k=0}^\infty \frac{(-1)^{n+k}}{k!} \left(\frac{v}{T}\right)^{k} \mathfrak{M}_{n+k} \, , \label{eq:equil-in-terms-of-series}
\end{align}
whenever both series are well-defined (and we have introduced $\mathfrak M_0 = 0$). Because the radius of convergence is $R_+ = (1-v)/2T$, one can check that both series on the left side of the equivalence above can be simultaneously evaluated for $-(1-v)< 2Tx<(1-3v)$ --- meaning that this way of writing the equilibration condition is, strictly speaking, only meaningful when $v<1/2$ (otherwise there is no $x$ for which~\eqref{eq:equil-in-terms-of-series} makes sense). Furthermore, note that for both series to be valid when $x=0$ we need $v<1/3$.

Then, one can check that if the leading behavior at large $n$ is of the form $\frac{1}{n!} \mathfrak{M}_n \sim  (R_+^{-1})^n + (-R_-^{-1})^n$, then this requires
\begin{align}
    (R_+^{-1})^n + (-R_-^{-1})^n &\sim \frac{1}{n!} \sum_{k=0}^\infty \frac{(-1)^{n+k} (n+k)!}{k!} \left(\frac{v}{T}\right)^{k} \left[ (R_+^{-1})^{n+k} + (-R_-^{-1})^{n+k} \right] \nonumber \\ &= \frac{(-R_+^{-1})^n}{(1+R_+^{-1} v/T)^{n+1}} + \frac{(R_-^{-1})^n}{(1-R_-^{-1} v/T)^{n+1}} \, ,
\end{align}
which means that, in order for these expressions to be consistent in their large $n$ behavior, $R_\pm$ must satisfy 
\begin{align}
    R_+ = R_- - \frac{v}{T} \, .
\end{align}
This therefore provides a valuable cross-check to our result.
The interpretation of these radii then becomes clear: the radius of convergence of the expansion of $K$ about $x=0$ is
$R(v) = R_+(v)$, setting the domain of $K$ towards positive $x$, whereas $R_-(v)$ signals the presence of the same singularity bounding the domain of $K$ at the KMS conjugate position $x \to -x - v/T$, sitting on the negative axis sits at $x=-R_-(v)$. Thus the longitudinal convergence interval is $-R_-<x<R_+$, with $R_--R_+=v/T$ as required by Eq.~\eqref{eq:K-analytic-property}.

It is worth looking more closely at the structure that underlies these radii, since it is in fact
more transparent than any of the individual limiting cases of physical interest. At large $|{\bf p}|$, each
momentum direction $\hat{\bf p}$ contributes to the hard kernel a factor given by the radial
exponential $\exp\{p[\hat{\bf p}\cdot{\bf x} - (1-{\bf v}\cdot\hat{\bf p})/(2T)]\}$,
so that uniform exponential convergence in \emph{every} direction requires
\begin{equation} \label{eq:convergence-ball}
    \big| 2T{\bf x} + {\bf v} \big| < 1 \, .
\end{equation}
That is to say, the interior of the domain of analyticity of the hard kernel is a ball of radius $1/(2T)$ centered at $-{\bf v}/(2T)$ --- manifestly invariant
under the map ${\bf x}\to-{\bf v}/T-{\bf x}$ as required by the equilibration condition. Its longitudinal
diameter reproduces $R_\pm(v)$, its intersection with the transverse line passing through the origin
gives $R_\perp(v) = \sqrt{1-v^2}/(2T)$
(see Appendix~\ref{app:large-order-transverse-arbitrary-v}), and consequently
\begin{equation}
    R_\perp(v)^2 = R_+(v) R_-(v) \, .
\end{equation}

This geometry also organizes the ultra-relativistic limit. The ball has fixed radius
$1/(2T)$, while the distance from the expansion point $x=0$ to its centre is $v/(2T)$;
as $v\to1$ the expansion point therefore migrates onto the boundary sphere itself. The positive longitudinal and transverse radii vanish, while the negative longitudinal radius remains finite:
\begin{align} \label{eq:radii-ultrarelativistic}
    R_+(v) \to \frac{1}{4\gamma^2 T} \, , & &
    R_\perp(v) = \frac{1}{2\gamma T} \, , & &
    R_-(v) \to \frac{1}{T} \, .
\end{align}
Momentum gain thus remains exponentially suppressed at the thermal scale at any boost,
whereas both forward longitudinal and transverse momentum transfer lose their
exponential suppression --- the latter more slowly by one power of $\gamma$. The
transverse channel is not needed for the equilibration dynamics studied here, which
involves only the longitudinal projection;
we shall comment further on what the next steps to explore the transverse dynamics would be in Sec.~\ref{sec:conclusions-outlook}.

\begin{figure}
    \centering
    \includegraphics[width=\linewidth]{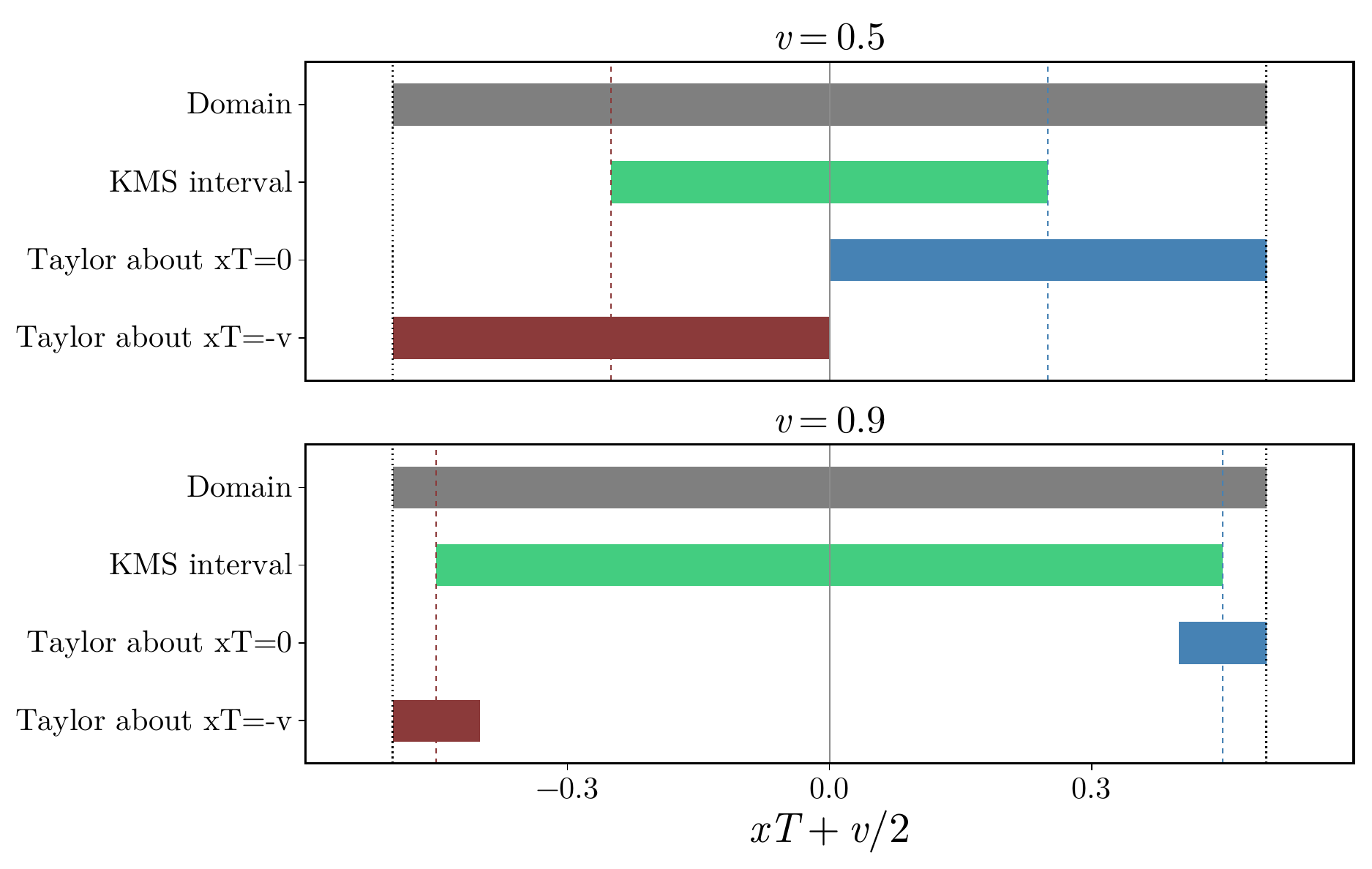}
    \caption{Longitudinal convergence domains in the centered coordinate $xT+v/2$, for $v=0.5$ and $v=0.9$. Gray shows the full convergence domain of the fixed-order integral, $|2Tx+v|<1$. Green marks the KMS interval $-v/T<x<0$, where direct quadrature (numerical integration) is used in our calculation. Blue and red show the convergence intervals of the Taylor expansions about $x=0$ and $x=-v/T$, respectively.}
    \label{fig:stripschematic}
\end{figure}

With the domain established, we now describe how we evaluate $K$ in practice. Inside the
interval $-v/T<x<0$ we obtain it directly from the integral representation in
Eq.~\eqref{eq:SFO-main}; for $x>0$ from the Taylor expansion about $x=0$;
and for
$x<-v/T$ from the corresponding expansion about $x=-v/T$ implied by
Eq.~\eqref{eq:K-analytic-property}, convergent out to $-R_-(v)$. Together with the direct integral representation on the KMS interval $-v/T<x<0$, the Taylor expansions about $x=0$ and $x=-v/T$ therefore cover the entire longitudinal analytic domain $-R_-(v)<x<R_+(v)$. Each Taylor expansion overlaps with the directly calculable KMS interval, and their agreement there provides a nontrivial check of the analytic continuation. We show the
overlapping regions schematically in Fig.~\ref{fig:stripschematic}.

Finally, the $n^{-2}$ prefactor in Eq.~\eqref{eq:Mn-large-n} fixes the \emph{type}
of the endpoint singularity. Up to less singular terms, we have
\begin{align} \label{eq:dilog-kernel}
    K_{\rm sing}(x,v) &= -A_+(v)\,{\rm Li}_2\!\left(\frac{x}{R_+(v)}\right)
    - A_-(v)\,{\rm Li}_2\!\left(-\frac{x}{R_-(v)}\right) \, ,
     \\ A_\pm(v) &\equiv \frac{g^4C_FN_c(N_X+2)}{32\pi^3}(1\mp v) \, ,
\end{align} 
and since ${\rm Li}_2(1-\epsilon)=\pi^2/6+\epsilon\ln\epsilon-\epsilon+\ldots$,
the endpoints are logarithmic branch points: $K$ stays finite there while its
first derivative diverges logarithmically. These are the branch points shown in Fig.~\ref{fig:complexKplot}.

Some properties of the momentum transfer distribution follow immediately.
At the level of the extensive exponent $\tilde{S}$, the large momentum transfer limit $C_L\to+\infty$ implies that the corresponding value of ${\bf x}$ (set by ${\bf C} + \frac{\partial K}{\partial {\bf x}} = 0$) gets pushed to $x\to R_+(v)$ due to the presence of the logarithmic divergence of $\partial_x K$ there. Meanwhile, $K$ approaches a finite value,
therefore $\tilde{S} = - x \frac{\partial K}{\partial x} + K \to R_+(v)\,C_L + {\rm const.}$, and consequently the tails of
Eq.~\eqref{eq:legendresaddledistribution} are \emph{purely} exponential,
\begin{align}
    &P(k_L) \propto e^{- R_+(v) k_L } & & {\rm as} \,\, k_L \to +\infty \, , \label{eq:tail-plus} \\
    &P(k_L) \propto e^{ + R_-(v) k_L } & &{\rm as} \,\, k_L \to -\infty \, , \label{eq:tail-minus}
\end{align}
with the two slopes differing by exactly $v/T$, as the equilibration
condition~\eqref{eq:K-analytic-property} requires. 

It is instructive to contrast this with the strongly coupled case~\cite{Rajagopal:2025ukd}. There the
cumulants grow as $\mathfrak{M}_n \sim a\,n!\,R^{-n}$ with no polynomial
suppression, so that the endpoint singularity is a simple pole rather than the logarithmic branch point found at weak coupling. In general, the location of the nearest singularity controls the leading exponential tail in both limits, while its order controls the subleading approach to this asymptotic behavior. We defer a more detailed investigation of this distinction to future work.

It is clear from Eqs.~\eqref{eq:tail-plus} and~\eqref{eq:tail-minus} that the convergence radii $R_\pm$ define the exponents of the exponential tails of the momentum transfer distribution. As such, we define the \emph{tail exponent} as $R(v)=R_+(v)$, which encodes the likelihood of experiencing rare large momentum kicks and which, both at weak and at strong coupling, is independent of the coupling at leading order. We have already compared the two in Fig.~\ref{fig:radconv}.

\subsection{Momentum Transfer Distributions across Gauge Theories}
\label{sec:result-numerics}

\begin{figure}
    \centering
    \includegraphics[width=0.48\linewidth]{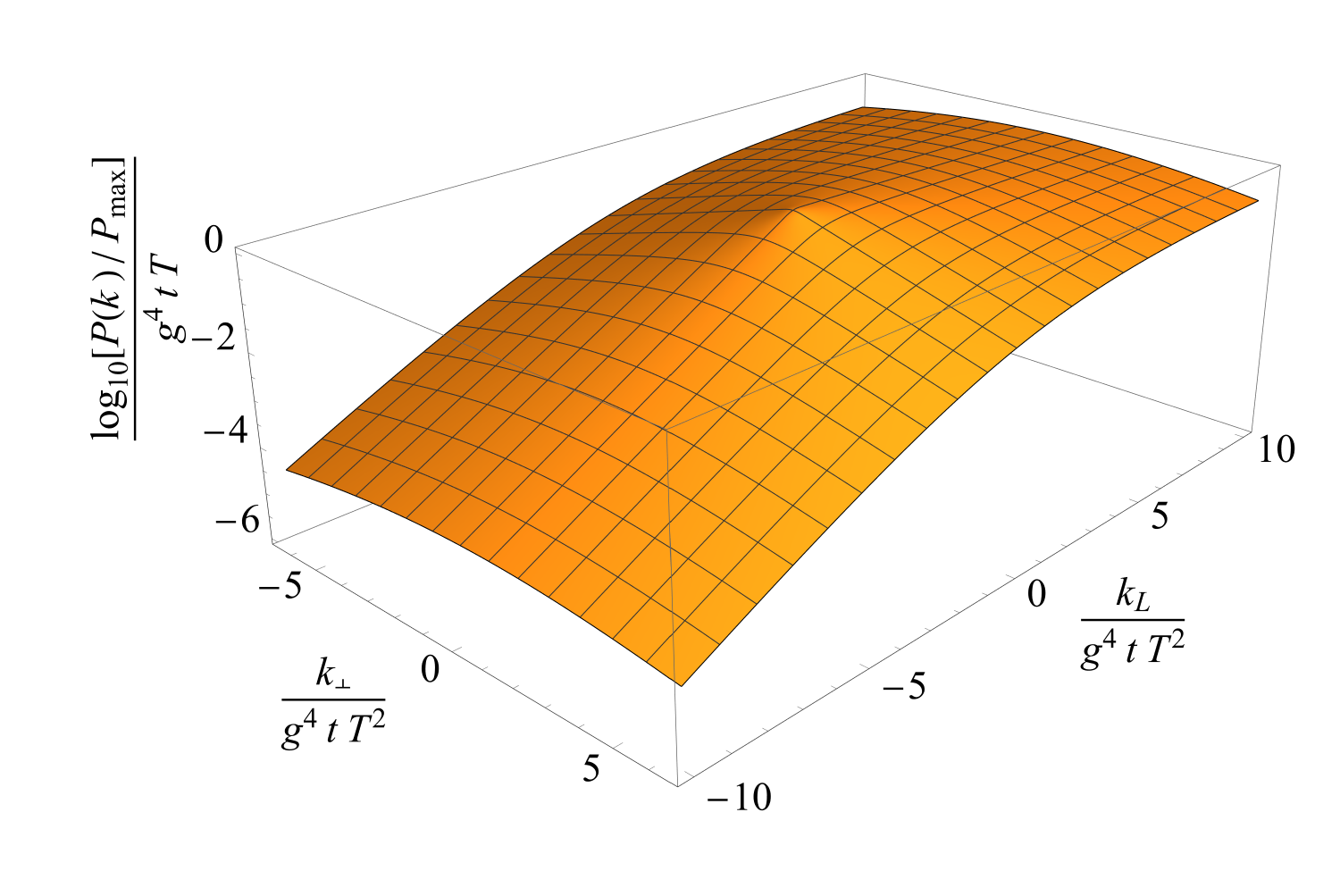}
    \includegraphics[width=0.48\linewidth]{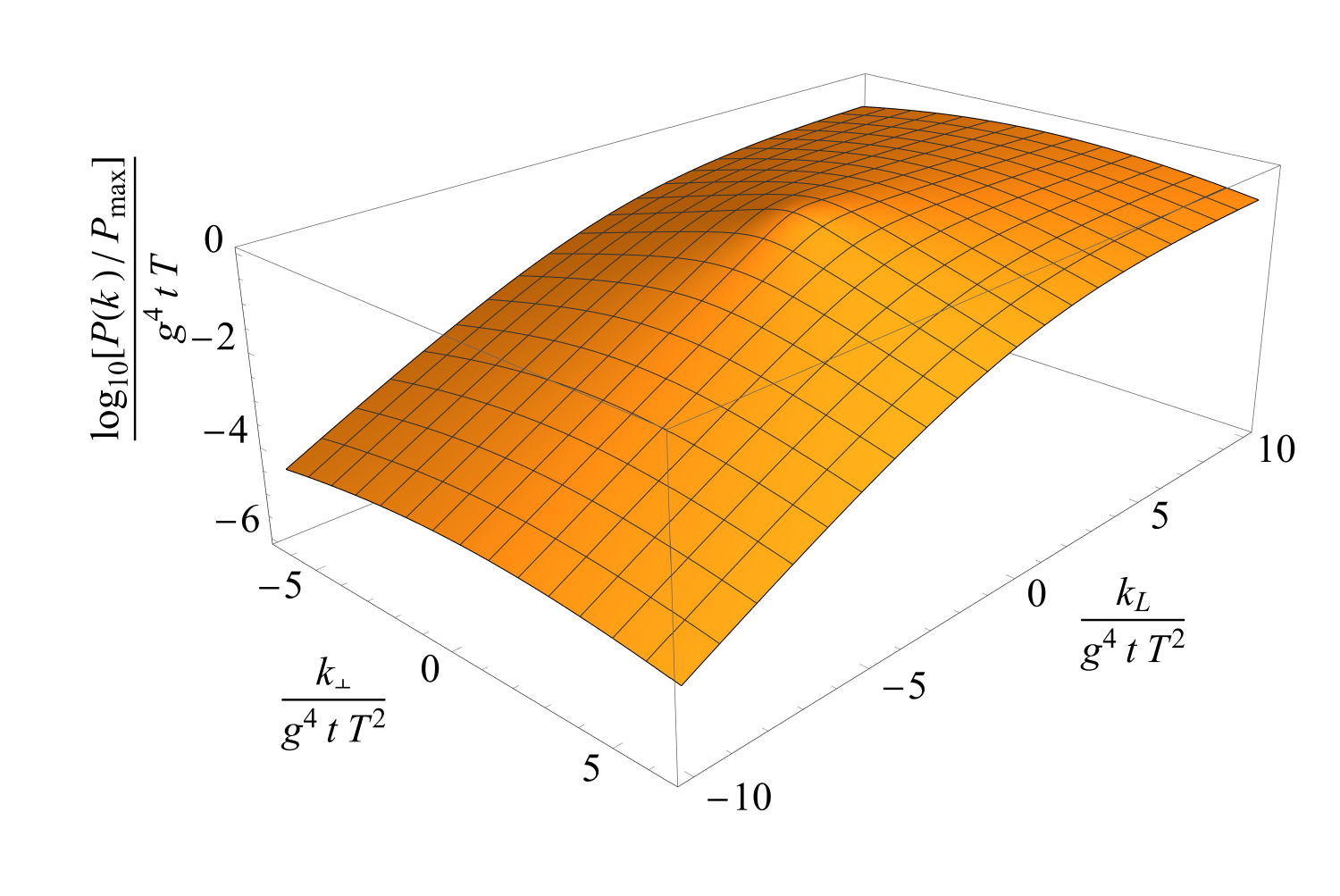}
    \caption{Momentum transfer distributions calculated from the Legendre transform of the logarithm of the Wilson loop, following Eqs.~\eqref{eq:P-of-S-Fourier}--\eqref{eq:tildeS-from-S}, relying on the large time limit $g^4 t T \to \infty$. Together with $T$, this large time sets the scales of the axes. Left: three-flavor QCD at $g=0.1$. Right: weakly coupled $\mathcal{N}=4$ SYM at $\lambda=0.03$. Both are evaluated at $v=0.9$. We show $\log_{10}[P({\bf k};v)/P_{\max}]/(g^4tT)$ as a function of the scaled longitudinal momentum transfer $k_L/(g^4tT^2)$ and one signed transverse Cartesian component $k_\perp/(g^4tT^2)$. The distribution is symmetric under $k_\perp\to-k_\perp$ but asymmetric under $k_L\to-k_L$, reflecting detailed balance between longitudinal momentum loss and gain.}
    \label{fig:Surfaces}
\end{figure}

\begin{figure}
    \centering
    \includegraphics[width=\linewidth]{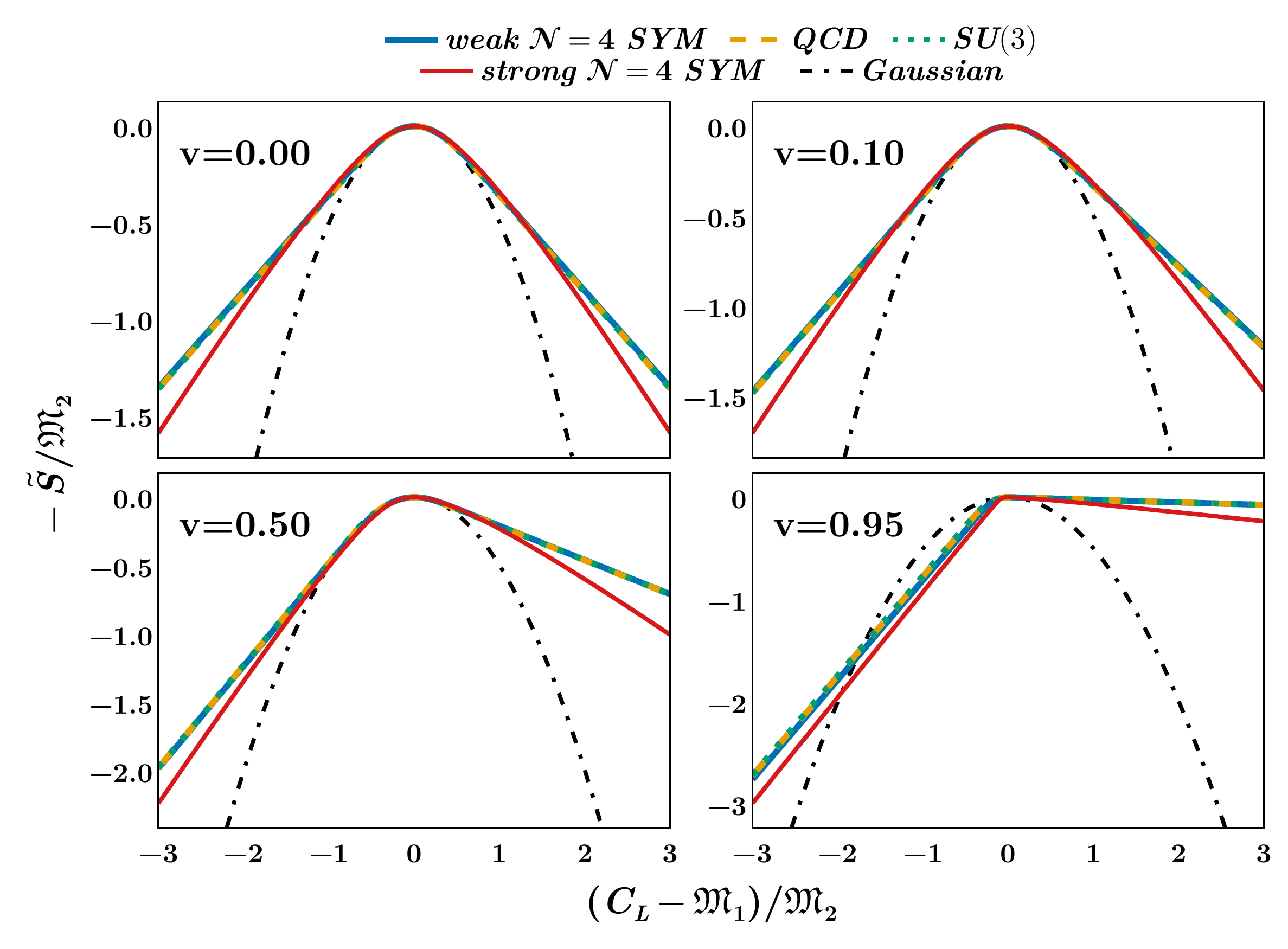}
    \caption{Normalized log-probability density for longitudinal momentum transfer, $-\tilde{S}/\mathfrak{M}_2$ as a function of $X=(C_L-\mathfrak{M}_1)/\mathfrak{M}_2$ for $v=0,0.1,0.5,0.95$. Weakly coupled $SU(3)$ Yang--Mills, QCD and $\mathcal{N}=4$ SYM are shown at $g=0.1$, together with strongly coupled $\mathcal{N}=4$ SYM at $\lambda=0.03$. The black dot-dashed curve, $-X^2/2$, is the universal Gaussian reference after centering and scaling by the first two cumulants. The weakly coupled theories exhibit very similar normalized shapes: an approximately Gaussian core gives way to linear wings corresponding to exponential momentum-transfer tails, with an increasingly pronounced asymmetry between momentum loss and momentum gain as $v$ increases.}
    \label{fig:StildeComparison}
\end{figure}

We now examine how the momentum transfer distribution depends on the matter content of
the plasma, and how it compares to its strongly coupled counterpart.
The object to compare is $\tilde{S} = K - {\bf x} \cdot \frac{\partial K}{\partial {\bf x}}$ as a function of ${\bf C} = - \frac{\partial K}{\partial {\bf x}}$ rather than $K$ itself: $\tilde{S}$ is the logarithm of the
momentum transfer probability, so its shape is directly physical, whereas $K$ is a
generating function, whose dynamical consequences are comparatively harder to read off. While it is possible to visualize the whole distribution as a function of transverse and longitudinal momentum simultaneously, as in Fig.~\ref{fig:Surfaces}, where many features are already apparent (e.g., that the distributions are highly skewed towards energy loss at a rather modest velocity), a cleaner comparison for our later purposes is to only visualize its longitudinal features.\footnote{Needless to say, transverse dynamics is also of high phenomenological interest. In this work, we choose to focus on the process of kinetic equilibration, and leave a detailed study of the transverse dynamics to future work.} In Fig.~\ref{fig:StildeComparison}
we plot $-\tilde{S}/\mathfrak{M}_2$ against $(C_L-\mathfrak{M}_1)/\mathfrak{M}_2$. This
normalization centers the distribution on its mean and rescales by its width, so that
\emph{any} Gaussian process collapses onto the single parabola $-\tfrac12 X^2$, shown as
the black dot-dashed curve. Departures from that parabola therefore isolate genuine non-Gaussian structure. Note that the scaling of the axes is different from that in Ref.~\cite{DuPlessis:2026pyr}.

Three features stand out. First, the three weakly coupled theories lie almost on top of
one another at every velocity: the matter content changes the overall normalization of
$K$, but has remarkably little effect on the normalized shape.
Second, the departure from Gaussianity is not solely a large-velocity effect. Already at $v=0$
the tails are approximately linear rather than parabolic --- and are exactly the exponential tails of
Eqs.~\eqref{eq:tail-plus}--\eqref{eq:tail-minus} --- a direct consequence of the large even cumulants of
Eq.~\eqref{eq:Mn-v0-closed}. Third, the asymmetry grows rapidly with velocity: by
$v=0.5$ the momentum-loss side is visibly flatter than the momentum-gain side, and by
$v=0.95$ the two are dramatically different, with the loss side almost flat. This is the
$v/T$ splitting between $R_+$ and $R_-$ demanded by
Eq.~\eqref{eq:K-analytic-property}, and it is precisely the feature a Gaussian truncation
cannot reproduce.

Figure~\ref{fig:StildeComparison} also includes the strongly coupled $\mathcal{N}=4$ SYM
result of~\cite{Rajagopal:2025ukd}. Once the Gaussian scale is divided out, the logarithm of the momentum transfer probability exhibits
the same qualitative profile: near-parabolic core, approximately linear tails, asymmetry
growing with velocity. Section~\ref{sec:cumulants} makes clear why this had to be so. In
both theories $K$ has a bounded domain of analyticity whose two edges are related by the
equilibration condition, and it is that boundedness --- not any detail of the microscopic
dynamics --- which produces exponential tails with slopes differing by exactly $v/T$.
Whether or not the medium is made of quasiparticles does not enter.

\begin{figure}
    \centering
    \includegraphics[width=\linewidth]{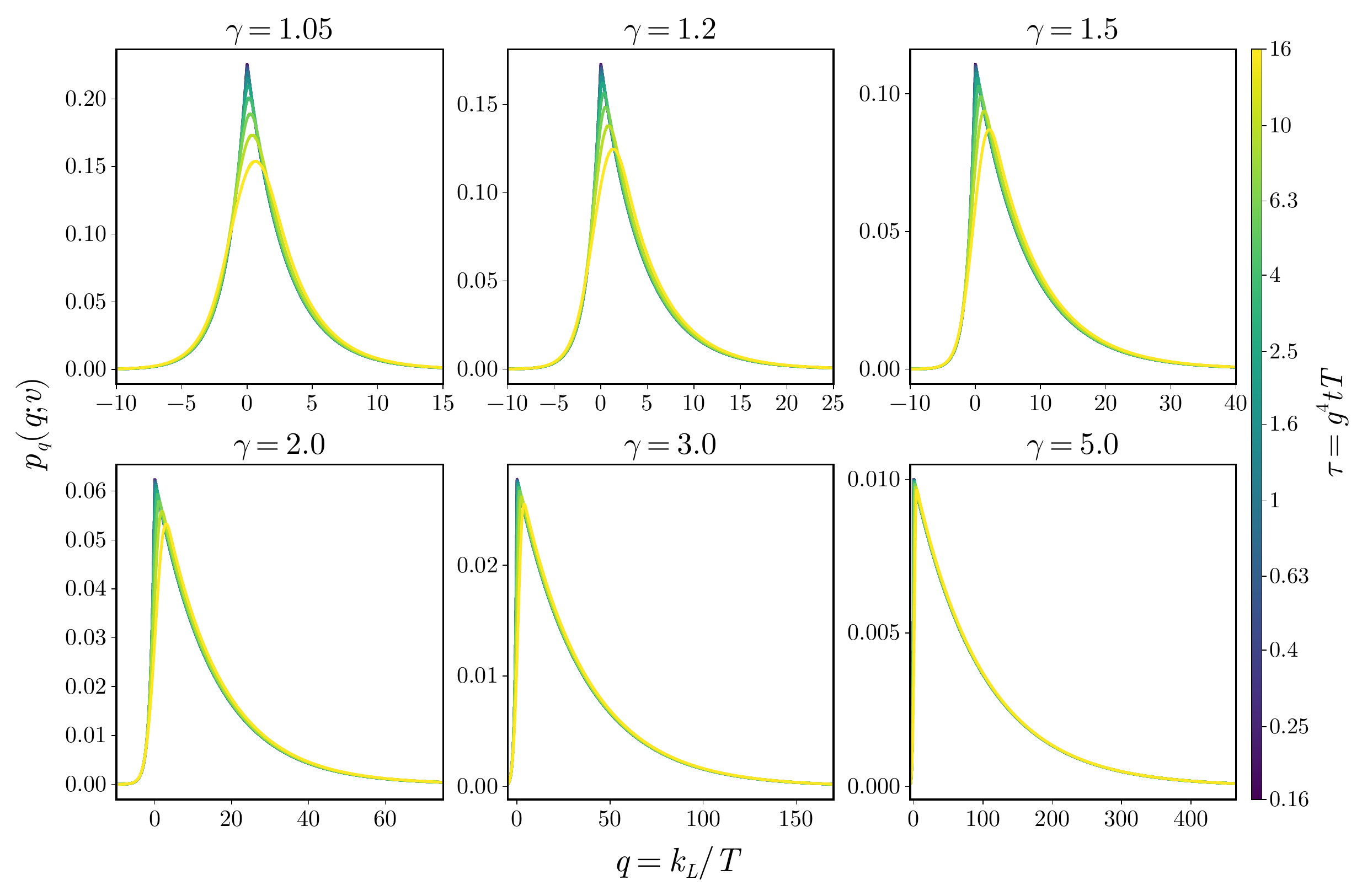}
    \caption{Momentum transfer distributions calculated from the Legendre transform of the logarithm of the Wilson loop, following Eqs.~\eqref{eq:P-of-S-Fourier}--\eqref{eq:tildeS-from-S}, relying on the large time limit $g^4 t T \to \infty$. We show $p_q(q;v) = T P(k_L=qT;v)$ in massless QCD ($N_c=N_f=3$, $N_X=0$, $g=0.1$) at the indicated boost factors. Colors denote the rescaled propagation time $g^4tT$. Positive $k_L$ corresponds to momentum transferred from the heavy quark to the medium. With increasing time the accumulated distribution broadens, while increasing $\gamma$ produces a progressively broader momentum-loss tail relative to the momentum-gain tail, reflecting the boost dependence of $R_+$ and $R_-$.}
    \label{fig:saddle-density}
\end{figure}

The differences are quantitative rather than qualitative. The dominant difference is where the singularities sit, and hence how steep the tails are. The strongly coupled tails
are steeper at every velocity --- $R(0)T \simeq 0.835$ against $1/2$ --- and
increasingly so at large boost, as Fig.~\ref{fig:radconv} shows. A finer difference is their analytic character --- a logarithmic branch point at weak coupling versus a simple pole at strong coupling --- which affects subleading details of the approach to the asymptotic tails but leaves no readily visible imprint on the curves shown here.

The picture that emerges is that the momentum transfer distribution in weakly coupled gauge theories share a common qualitative structure that is also present in the strongly coupled counterpart: exponential tails whose slopes differ by $v/T$. This arises due to a bounded domain of analyticity of the evolution kernel $K$. The differences between the gauge theories, instead, is quantitative --- where the singularities sit, and hence how steep the tails are --- and, at a finer (sub-leading) level, what type of singularity they are. 

\paragraph{A closer look at QCD.}
 Fig.~\ref{fig:saddle-density} shows the weakly coupled massless QCD distributions themselves, obtained from
Eq.~\eqref{eq:legendresaddledistribution}, at several boost factors and over a range of rescaled times $g^4tT$. As the propagation time increases, the distribution broadens and accumulates positive longitudinal momentum transfer, corresponding to momentum loss by the heavy quark. The velocity dependence is considerably more striking. With increasing $\gamma$, the positive-$k_L$ tail broadens rapidly while the momentum-gain side remains strongly suppressed. This is the finite-time manifestation of the hierarchy $R_+\sim1/(4\gamma^2T)$ and $R_-\to1/T$: at large boost, rare momentum-loss fluctuations can extend over parametrically larger momentum transfers even though comparably large momentum gains remain thermally costly. The independence of the exponential tails is also very visible, contrasted by the mean and Gaussian width scaling linearly with time.

\section{How Do Relativistic Heavy Quarks Thermalize?}
\label{sec:Kolmogorov}

With the complete evolution kernel in hand, we are now in a position to explore the equilibration dynamics of heavy quarks. In addition to a more detailed understanding of how equilibration proceeds, the results of our calculation also allows us to compare the full dynamics to commonly used truncations in phenomenological practice, and assess how different the dynamics is from the strongly coupled counterpart that~\cite{Rajagopal:2026urq} explored.

The first two cumulants calculated in Section~\ref{sec:cumulants} do not satisfy the Einstein relation at finite velocity, either at weak or at strong coupling. A Fokker--Planck description that retains these microscopic transport coefficients therefore does not have the Boltzmann distribution as its stationary solution. Nevertheless, as we saw, the complete evolution kernel does satisfy the equilibration condition. As such, a comparison between these will not only reveal how different assumptions affect the approach to equilibrium, but also how the non-Gaussian fluctuations reconcile the breaking of the Einstein relation with equilibration.

We follow the Hamilton--Jacobi formulation developed for strongly coupled $\mathcal N=4$ SYM in Ref.~\cite{Rajagopal:2026urq}. Formally, this formulation contains all of the information of the dynamics at leading order in a $1/M$ expansion.
We first consider the dynamics following from the Fokker--Planck truncation with the microscopic drag and diffusion coefficients, then we consider dynamics where we impose the Einstein relation by keeping the drag coefficient and setting the value of the diffusion coefficient by hand, and finally we restore the complete non-Gaussian kernel and study equilibration in the presence of non-Gaussian fluctuations. This sequence of steps allows us to tackle two separate questions: whether a given dynamical evolution has the correct equilibrium solution, and, regardless of its correctness, how the distribution of heavy quarks approaches each respective solution. Throughout this section we consider an infinite medium held at constant temperature.

\subsection{Fokker--Planck evolution with microscopic transport coefficients}

We follow the notation of Ref.~\cite{Rajagopal:2026urq} and write the heavy quark momentum distribution as $\mathscr{P}$. This is not to be confused with the momentum transfer probability $P$, which does not feature in this section. For a spherically symmetric distribution, writing $f\equiv f_{-1}$ in the expansion introduced in Section~\ref{sec:Kolmogorov-setup}, the leading-order evolution equation for $f$ written in terms of dimensionless variables is
\begin{equation}
\partial_{\tau_d}f(u,\tau_d)=-\frac{T^2}{M\eta_D(0)}K\!\left(\frac{1}{T}\partial_u f;v(u)\right) \, ,
\label{eq:weak-evolution-HJ}
\end{equation}
where
\begin{align}
    u=\frac{p}{M} \, , & & v(u)=\frac{u}{\sqrt{1+u^2}} \, , & & \tau_d=\eta_D(0)t \, .
\end{align}
Here $\eta_D(0)$ is the zero-velocity drag coefficient in each theory. Following the evolution in terms of $\tau_d$ allows one to scale out from the dynamics the difference in the characteristic time scales of each theory, and focus on the velocity-dependent features with a common non-relativistic baseline. 
At strong coupling the drag coefficient is given by $\eta_D = \pi\sqrt{\lambda}T^2/(2M)$, so $\tau_d$ is $\pi/2$ times the time variable used in Ref.~\cite{Rajagopal:2026urq}.

We solve the Hamilton--Jacobi equation~\eqref{eq:weak-evolution-HJ} by the method of characteristics, which is equivalent to solving Hamilton's equations in terms of the coordinate variable $u$ and its canonical conjugate momentum (which is not a physical kinetic momentum), equal to $\partial_u f$ on the solution. In terms of the dimensionful variable $x=T^{-1}\partial_u f$,  
the evolution equations are
\begin{equation}
\frac{du}{d\tau_d}=\frac{T}{M\eta_D(0)}\partial_x K,
\qquad
\frac{dx}{d\tau_d}=-\frac{T}{M\eta_D(0)}\partial_u K,
\label{eq:weak-Hamilton-equations}
\end{equation}
where $\partial_u K$ is evaluated at fixed $x$. Even without the earlier discussion, it is clear from these equations that $Tx$ corresponds to the momentum conjugate to $u$, and that $K$ plays the role of a Hamiltonian. 

We use the FONLL-motivated initial condition of the bottom-like example in Ref.~\cite{Rajagopal:2026urq}, specified by
\begin{equation}
\mathscr P_0(u)=\frac{A}{[1+(u/s_0)^2]^3},
\qquad s_0=1.7,\qquad M=4.75~{\rm GeV},\qquad T=0.20~{\rm GeV}.
\label{eq:weak-initial-condition}
\end{equation}
Its normalization does not enter the characteristic equations. The initial conditions for $(u,x)$ are set by sampling $(u,\partial_u f)$ along the (positive) real $u$ line. That is to say, for a given value of $u$ the initial condition for $x$ follows from $Tx = \partial_u f = -6Tu/[M(s_0^2+u^2)]$. In what follows, the weak-coupling comparisons use QCD at $g=0.1$ and $\mathcal N=4$ SYM at $\lambda=0.03$, corresponding to $g=0.1$ for $N_c=3$. We also include QCD at $g=0.01$ to exhibit the coupling dependence, which quantifies the relative importance of soft (HTL-resummed) physics. This is because once the equations are written in terms of $\tau_d$, the only remaining coupling dependence appears through the ratio of the Debye screening scale $m_D$ and the temperature $T$.

The direct Fokker--Planck truncation is obtained by truncating the series in Eq.~\eqref{eq:Kxv} to the first two cumulants, giving the Gaussian kernel
\begin{equation}
K_{\rm FP}(x;v)=-\mathfrak M_1(v)x-\frac{\mathfrak M_2(v)}{2}x^2 \, ,
\label{eq:weak-unprojected-FP}
\end{equation}
and nothing else.
As per our calculation in Section~\ref{sec:calculation}, these cumulants are obtained from the complete fixed-order kernel, including the matching contributions necessary to fully define the weakly coupled result. This evolution kernel possesses a nonzero root as a function of $x$ given by $x=-2\mathfrak M_1/\mathfrak M_2$. Note that, if the equilibration condition $K(x) = K(-x-v/T)$ held true, this root would be constrained to being at $x=-v/T$. However, a direct truncation of the series does not guarantee this. Consequently, the nontrivial stationary solution of Eq.~\eqref{eq:weak-evolution-HJ} is given by
\begin{equation}
f_{\rm FP,eq}(u)-f_{\rm FP,eq}(0)=-2T\int_0^u dw\,\frac{\mathfrak M_1(v(w))}{\mathfrak M_2(v(w))},
\label{eq:weak-unprojected-stationary}
\end{equation}
whereas the equilibration condition guarantees that the equilibrium distribution is given by Boltzmann weights, i.e., $f_{\rm eq}(u)-f_{\rm eq}(0)=1-\sqrt{1+u^2}$. These agree only when the Einstein relation holds.

\begin{figure}[t]
\centering
\includegraphics[width=0.95\linewidth]{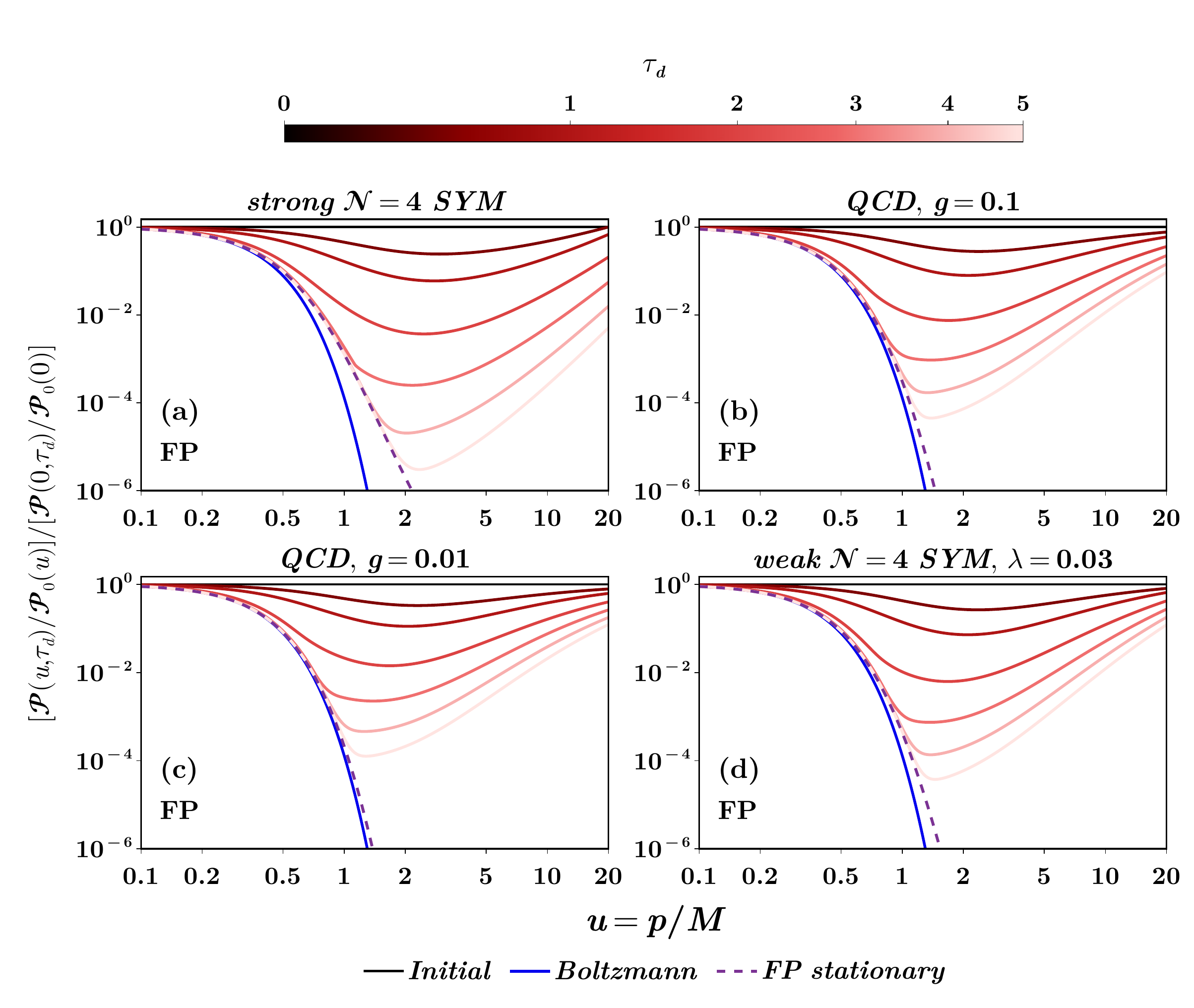}
\caption{Fokker--Planck evolution with the drag and longitudinal diffusion coefficients as determined by the field theory calculation. We show $[\mathscr P(u,\tau_d)/\mathscr P_0(u)]/[\mathscr P(0,\tau_d)/\mathscr P_0(0)]$, which compares shapes while scaling out the overall particle-number normalization. Black denotes the initial shape and blue the Boltzmann reference. The purple dashed curve denotes the stationary distribution of the process, set by the drag and diffusion coefficients. Red curves, from darker to lighter, correspond to $\tau_d=0.5,1,2,3,4,5$. Panels show (a) strongly coupled $\mathcal N=4$ SYM, (b) QCD at $g=0.1$, (c) QCD at $g=0.01$, and (d) weakly coupled $\mathcal N=4$ SYM at $\lambda=0.03$. All calculations are accurate to the leading-$T/M$ approximation, beyond which corrections outside of the Hamilton-Jacobi framework presented here (as well as the HQET setup used for the calculation) come into play.
}
\label{fig:weak-unprojected-evolution}
\end{figure}

The difference can be quite large. A famous example that can be worked out analytically is the strongly coupled case~\cite{Gubser:2006nz}, 
where $2T\mathfrak M_1/\mathfrak M_2=v/\gamma^{3/2}$. In this case
\begin{equation}
f_{\rm FP,eq}(u)-f_{\rm FP,eq}(0)=-2\left[1-(1+u^2)^{-1/4}\right] \, .
\label{eq:strong-unprojected-stationary}
\end{equation}
As such, the logarithm of the heavy quark distribution approaches a finite constant at large momentum, rather than the Boltzmann form. The problem is rather serious, because exponentiating it alone does not give a normalizable stationary distribution. While this statement concerns the leading-$T/M$ approximation used here, not the stationary solution of a finite-mass Fokker--Planck equation including subleading terms, which could, in principle, provide corrections to the dynamics such that a sensible equilibrium distribution exists, these would not change the fact that there is a serious consistency problem at leading order. While we don't have an analytic formula, this problem is also present at weak coupling, since $2T\mathfrak M_1/\mathfrak M_2 \propto (\ln \gamma)/\gamma^{2}$ at large $\gamma$.

Figure~\ref{fig:weak-unprojected-evolution} shows the time-dependent consequences of this truncation, for the three different weakly coupled theories described above, and compares it to the strongly coupled example that was already studied in~\cite{Rajagopal:2026urq}. We show results in terms of the ratio between the heavy quark distribution at time $\tau_d$ and the initial time as a function of the dimensionless momentum $u = p/M$, normalized by the same ratio at $u = 0$, such that all curves start at one at $u = 0$. This double ratio is insensitive to the overall normalization of the distribution, and allows for a clean comparison of the shape of the heavy quark distribution at leading order in $T/M$, encoded in $f$. 

We see from Fig.~\ref{fig:weak-unprojected-evolution} that, consistent with the fact that the Einstein relation is enforced by the fluctuation-dissipation theorem at zero momentum, the distributions develop a low-momentum region that closely follows the Boltzmann thermal shape. However, as one goes to higher momentum it becomes clear that the emergent equilibrium distribution starts to deviate from the thermal prediction. Indeed, the stationary condition~\eqref{eq:weak-unprojected-stationary} establishes that this evolution cannot approach the Boltzmann shape at all momenta.

\subsection{Imposing the Einstein relation}

\begin{figure}[t]
\centering
\includegraphics[width=0.95\linewidth]{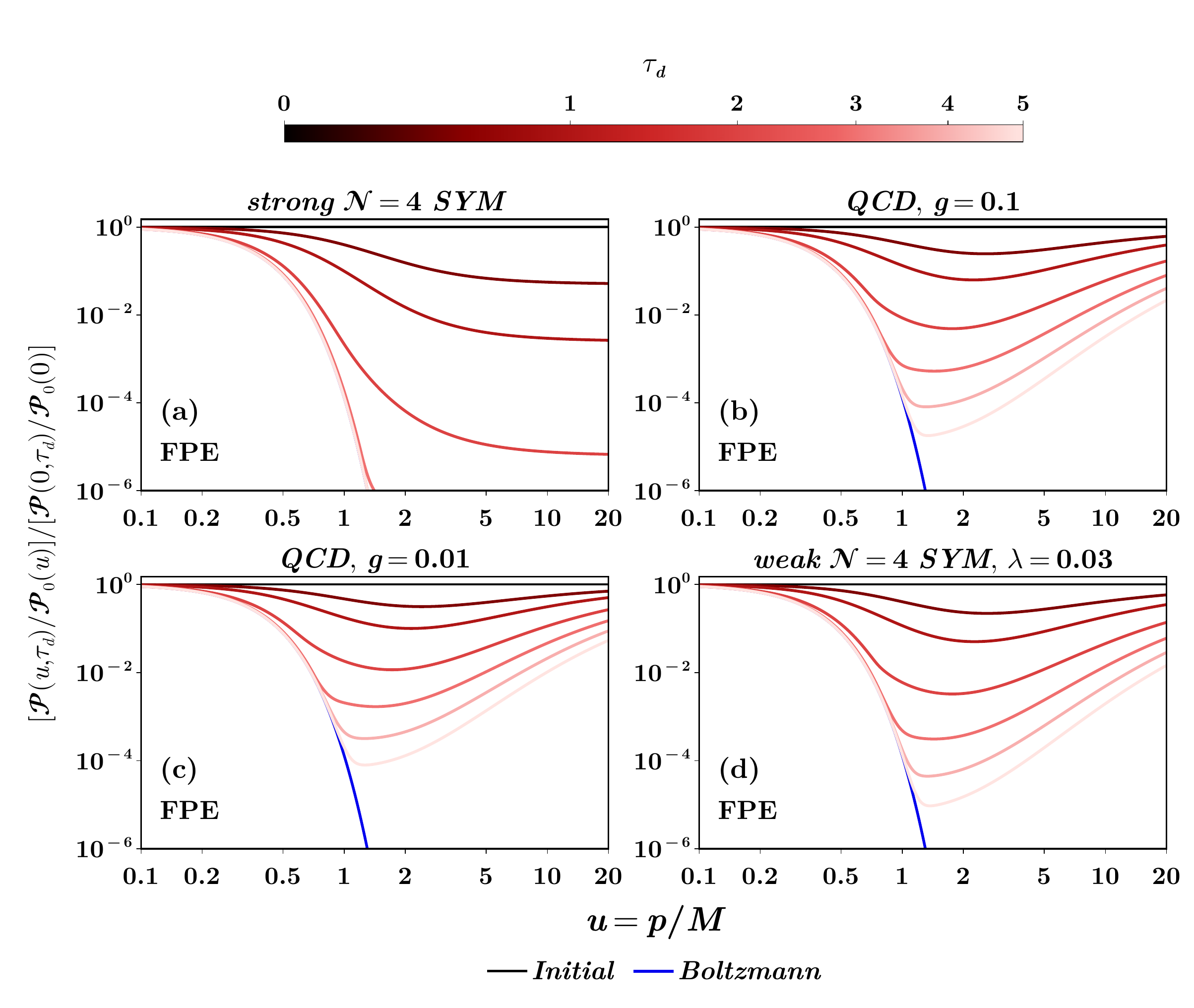}
\caption{Fokker--Planck evolution with the Einstein relation imposed, shown as $[\mathscr P(u,\tau_d)/\mathscr P_0(u)]/[\mathscr P(0,\tau_d)/\mathscr P_0(0)]$ for the same initial condition, theories and times as Fig.~\ref{fig:weak-unprojected-evolution}. Black denotes the initial distribution and blue the Boltzmann distribution, each expressed relative to $\mathscr P_0$. The  overall normalization is set in the same way as in Fig.~\ref{fig:weak-unprojected-evolution}. The microscopic drag coefficient is retained, while the longitudinal diffusion coefficient is replaced as specified by Eq.~\eqref{eq:weak-FP-E}.}
\label{fig:weak-FP-E-evolution}
\end{figure}

An economic way to fix the failure of the Fokker-Planck truncation to equilibrate correctly is to adjust---by hand---the second cumulant of the expansion to make it satisfy the Einstein relation. 
We emphasize that this is an operationally motivated -- and explicitly not microscopically motivated -- procedure, which preserves the characteristic time scale of equilibration and the correct equilibrium distribution, but neglects all of the information encoded in the second and higher cumulants. We will refer to this prescription as a ``projection,'' owing to the fact that it takes a general vector of cumulants and projects it onto the subspace that satisfies the Einstein relation with the same drag coefficient and vanishing higher cumulants. The resulting Einstein-projected kernel is
\begin{equation}
K_{\rm FP-E}(x;v)=-\frac{T\mathfrak M_1(v)}{v}x\left(x+\frac{v}{T}\right),
\qquad \mathfrak M_2^{\rm FP-E}(v)=\frac{2T\mathfrak M_1(v)}{v},
\label{eq:weak-FP-E}
\end{equation}
which is smooth in the limit $v\to0$ because $\mathfrak M_1$ goes to zero linearly with $v$. The ${\rm FP-E}$ sub-script denotes that this is the Einstein-projected Fokker-Planck kernel. Consistent with the equilibration condition, this kernel has the required zero at $x=-v/T$, so the Boltzmann distribution is now stationary. The price to pay is that the diffusion coefficient necessarily departs from the first-principles result at finite velocity.

Figure~\ref{fig:weak-FP-E-evolution} shows the same comparison as Fig.~\ref{fig:weak-unprojected-evolution}, but now with dynamics determined by~\eqref{eq:weak-FP-E}, which is guaranteed to approach towards the correct equilibrium profile. Indeed, at low momentum the reconstructed distributions approach the Boltzmann reference. Although it will eventually reach equilibrium, the high-momentum region retains a nontrivial amount of memory of the initial condition for the times we show in the figure.

The larger qualitative difference between weak and strong coupling in Fig.~\ref{fig:weak-FP-E-evolution} compared to Fig.~\ref{fig:weak-unprojected-evolution} warrants further discussion.
At strong coupling, the drag coefficient is velocity independent. At weak coupling it decreases at large momentum, so energetic quarks lose a smaller fraction of their momentum over the same nonrelativistic drag time. Equal $\tau_d$ therefore does not imply equal relaxation of the relativistic part of the spectrum. In particular, the weak-coupling tails have not equilibrated by $\tau_d=5$. This shows that, contrary to what one might have been tempted to conclude from Fig.~\ref{fig:weak-unprojected-evolution}, the qualitative difference between weak and strong coupling cannot simply be absorbed as a rescaling of timescales when considering the full relativistic spectrum. In fact, the qualitative similarity between all panels in Fig.~\ref{fig:weak-unprojected-evolution} is an apparent coincidence, because the drag and diffusion coefficients scale with $\gamma$ very differently in the ultra-relativistic limit.

\subsection{The role of drag and non-Gaussian fluctuations in equilibration}
\label{sec:weak-survivor-bias}

We now turn to exploring the dynamics of the complete evolution kernel, keeping the entire infinite tower of cumulants, exactly as determined by the first-principles calculation in terms of the Wilson loop. Because it follows directly from a field theory calculation, it retains the calculated drag and diffusion while satisfying $K(-v/T;v)=0$ through the balance among all cumulants. Thus, it has the same (thermal) equilibrium distribution as $K_{\rm FP-E}$, but features different dynamics in its approach to equilibrium.

For the initial condition used here, it turns out that the Hamilton-Jacobi formulation of the dynamics is only sensitive to the interval $-v/T\leq x\leq 0$. This is so because i) any instance of the initial condition $(u,x)$ from which Hamilton's equations are solved features $x = T^{-1} \partial_u f$ in this interval due to the large separation of scales $T \ll M$ together with $f' \propto - 1/u$ at large $u$, and ii) because the trajectories $(u(t),x(t))$ that follow from solving Hamilton's equations never cross the configurations that can describe an equilibrium solution. Conversely, a point starting at $(u_0,0)$ or $(u_0,-v(u_0)/T)$ will move along trajectories described by changing the value of $u_0$ in their expressions, but will still lie along curves in the $(u,x)$ plane that describe stationary solutions. See the discussion in section 3 of~\cite{Rajagopal:2026urq} for a detailed accounting of this in the strongly coupled case.

To obtain the evolution kernel $K$, we evaluate the
weak-coupling kernel 
using the fixed-order integral representation of Appendix~\ref{app:coord_transforms}.
Because these evolutions do not require continuation beyond the directly evaluated interval, there is no need to resort to the series representation of $K$ in terms of a moment expansion as in Eq.~\eqref{eq:Kxv}.

\begin{figure}[t]
\centering
\includegraphics[width=0.95\linewidth]{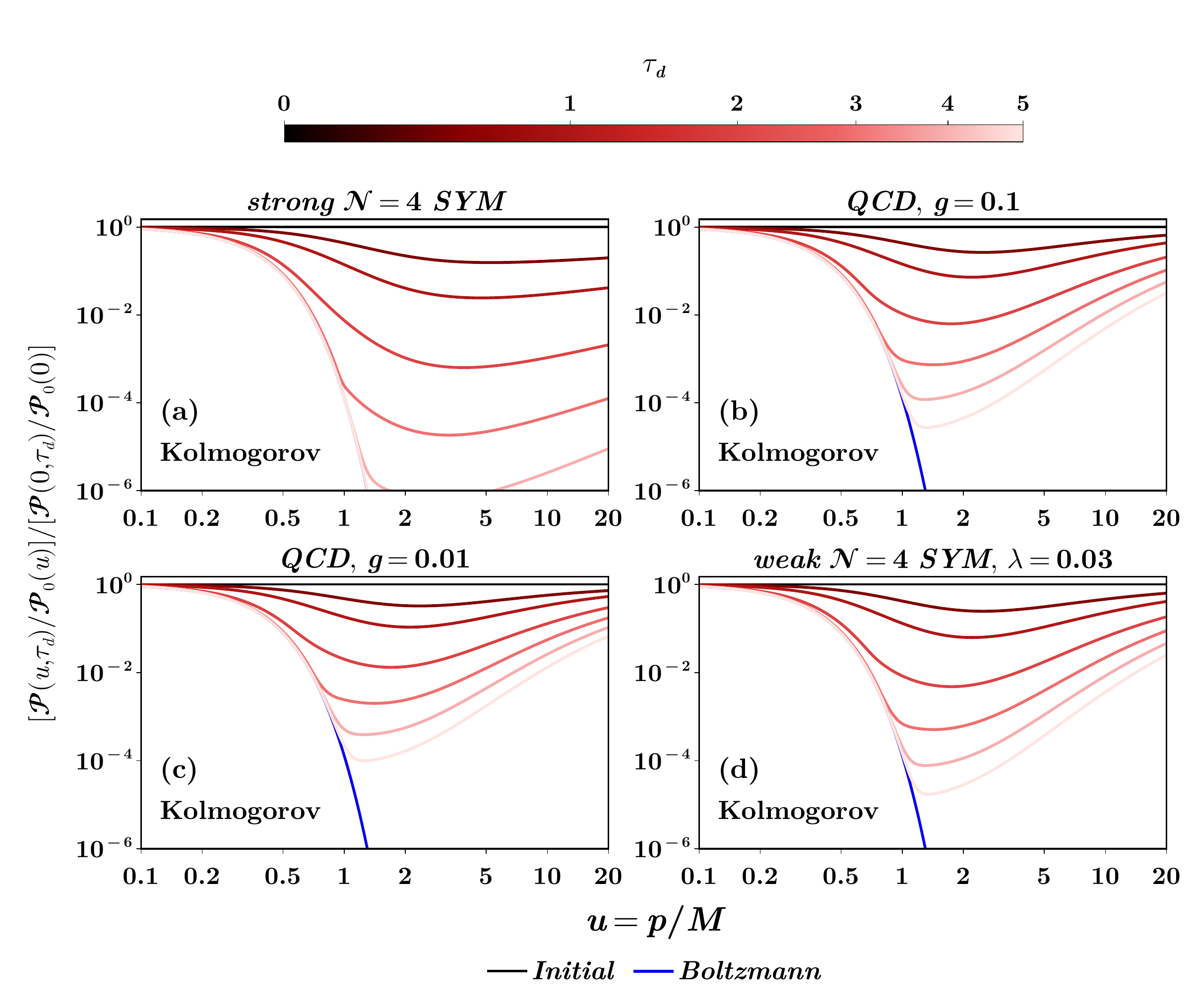}
\caption{Evolution with the complete non-Gaussian evolution kernel, shown as $[\mathscr P(u,\tau_d)/\mathscr P_0(u)]/[\mathscr P(0,\tau_d)/\mathscr P_0(0)]$ for the same initial condition, theories, times and normalization as Figs.~\ref{fig:weak-unprojected-evolution} and~\ref{fig:weak-FP-E-evolution}. Compared with the Einstein-projected Fokker--Planck evolution, the full kernel preserves more of the high-momentum tail, particularly at strong coupling.}
\label{fig:weak-NG-evolution}
\end{figure}

\begin{figure}[t]
\centering
\includegraphics[width=0.95\linewidth]{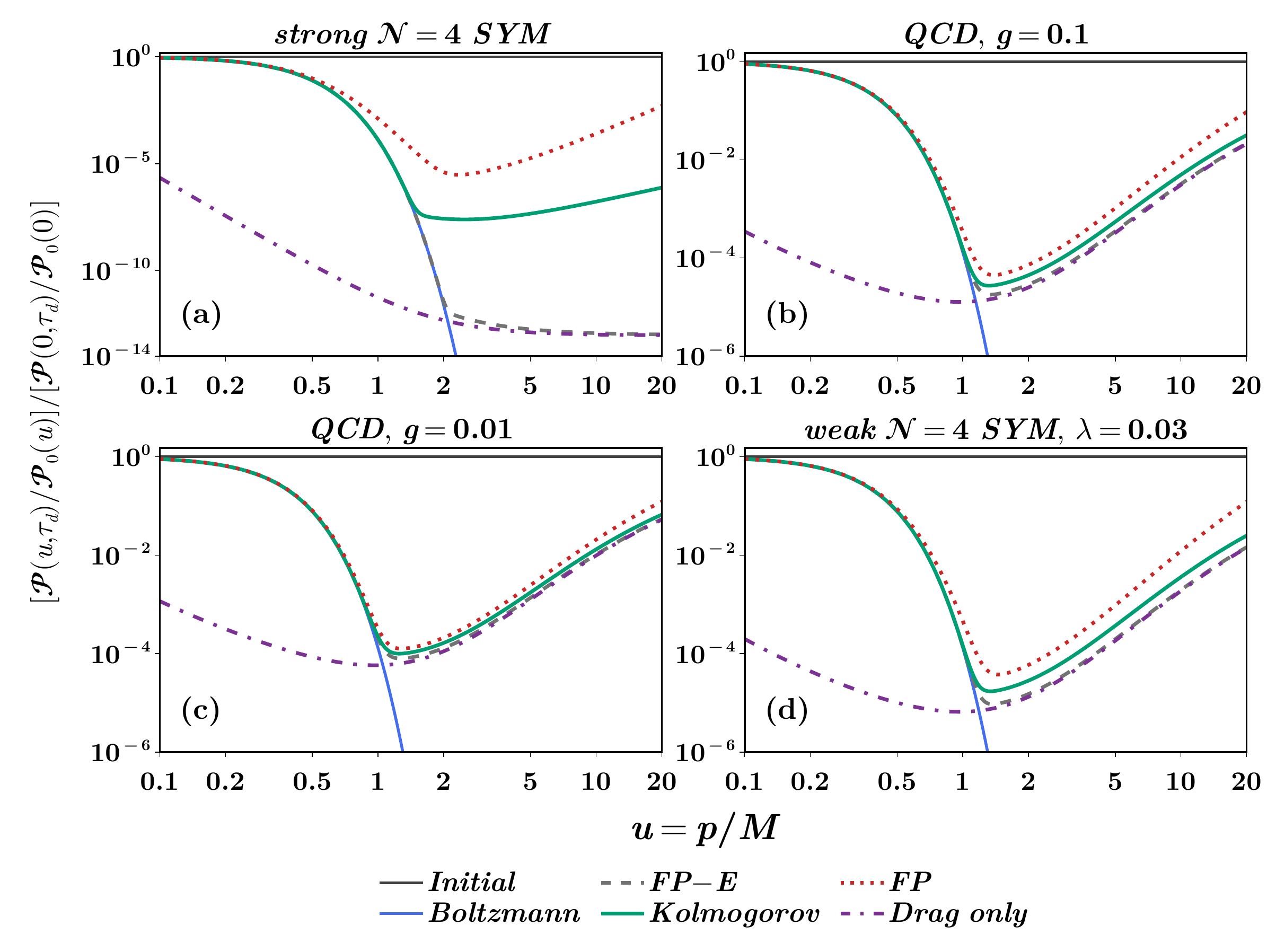}
\caption{Comparison at $\tau_d=5$ of the complete Kolmogorov evolution (green), the Fokker--Planck truncation with the Einstein relation imposed (FP-E, gray dashed), the truncation retaining the microscopic drag and diffusion coefficients (FP, red dotted), and drag-only evolution (purple dot-dashed). All curves use the same initial condition and the shape ratio $[\mathscr P(u,\tau_d)/\mathscr P_0(u)]/[\mathscr P(0,\tau_d)/\mathscr P_0(0)]$, removing differences in overall normalization. Black denotes the initial shape and blue the Boltzmann reference. Panels show strongly coupled $\mathcal N=4$ SYM, QCD at $g=0.1$, QCD at $g=0.01$, and weakly coupled $\mathcal N=4$ SYM at $\lambda=0.03$. The strong-coupling panel spans a larger ordinate range.}
\label{fig:weak-four-way-evolution}
\end{figure}

Figure~\ref{fig:weak-NG-evolution} shows the same comparison as Figs.~\ref{fig:weak-unprojected-evolution} and~\ref{fig:weak-FP-E-evolution}, but now without discarding any of the information of the full non-Gaussian dynamics. We see that equilibration proceeds consistently towards the Boltzmann distribution, with the biggest visible difference among the four panels being that the strongly coupled case equilibrates more rapidly relative to the time scale set by the drag coefficient, but are overall qualitatively similar otherwise. However, a most striking pattern appears when we compare all of the figures mentioned in this paragraph to each other. We present this comparison in Fig.~\ref{fig:weak-four-way-evolution}, where,
to distinguish the effect of fluctuations from that of the mean momentum loss, we also consider evolution with a purely deterministic drag force,
\begin{equation}
K_{\rm drag}(x;v)=-\mathfrak M_1(v)x,
\qquad
\frac{du}{d\tau_d}=-\frac{T\mathfrak M_1(v(u))}{M\eta_D(0)}
=-\frac{\eta_D(v(u))}{\eta_D(0)}u.
\label{eq:weak-drag-only}
\end{equation}
This calculation, by construction, features the same mean energy loss as the other evolutions but discards all fluctuations. 
In this evolution equation, the initial condition is transported along deterministic momentum-loss trajectories. At strong coupling these trajectories are simply $u(\tau_d)=u_0e^{-\tau_d}$. We emphasize that a drag force alone does not lead the heavy quark distribution to approach a Boltzmann distribution: fluctuations that balances out the effects of the drag force are required to sustain a finite thermal width.

Figure~\ref{fig:weak-four-way-evolution} reveals a marked difference between weak and strong coupling. There, we compare the dynamics presented in all previous panels in Figs.~\ref{fig:weak-unprojected-evolution}--\ref{fig:weak-NG-evolution} and the drag-only calculation with a snapshot at the characteristic time $\tau_d = 5$. In the weak-coupling examples, the high-momentum evolution can be qualitatively understood almost entirely in terms of the drag force: the Einstein-projected Fokker-Planck (FP-E) curve approaches the drag-only result, and restoring the complete kernel only produces a comparatively modest enhancement of the tail. At strong coupling FP-E also approaches drag-only evolution at sufficiently large momentum, but the complete Kolmogorov result remains far above both. The direct Fokker-Planck (FP) truncation is always above the other curves, but, as for the FP-E curve, the difference with the complete kernel is much more pronounced in the strongly coupled case, whereas it roughly follows the trend set by the drag-only calculation in the large momentum regime. At low momentum, by contrast, drag-only evolution differs substantially from all three calculations that retain fluctuations, since it does not have the required fluctuations to equilibrate.

It is quite remarkable that while all of the evolution equations lead to a high-momentum tail that takes a long time to equilibrate, the underlying mechanism that sustains the tails in the weakly and strongly coupled examples is quite different. In the strongly coupled case, it is the large fluctuations that sustain the high-momentum population much above what a drag-only calculation would allow. In the weakly coupled case, the effect of fluctuations makes no large qualitative difference, and the shape of all the curves follows that of the drag-only calculation. These differences can be understood quantitatively. At strong coupling, $\eta_D$ is independent of velocity, whereas the weak-coupling darg coefficient decays at large $\gamma$, giving
\begin{equation}
\eta_D^{\rm weak}\propto\frac{g^4T^2}{M}\frac{\ln\gamma}{\gamma} \, ,
\qquad
\eta_D^{\rm strong}=\frac{\pi\sqrt{\lambda}T^2}{2M} \, .
\label{eq:weak-strong-drag-asymptotics}
\end{equation}
Energetic quarks consequently lose a smaller fraction of their momentum over a fixed non-relativistic drag time $\tau_d$ at weak coupling. The persistence of a weak-coupling high-momentum tail is therefore not, by itself, evidence for a large fluctuation effect: that persistence is already present with drag alone.

The velocity dependence of the longitudinal diffusion provides a second distinction. At fixed coupling, the leading large-$\gamma$ behavior of the microscopic coefficients is
\begin{equation}
\kappa_L^{\rm weak}\propto g^4T^3\gamma^2,
\qquad
\kappa_L^{\rm strong}=\pi\sqrt{\lambda}T^3\gamma^{5/2},
\label{eq:weak-strong-kappa-asymptotics}
\end{equation}
with color and matter-content factors suppressed in the weak-coupling expression. At strong coupling the diffusion coefficient exceeds its Einstein-relation value by
\begin{equation}
\frac{\kappa_L^{\rm strong}}{2TE\eta_D^{\rm strong}}=\gamma^{3/2}.
\label{eq:strong-diffusion-enhancement}
\end{equation}
Imposing the Einstein relation therefore removes an increasingly large part of the microscopic fluctuations as the boost increases. As emphasized in Ref.~\cite{Rajagopal:2026urq}, retaining these fluctuations together with the complete non-Gaussian structure allows the relativistic tail to remain populated much longer than in the Einstein-projected Gaussian evolution.

At strong coupling, the physical interpretation is the survivor bias discussed in Ref.~\cite{Rajagopal:2026urq}. The typical momentum transfer that a heavy quark arriving at the momentum bin labeled by $u$ experienced in the time step leading to its arrival is selected by the convolution between the transition probability and the heavy quark distribution. This typical transfer is selected according to
\begin{equation}
\left.\frac{\partial\tilde S}{\partial C_L}\right|_{\bar C_L}=\frac{1}{T}\partial_u f=x \, ,
\qquad
\bar C_L=-\partial_xK(x;v) \, .
\label{eq:weak-selected-transfer}
\end{equation}
The selected momentum transfer rate $\bar C_L$ characterizes the momentum loss conditioned on arrival in a given momentum bin, rather than the mean loss $\langle C_L\rangle=\mathfrak M_1(v)$ of a quark initially occupying that bin. For a falling spectrum, quarks that lost less momentum represent most of the population remaining in any given high-momentum bin. Concavity of $K$ gives $\bar C_L<\mathfrak M_1$ for $x<0$, so fluctuations slow the evolution of the spectrum relative to the momentum loss of a typical quark. This selection occurs even for Gaussian fluctuations; the full kernel determines its magnitude beyond the Gaussian approximation.

The four calculations that we compared in Fig.~\ref{fig:weak-four-way-evolution} separate these effects. Drag only versus Einstein-projected FP identifies the effect of adding Gaussian fluctuations consistent with equilibration at a fixed drag coefficient. Einstein-projected FP versus Kolmogorov compares two kernels with the correct Boltzmann stationary distribution, but different diffusion coefficient and higher cumulants. Unprojected FP versus Kolmogorov instead retains the same microscopic drag and diffusion coefficients and isolates the effect of restoring the higher cumulants. The unprojected FP curve lies above the complete result in the relativistic regime: preserving a large microscopic diffusion coefficient alone is not sufficient to accurately describe the dynamics, and moreover, retaining it without all of the other higher cumulants fails to generate consistent equilibration dynamics.

At weak coupling, the differences are smaller but remain finite --- even visible with the logarithmic scale used in Fig.~\ref{fig:weak-four-way-evolution}. At $u\simeq10$ and $\tau_d=5$, the Kolmogorov curve exceeds the FP-E curve by factors of approximately $1.5$ in QCD at $g=0.1$, $1.3$ in QCD at $g=0.01$, and $1.8$ in weakly coupled $\mathcal N=4$ SYM at $\lambda=0.03$. In the strongly coupled example the corresponding separation is about six orders of magnitude.
We emphasize that the relatively drag-dominated weak-coupling evolution is consistent with the fact that the drag and diffusion coefficients are logarithmically enhanced by the value of the coupling. At fixed velocity this leading-logarithmic part is Gaussian and satisfies the Einstein relation, while both the higher cumulants and the violation of that relation enter beyond leading logarithm. The relative importance of the latter kind of contributions thus decreases as the coupling is reduced. That being said, this hierarchy is not uniform as $v\to1$. In particular, the weak-coupling $\gamma^2$ growth in Eq.~\eqref{eq:weak-strong-kappa-asymptotics} comes from the hard contribution, and its ratio to the Einstein-relation value also grows at sufficiently large boost. This means that the larger fluctuation effect seen at strong coupling in Fig.~\ref{fig:weak-four-way-evolution} compared to that at weak coupling is therefore \textit{not} a statement that weak-coupling fluctuations remain negligible at arbitrarily large $\gamma$. Nor can the magnitude of the survivor bias be inferred from $\kappa_L$ alone: Eq.~\eqref{eq:weak-selected-transfer} samples the complete kernel. It is a statement that, for the chosen parameters and displayed kinematic range, the qualitative features of the equilibration dynamics are driven by the drag coefficient.

To reach a conclusion about a closer-to-realistic scenario, one should explore the outcomes of carrying out the above comparison for larger values of the coupling $g$. However, the notoriously bad convergence of the weak coupling transport coefficients as the coupling is increased, together with the non-uniformity of the $v\to1$ and small coupling limits, makes this extrapolation somewhat precarious. It certainly leaves open the possibility that even weakly coupled plasma at non-negligible coupling (say, $g=1$, i.e., $\alpha_s = 1/(4\pi)$) could display comparatively large signatures of non-Gaussianity. Given the past qualitative successes of strongly coupled $\mathcal{N}=4$ SYM in modeling QGP, the strong coupling result leads one to suspect that realistic QGP formed in heavy ion collisions could display this to an experimentally-accessible extent.

\section{Conclusions and Outlook} \label{sec:conclusions-outlook}

In this work we have calculated the complete non-Gaussian momentum-transfer kernel for a heavy quark of arbitrary velocity in weakly coupled non-Abelian plasmas through strict $\mathcal{O}(g^4)$, and used it to study, for the first time from a first principles perturbative field theory approach beyond leading logarithm, how relativistic heavy quarks approach equilibrium. At leading logarithmic order the kernel is quadratic and the dynamics is Gaussian, with the drag and longitudinal momentum-diffusion coefficients satisfying the Einstein relation. Beyond leading logarithm this relation is violated at finite velocity, but heavy quarks nevertheless equilibrate: detailed balance is instead realized through the complete momentum-transfer distribution, requiring a balance among the entire tower of cumulants. The higher cumulants are therefore not merely small corrections to a Gaussian process, but an essential part of relativistic heavy-quark equilibration.

The fixed-order kernel makes this non-Gaussian structure particularly transparent. At large longitudinal momentum transfer the distribution develops asymmetric exponential tails,
\begin{equation}
    P(k_L)\propto e^{-R_+(v)k_L}\quad (k_L\to+\infty), \qquad P(k_L)\propto e^{+R_-(v)k_L}\quad (k_L\to-\infty),
\end{equation}
with
\begin{equation}
    R_+(v)=\frac{1-v}{2T}, \qquad R_-(v)=\frac{1+v}{2T}.
\end{equation}
These tail exponents follow from the bounded domain of analyticity $|2T{\bf x}+{\bf v}|<1$ of the weak-coupling evolution kernel. Their difference, $R_-(v)-R_+(v)=v/T$ (the asymmetry of the tails), is fixed exactly by the equilibration condition, while the forward exponent decreases as $R_+\sim1/(4\gamma^2T)$ at large boost. The same analytic domain also fixes the transverse tail exponent $R_\perp=\sqrt{1-v^2}/(2T)$.

One of the more striking results of our comparison is how much of this structure is shared between very different gauge theory plasmas. Once their Gaussian scales are divided out, the weakly coupled versions of Yang--Mills theory, QCD and $\mathcal{N}=4$ SYM all exhibit very similar non-Gaussian momentum-transfer distributions, and the same qualitative structure is present in strongly coupled $\mathcal{N}=4$ SYM. The differences are nevertheless quantitatively important. Both the drag and diffusion coefficients and the locations of the singularities, and therefore the tail exponents, have different velocity dependence in the different theories. 

The importance of these differences becomes particularly clear once the kernels are used to evolve a steeply falling heavy-quark distribution. The exponential tails lead to a strong survivor bias compared to Gaussian models that retain the drag coefficient and fix the diffusion coefficient by hand: heavy quarks which fluctuate toward unusually small momentum loss are preferentially represented in the population that remains at large momentum, resulting in less suppression than in a Gaussian evolution compatible with equilibration and with comparable low-order transport properties. The tail exponent consequently plays a direct and impactful role in equilibration at nonzero velocity. At the same time, the different velocity dependence of the drag and diffusion coefficients and the tail exponent in the theories studied here lead to visibly different approaches to equilibrium despite their common non-Gaussian structure. This is encouraging phenomenologically: the analytic structure provides a relatively constrained handle on what would otherwise appear to require an infinite set of independent transport coefficients, while the remaining velocity dependence retains substantial sensitivity to the microscopic dynamics.

\paragraph{Scope and limitations.}
Several limitations of our calculation should be kept in mind. We have worked throughout at leading order in the infinite mass expansion and at strict $\mathcal{O}(g^4)$. The medium is taken to be infinite and in thermal equilibrium at a fixed temperature, and we assume the Markovian regime in which the Wilson loop exponentiates extensively in time. Our dynamical analysis has focused on the longitudinal kernel relevant for equilibration at leading order in $T/M$. We have also characterized the transverse tail structure, but its implications for momentum broadening and its sensitivity to hard quasiparticle scattering remain to be explored. Finally, we stress that the ultrarelativistic limit is nonuniform in $\gamma$, $g$ and $M$, so the strict $v\to1$ limit lies outside the regime in which the present approximations can be naively extrapolated. Subleading corrections in $T/M$ also remain as a natural future direction.

\paragraph{Phenomenological implications and outlook.}
The survivor bias effect we discussed also has an immediate implication for the extraction of heavy-quark transport properties from data. Since the non-Gaussian evolution produces less suppression at fixed interaction strength, including the exponential tails should, all else being equal, increase the effective drag required to describe a given amount of observed suppression and correspondingly decrease the inferred spatial diffusion coefficient. Although the lattice spatial diffusion coefficient is a $v=0$ quantity, phenomenological determinations of it are constrained largely by heavy quarks at finite momentum and therefore depend on the (implicit or explicit) assumptions about the dynamics away from $v=0$ \cite{Das:2013kea,Scardina:2017ipo,Cao:2018ews,Sambataro:2025obe,Beraudo:2025nvq}. Accounting for non-Gaussian transport may consequently be essential for a controlled comparison between phenomenologically extracted diffusion coefficients and recent lattice-QCD determinations such as those of the HotQCD Collaboration~\cite{HotQCD:2025fbd}.

The immediate phenomenological question is whether heavy-flavor data can separately constrain the velocity dependence of the drag, longitudinal momentum diffusion and tail exponent, and how the inferred low-order transport properties change once the Gaussian approximation is relaxed. A Bayesian analysis in which these three ingredients are allowed to vary provides a direct way to answer this question and, in particular, to quantify the resulting shift in the extracted drag coefficient --- and hence in the inferred spatial diffusion coefficient $2\pi T D_s$ --- induced by non-Gaussian effects. Existing observables such as open heavy-flavor $R_{AA}$ and $v_2$ already probe different parts of the heavy-quark momentum distribution and therefore contain sensitivity to these effects. More precise measurements exploiting the full Run~3 dataset, and eventually the HL-LHC and proposed ALICE~3 program~\cite{ALICE:2022wwr,ALICE3Scoping2025}, should sharpen this sensitivity substantially, while measurements across collision systems with different sizes and lifetimes, including O+O, provide complementary sensitivity to the accumulated transport dynamics. Similarly, since the tail exponent is mass-independent at leading order, simultaneous charm- and bottom-flavor measurements provide complementary sensitivity for disentangling the non-Gaussian tail structure from mass-dependent transport effects, particularly as more precise non-prompt and bottom-hadron data become available.

\paragraph{Theoretical outlook.}
There are several complementary theoretical questions. The KMS symmetry nonperturbatively requires the analytic structure of $K$ to be consistent with a $\mathbb{Z}_2$ symmetry ${\bf x}\to-{\bf v}/T-{\bf x}$, but it does not by itself explain how generic a finite domain of analyticity is in thermal quantum field theory or what microscopic physics determines the nature of its boundary. The natural question is then whether the existence of finite-distance singularities of the Wilson loop $\langle W_{\bf v} \rangle({\bf L})$ as a function of complex ${\bf L}$ can be established under more general assumptions, and which aspects of the weak- and strong-coupling structures found here survive at all couplings.

On the weakly coupled side, the next perturbative order is a particularly concrete direction. Our leading fixed-order result gives coupling-independent tail exponents and logarithmic branch points. Calculating the next order would test whether the endpoint singularity class is stable under perturbative corrections, determine the first coupling-dependent correction to the tail exponent $R(v)$, and clarify how the large-momentum-transfer asymptotics evolve under renormalization-group running. Understanding the interplay of the coupling, mass and large-velocity limits is another natural extension, particularly for determining how far toward the ultra-relativistic regime the effective description remains useful, and how that depends on survivor bias effects.

Finally, the analytic structure carries information about the microscopic constituents of the plasma. At weak coupling, we anticipate that the $n^{-2}$ large-order behavior and the resulting logarithmic branch point are indicative of Rutherford-like scattering from point-like quasiparticles; whereas the strongly coupled kernel has a simple pole and no such interpretation. The transverse momentum broadening channel is the natural place to sharpen this. 
Computing the transverse momentum change distribution and locating the crossover into the Moli\`ere regime would establish whether heavy quarks can serve as a probe of quasiparticle degrees of freedom in the QGP, complementary to jets~\cite{Kudinoor:2026wcs} and with very different kinematics and systematics.

\acknowledgments
We gratefully acknowledge helpful conversations with Dani Pablos, Krishna Rajagopal, Rachel Steinhorst, Eamonn Weitz, Urs Wiedemann, and Xiaojun Yao.
Research supported in part by the U.S.~Department of Energy, Office of Science, Office of Nuclear Physics under grant Contract Number DE-SC0011090, by grant NSF PHY-2309135 to the Kavli Institute for Theoretical Physics (KITP), and by grant 994312 from the Simons Foundation. This research has benefited from the Mitchel Postdoctoral Scholar Career Development Fund at KITP.

\appendix

\section{Properties of the Wilson Loop and the Stochastic Process it Defines}

\subsection{A Bound on the Magnitude of the Wilson loop} \label{app:proof-maximum}

In what follows, we show that the Hessian matrix of the real part of the kernel
\begin{equation}
    \frac{\partial^2 {\rm Re}\{S\} }{\partial L_i \partial L_j}({\bf L}={\bf 0},{\bf x}) \succeq 0
\end{equation}
is a positive definite matrix for all imaginary separations ${\bf x}$. We do so by showing that the magnitude of the Wilson loop $\langle W_{\bf v} \rangle ({\bf L} - i {\bf x}) $ has a maximum at ${\bf L} = 0$ for all ${\bf x}$.

We start from the operator definition in~\eqref{eq:Wloop-Wlines}
\begin{align}
    \langle W_{\bf v} \rangle ({\bf L}) = \frac{1}{Z} {\rm Tr}_{\mathcal{H}} \! \left[ W^{lk}_{[(0, {\bf L}),(\Delta t, {\bf v} \Delta t + {\bf L} )]} \tilde{\rho}^Q_{ki} W^{ij}_{[(\Delta t, {\bf v} \Delta t ),(0,{\bf 0})]} e^{-\beta H}  \rho_{jl}^Q \right] \, , 
\end{align}
and we note that because $S$ is only determined by the time-extensive contributions, it should be independent of our choice for initial $\rho^Q$ and final $\tilde{\rho}^Q$ state projectors. As such, in order to show that the Hessian derived from $S$ is positive definite, we may choose a specific form for them. 

The simplest possibility is to consider initial and final state projections to pure states describing hadrons with a single heavy quark in them. In this case, Eq.~\eqref{eq:Wloop-Wlines} can be written as 
\begin{align}
    \langle W_{\bf v} \rangle ({\bf L}) = \frac{1}{Z} {\rm Tr}_{\mathcal{H}}  \Big[ M_l^Q({\bf L})^\dagger W^{lk}_{[(0, {\bf L}),(\Delta t, {\bf v} \Delta t + {\bf L} )]} \tilde{M}_k^Q({\bf v} \Delta t + {\bf L})^\dagger & \nonumber \\ 
    \tilde{M}_i^Q({\bf v} \Delta t) W^{ij}_{[(\Delta t, {\bf v} \Delta t ),(0,{\bf 0})]}  M_j^Q({\bf 0}) e^{-\beta H} & \Big] \, , 
\end{align}
where $M^Q$ and $\tilde{M}^Q$ are the ``brown muck'' part of the wave-functional of the heavy hadron in the initial and final state, respectively, centered around the position specified by their arguments.

It is then natural to insert translation operators (of the light QCD degrees of freedom) in this observable as
\begin{align}
    \langle W_{\bf v} \rangle ({\bf L}) = \frac{1}{Z} {\rm Tr}_{\mathcal{H}}  \Big[ & e^{-i {\bf P} \cdot {\bf L} } M_l^Q({\bf 0})^\dagger W^{lk}_{[(0, {\bf 0}),(\Delta t, {\bf v} \Delta t )]} \tilde{M}_k^Q({\bf v} \Delta t)^\dagger  \nonumber \\ 
    & e^{+i {\bf P} \cdot {\bf L} } \tilde{M}_i^Q({\bf v} \Delta t) W^{ij}_{[(\Delta t, {\bf v} \Delta t ),(0,{\bf 0})]}  M_j^Q({\bf 0}) e^{-\beta H}  \Big] \, , 
\end{align}
where ${\bf P}$ is the total momentum operator of the QGP degrees of freedom. Then, inserting eigenstates of the QGP degrees of freedom $4$-momentum operator, the trace can be converted into a formal sum over these states
\begin{align}
    \langle W_{\bf v} \rangle ({\bf L}) = \frac{1}{Z} \sum_{n,m}  & e^{-i {\bf P}_n \cdot {\bf L} } \bra{n} M_l^Q({\bf 0})^\dagger W^{lk}_{[(0, {\bf 0}),(\Delta t, {\bf v} \Delta t )]} \tilde{M}_k^Q({\bf v} \Delta t)^\dagger \ket{m} \nonumber \\ 
    & e^{+i {\bf P}_m \cdot {\bf L} } \bra{m} \tilde{M}_i^Q({\bf v} \Delta t) W^{ij}_{[(\Delta t, {\bf v} \Delta t ),(0,{\bf 0})]}  M_j^Q({\bf 0}) \ket{n} e^{-\beta E_n} \nonumber \\
    =\frac{1}{Z} \sum_{n,m}  & e^{-i ({\bf P}_n - {\bf P}_m) \cdot {\bf L} }  \left| \bra{m} \tilde{M}_i^Q({\bf v} \Delta t) W^{ij}_{[(\Delta t, {\bf v} \Delta t ),(0,{\bf 0})]}  M_j^Q({\bf 0}) \ket{n} \right|^2 e^{-\beta E_n} \, . 
\end{align}
Because the weights of $e^{-i ({\bf P}_n - {\bf P}_m) \cdot {\bf L} }$ are real and positive for all $n,n$, it immediately follows that
\begin{equation}
    \left|\langle W_{\bf v} \rangle ({\bf L} - i{\bf x}) \right| \leq \langle W_{\bf v} \rangle ( - i{\bf x}) 
\end{equation}
for all ${\bf x}$, because a nonzero ${\bf L}$ can only generate destructive interference in a sum that otherwise consists purely of positive definite terms.

An even simpler way of verifying this is using that the Wilson loop is the Fourier transform of a probability distribution
\begin{equation}
    \langle W_{\bf v} \rangle({\bf L}) = \int d^3{\bf k} \, e^{i {\bf k} \cdot {\bf L} } P({\bf k};{\bf v}) \, ,
\end{equation}
which means that, whenever the integral is well-defined,
\begin{equation}
    \left| \langle W_{\bf v} \rangle({\bf L} - i {\bf x}) \right| = \left| \int d^3{\bf k} \, e^{i {\bf k} \cdot {\bf L} + {\bf k} \cdot {\bf x} } P({\bf k};{\bf v}) \right| \leq  \int d^3{\bf k} \, e^{ {\bf k} \cdot {\bf x} } P({\bf k};{\bf v}) = \langle W_{\bf v} \rangle( - i {\bf x}) \, ,
\end{equation}
which verifies that the magnitude of the Wilson loop is indeed maximized at ${\bf L} = 0$ for any ${\bf x}$, and as such the kernel $S$ is necessarily convex --- i.e., it has a positive definite Hessian.

\subsection{Large Deviation Theory} \label{app:large-deviation}

For a heavy quark moving with fixed velocity ${\bf v}$ in the Markovian regime, the total momentum transfer accumulated over a time $t$ may be viewed as an additive random variable. We can therefore interpret many of these results in the language of large deviation theory \cite{TOUCHETTE20091}. The analytically continued (fixed order) Wilson loop then defines the associated scaled cumulant generating function in the sense of large deviation theory,
\begin{equation}
    \lambda_{\bf v}({\bf x})
    \equiv
    \lim_{t\to\infty}\frac{1}{tT}
    \log \left\langle e^{{\bf x}\cdot{\bf k}_t}\right\rangle
    = -K({\bf x};{\bf v}) \, .
\end{equation}
Therefore, whenever $K({\bf x};{\bf v})$ is finite and differentiable on the relevant real domain\footnote{With the origin in the interior of its domain, and with its first derivative diverging at the edges of the domain.}, the G\"artner--Ellis theorem implies that the momentum-transfer rate ${\bf C}={\bf k}_t/(tT)$ satisfies a large deviation principle with rate function
\begin{equation}
    \tilde S({\bf C};{\bf v})
    =
    \sup_{{\bf x}}
    \left[
        {\bf x}\cdot{\bf C}+K({\bf x};{\bf v})
    \right].
\end{equation}
In this sense, the Legendre--Fenchel transform of $K$ is the natural object controlling the asymptotic statistics of momentum transfer. Since $K$ is concave, $\tilde S$ is convex; moreover, wherever the Hessian of $K$ is negative definite, the saddle is unique and the map ${\bf C}=-\partial_{\bf x}K$ is one-to-one. The minimum of $\tilde S$ is reached at the typical momentum-transfer rate
\begin{equation}
    {\bf C}_*({\bf v})=-\partial_{\bf x}K({\bf 0};{\bf v}),
\end{equation}
for which $\tilde S({\bf C}_*;{\bf v})=0$. The Gaussian/Langevin approximation is then simply the local quadratic approximation of $\tilde S$ around this minimum, while the higher moments of $K$ determine the genuine non-Gaussian large deviations away from it.

Finally, if one is interested in a derived observable such as the longitudinal momentum-transfer rate
$C_\parallel=\hat{\bf v}\cdot{\bf C}$, the contraction principle implies that this observable also satisfies a large deviation principle, with rate function
\begin{equation}
    \tilde S_\parallel(c;{\bf v})
    =
    \inf_{{\bf C}_\perp}
    \tilde S \big(c\,\hat{\bf v}+{\bf C}_\perp;{\bf v}\big).
\end{equation}
Thus the one-dimensional longitudinal rate function studied in the main text is the contraction of the full vector large deviation principle.

\section{Calculation of the Evolution Kernel} \label{app:calculation-evol}

In this appendix we discuss with more detail the individual elements of the calculation. Appendix~\ref{app:calculation-structure} presents the structural aspects thereof, and Appendix~\ref{app:calculation-perturbative} presents the explicit expressions of intermediate steps that lead to the result in the main text.

\subsection{Structural aspects of the calculation} \label{app:calculation-structure}

Equation~\eqref{eq:W-expansion} organizes the calculation into three distinct nonzero pieces: $W_2$, $W_3$ and $W_4$, corresponding each to the term in the expansion of the Wilson loop with $n$ explicit powers of the coupling $g$ in its expression in terms of correlation functions (also referred to as ``correlators'') of gauge fields. The gauge field correlators themselves can be expanded in powers of the coupling constant, each contributing to the result at all orders higher than $n$ (the number of gauge fields in the correlator). Because we are interested in the result up to $\mathcal{O}(g^4)$, it is necessary to determine $W_2$ up to $\mathcal{O}(g^2)$, $W_3$ up to $\mathcal{O}(g)$, and $W_4$ up to $\mathcal{O}(g^0)$.

\subsubsection{$2$-point contribution}
 \label{app:calculation-structure-2}

After setting the $\rho$ matrices to unity, expanding out the path-ordered exponential up to quadratic order leads to
\begin{align}
    W_2({\bf L};{\bf v},t) = \frac{1}{N_c} n^\mu n^\nu \, {\rm Tr}_c \left[ T^a T^b \right] \bigg\{ &-  \int_0^t dt' \int_0^{t'} dt'' \langle  A_\mu^a(t',{\bf v} t' ) A_\nu^b(t'',{\bf v} t'') \rangle_T \nn \\ &+  \int_0^t dt' \int_0^{t} dt'' \langle  A_\mu^a(t',{\bf v} t' + {\bf L} ) A_\nu^b(t'',{\bf v} t'') \rangle_T \nn \\ &-  \int_0^t dt' \int_0^{t'} dt'' \langle  A_\mu^a(t'',{\bf v} t''+ {\bf L}) A_\nu^b(t',{\bf v} t'+ {\bf L}) \rangle_T \bigg\} \, .
\end{align}
Taking the trace over the color matrices and using that the correlators are translationally invariant, we arrive at Eq.~\eqref{eq:W2}, reproduced here:
\begin{align}
    W_2({\bf L};{\bf v},t) = \frac{T_F}{N_c}  n^\mu n^\nu \int_0^t dt' \int_0^t dt'' \Big[  \langle A^a_\mu(t'-t'',(t'-t''){\bf v} + {\bf L} ) A^a_\nu(0,{\bf 0} ) \rangle_T &\nonumber \\ - \langle A^a_\mu(t'-t'',(t'-t''){\bf v}  ) A^a_\nu(0,{\bf 0} ) \rangle_T &\Big] \, . \label{eq:W2-app-0}
\end{align}

It is clear that the second line is an ${\bf L}$-independent additive constant whose role is to enforce that $W_2$ vanishes if ${\bf L} = {\bf 0}$. Therefore, we may focus on the ${\bf L}$-dependent contribution. Like any $2$-point function of local operators in thermal equilibrium, it can be written in terms of a spectral function as
\begin{equation}
    \langle A^a_\mu(t,{\bf x}) A^a_\nu(0,{\bf 0} ) \rangle_T = (N_c^2 - 1) \int \frac{d^3 k}{(2\pi)^3} \frac{d\omega}{(2\pi)} e^{-i\omega t} e^{i {\bf k} \cdot {\bf x} } \frac{e^{\omega/T} }{e^{\omega/T} - 1 } \varrho_{\mu \nu}(\omega,{\bf k}) \, ,
\end{equation}
where we have used that $\langle A^a A^b \rangle \propto \delta^{ab}$ to do the sum over color indices.

It then follows that
\begin{align}
    W_2({\bf L};{\bf v},t) = C_F \, n^\mu n^\nu & \int \frac{d^3 k}{(2\pi)^3} \frac{d\omega}{(2\pi)} \frac{e^{\omega/T} }{e^{\omega/T} - 1 } \varrho_{\mu \nu}(\omega,{\bf k}) \left[ e^{ i {\bf k} \cdot {\bf L} } - 1 \right] \nonumber \\
    \times & \int_0^t dt' \int_0^t dt''  e^{-i\omega (t'-t'') } e^{i (t'-t'') {\bf k} \cdot {\bf v}  } \, , 
\end{align}
where the minus one in the square brackets comes from the ${\bf L}$-independent term discussed below~\eqref{eq:W2-app-0}.

In the large $t$ limit, we have
\begin{equation}
    \int_0^t dt' \int_0^t dt''  e^{-i\omega (t'-t'') } e^{i (t'-t'') {\bf k} \cdot {\bf v}  } \to t (2\pi) \delta(\omega - {\bf k} \cdot {\bf v}) \, ,
\end{equation}
meaning that
\begin{align}
    W_2({\bf L};{\bf v},t) = t \, C_F \, n^\mu n^\nu & \int \frac{d^3 k}{(2\pi)^3} \frac{e^{{\bf k} \cdot {\bf v}/T} }{e^{{\bf k} \cdot {\bf v}/T} - 1 } \varrho_{\mu\nu}({\bf k} \cdot {\bf v},{\bf k}) \left[ e^{ i {\bf k} \cdot {\bf L} } - 1 \right] \, . \label{eq:W2-app}
\end{align}

For our purposes, we need to determine $W_2$ up to order $g^2$, which means that we need the tree-level and the 1-loop contributions to $\varrho_{\mu\nu}$. Because of the many contributions it receives, we discuss the 1-loop calculation in Appendix~\ref{app:calculation-perturbative-2}. The tree-level contribution is direct to obtain, and is simply given by inserting
\begin{equation}
    \varrho_{\mu\nu}(\omega,{\bf k}) = - g_{\mu \nu} (2\pi) {\rm sgn}(\omega) \delta(\omega^2 - {\bf k}^2 ) \, , 
\end{equation}
leading to
\begin{align}
    W_2({\bf L};{\bf v},t)|_{\mathcal{O}(g^0)} &= - t (1-{\bf v}^2) C_F   \int \frac{d^3 k}{(2\pi)^3} \frac{e^{{\bf k} \cdot {\bf v}/T} }{|e^{{\bf k} \cdot {\bf v}/T} - 1| } (2\pi) \delta(({\bf k} \cdot {\bf v})^2 - {\bf k}^2 ) \left[ e^{ i {\bf k} \cdot {\bf L} } - 1 \right] \nonumber \\
    &= - t \frac{C_F}{\pi}  \int \frac{d^2\hat{k}}{4\pi} \frac{ 1-{\bf v}^2}{1 - ( \hat{k} \cdot {\bf v})^2} \int_0^\infty  dk \, k^2 \frac{e^{k \hat{k} \cdot {\bf v} /T} }{|e^{k  \hat{k} \cdot {\bf v}/T} - 1| } \delta( k^2 ) \left[ e^{ i k {\hat k} \cdot {\bf L} } - 1 \right] \nonumber \\
    &= 0 \, , 
\end{align}
which vanishes because $dk = d(k^2)/(2k)$ and the combination
\begin{equation}
    \frac{1}{2k} k^2 \frac{e^{k \hat{k} \cdot {\bf v} /T} }{|e^{k  \hat{k} \cdot {\bf v}/T} - 1| } \left[ e^{ i k {\hat k} \cdot {\bf L} } - 1 \right] 
\end{equation}
vanishes as $k^2 \to 0$. Note that this is only identically zero because of the subtraction by minus one, the same that ensures that $W_2({\bf L} = {\bf 0};{\bf v},t) = 0$. Without this subtraction, the result is nonzero (and actually encodes a correction to the heavy quark mass at order $g^2$).

It follows from here that all we need to determine are the contributions to the spectral function of the gauge field $\varrho_{\mu\nu}$ at order $g^2$. As discussed in the main text around Eq.~\eqref{eq:Wightman-decomp}, for a complete fixed-order calculation of this spectral function (equivalently, the Wightman function), it is necessary to consider the contributions coming from HTL resummation at the same time as one considers the perturbative contributions, while making sure not to double-count. The discussion in the main text covers all of these aspects. Appendix~\ref{app:calculation-perturbative-2} discusses the calculation of the perturbative piece. Appendix~\ref{app:calculation-HTL} presents the HTL-resummed expressions in full detail and points the reader to references where the derivations thereof can be found. The contribution that needs to be subtracted to match these two is discussed in detail in the main text.

\subsubsection{$3$-point contribution}
 \label{app:calculation-structure-3}

Expanding out the Wilson lines up to cubic order, one arrives at
\begin{align} \label{eq:W3}
    W_3({\bf L};{\bf v},t) &= \frac{i^3}{N_c} {\rm Tr}_c \! \left[ T^a T^b T^c \right] n^\mu n^\nu n^\rho  \\
    & \quad \times \Bigg[ -\int_{0}^t \!\! dt_3 \! \int_{t_3}^t \!\! dt_2 \! \int_{t_2}^t \!\! dt_1  \langle A_\mu^a(h_{{\bf v}}^{{\bf L}}(t_3)) A_\nu^b(h_{{\bf v}}^{{\bf L}}(t_2)) A_\rho^c(h_{{\bf v}}^{{\bf L}}(t_1)) \rangle_T \nonumber \\
    & \quad\quad\quad +\int_0^t \!\! dt_3 \! \int_{t_3}^t \!\! dt_2 \! \int_{0}^t \!\! dt_1 \langle A_\mu^a(h_{{\bf v}}^{{\bf L}}(t_3)) A_\nu^b(h_{{\bf v}}^{{\bf L}}(t_2)) A_\rho^c(h_{{\bf v}}^{{\bf 0}}(t_1)) \rangle_T \nonumber \\
    & \quad\quad\quad - \int_0^t \!\! dt_3 \! \int_{0}^t \!\! dt_2 \! \int_{0}^{t_2} \!\! dt_1 \langle A_\mu^a(h_{{\bf v}}^{{\bf L}}(t_3)) A_\nu^b(h_{{\bf v}}^{{\bf 0}}(t_2)) A_\rho^c(h_{{\bf v}}^{{\bf 0}}(t_1)) \rangle_T \nonumber \\
    & \quad\quad\quad + \int_{0}^t \!\! dt_3 \! \int_{0}^{t_3} \!\! dt_2 \! \int_{0}^{t_2} \!\! dt_1 \langle A_\mu^a(h_{{\bf v}}^{{\bf 0}}(t_3)) A_\nu^b(h_{{\bf v}}^{{\bf 0}}(t_2)) A_\rho^c(h_{{\bf v}}^{{\bf 0}}(t_1)) \rangle_T  \nonumber
    \Bigg] \, ,
\end{align}
which is simply a functional of the gauge field $3$-point function.

Perturbatively, the time-ordered $3$-gluon vertex is equal to (with an overall minus sign in the anti-time-ordered case)
\begin{equation}
    V_{\mu \nu \rho}^{(3),abc}(k_1,k_2,k_3) g f^{abc} \left[ g_{\mu \nu} (k_1 - k_2)_\rho + g_{\nu \rho} (k_2 - k_3)_\mu + g_{\rho \mu} (k_3 - k_1)_\nu \right]
\end{equation}
where the external particle labels are grouped as $(k_1,a,\mu)$, $(k_2,b,\nu)$, $(k_3,c,\rho)$, and all momenta are ingoing. In Feynman gauge, the propagators connecting each of the gauge field insertions to the vertex is proportional to $g_{\mu \nu}$, meaning that the direction $4$-vectors $n^\mu$ get directly contracted with the indices of the $3$-gluon vertex. Because the vertex is antisymmetric and it is being contracted with $n^\mu n^\nu n^\rho $, we conclude that
\begin{equation}
    W_3({\bf L}; {\bf v},t) = 0 
\end{equation}
in Feynman gauge.

We emphasize that this is a consequence of our gauge choice; other gauges will feature a momentum-dependent tensor in the place of the $g_{\mu \nu}$ in the numerator of the Feynman gauge propagator, which can lead to a nonzero contribution from this term.

\subsubsection{$4$-point contribution}
 \label{app:calculation-structure-4}

The explicit expression for $W_4$ after expanding out the Wilson loop in terms of gauge fields was already given in the main text in Eq.~\eqref{eq:W4}. For purposes of careful examination, we reproduce it here:
\begin{align} \label{eq:W4-app}
    W_4({\bf L};{\bf v},t) &= \frac{1}{N_c} {\rm Tr}_c \! \left[ T^a T^b T^c T^d \right] n^\mu n^\nu n^\rho n^\sigma  \\
    & \quad \times \Bigg[ \int_0^t \!\! dt_4 \! \int_{t_4}^t \!\! dt_3 \! \int_{t_3}^t \!\! dt_2 \! \int_{t_2}^t \!\! dt_1  \langle A_\mu^a(h_{{\bf v}}^{{\bf L}}(t_4)) A_\nu^b(h_{{\bf v}}^{{\bf L}}(t_3)) A_\rho^c(h_{{\bf v}}^{{\bf L}}(t_2)) A_\sigma^d(h_{{\bf v}}^{{\bf L}}(t_1)) \rangle_T \nonumber \\
    & \quad \,\,\, -\int_0^t \!\! dt_4 \! \int_{t_4}^t \!\! dt_3 \! \int_{t_3}^t \!\! dt_2 \! \int_{0}^t \!\! dt_1 \langle A_\mu^a(h_{{\bf v}}^{{\bf L}}(t_4)) A_\nu^b(h_{{\bf v}}^{{\bf L}}(t_3)) A_\rho^c(h_{{\bf v}}^{{\bf L}}(t_2)) A_\sigma^d(h_{{\bf v}}^{{\bf 0}}(t_1)) \rangle_T \nonumber \\
    & \quad \,\,\, + \int_0^t \!\! dt_4 \! \int_{t_4}^t \!\! dt_3 \! \int_{0}^t \!\! dt_2 \! \int_{0}^{t_2} \!\! dt_1 \langle A_\mu^a(h_{{\bf v}}^{{\bf L}}(t_4)) A_\nu^b(h_{{\bf v}}^{{\bf L}}(t_3)) A_\rho^c(h_{{\bf v}}^{{\bf 0}}(t_2)) A_\sigma^d(h_{{\bf v}}^{{\bf 0}}(t_1)) \rangle_T \nonumber \\
    & \quad \,\,\, - \int_0^t \!\! dt_4 \! \int_{0}^t \!\! dt_3 \! \int_{0}^{t_3} \!\! dt_2 \! \int_{0}^{t_2} \!\! dt_1 \langle A_\mu^a(h_{{\bf v}}^{{\bf L}}(t_4)) A_\nu^b(h_{{\bf v}}^{{\bf 0}}(t_3)) A_\rho^c(h_{{\bf v}}^{{\bf 0}}(t_2)) A_\sigma^d(h_{{\bf v}}^{{\bf 0}}(t_1)) \rangle_T \nonumber \\
    & \quad \,\,\, + \int_0^t \!\! dt_4 \! \int_{0}^{t_4} \!\! dt_3 \! \int_{0}^{t_3} \!\! dt_2 \! \int_{0}^{t_2} \!\! dt_1 \langle A_\mu^a(h_{{\bf v}}^{{\bf 0}}(t_4)) A_\nu^b(h_{{\bf v}}^{{\bf 0}}(t_3)) A_\rho^c(h_{{\bf v}}^{{\bf 0}}(t_2)) A_\sigma^d(h_{{\bf v}}^{{\bf 0}}(t_1)) \rangle_T  \nonumber
    \Bigg] \, .
\end{align}

In the main text, we argued that only the middle line gives rise to a genuine ${\bf L}$ dependence. We now review this in more detail. 
It is clear that an overall coordinate shift can remove the explicit ${\bf L}$ dependence from the first line; hence this is indeed just a constant. Similarly, the fifth line is explicitly ${\bf L}$-independent. The second and fourth lines require a somewhat closer look to reach this conclusion: while the two-point function does not give rise to a genuine ${\bf L}$ dependence even for a contraction of gauge fields coming from different Wilson lines, one should still check that there are no subtleties induced by the extra time integrations or the extra number of fields.

We proceed via explicit calculation. Consider first the second line in Eq.~\eqref{eq:W4-app} and carry out all of the Wick contractions in the $4$-point function. One obtains
\begin{align} \label{eq:W4-second-line}
    \int_0^t \!\! dt_4 \! \int_{t_4}^t \!\! dt_3 \! \int_{t_3}^t \!\! dt_2 \! \int_{0}^t \!\! dt_1 & \langle A_\mu^a(h_{{\bf v}}^{{\bf L}}(t_4)) A_\nu^b(h_{{\bf v}}^{{\bf L}}(t_3)) A_\rho^c(h_{{\bf v}}^{{\bf L}}(t_2)) A_\sigma^d(h_{{\bf v}}^{{\bf 0}}(t_1)) \rangle_T \\
    = \int_0^t \!\! dt_4 \! \int_{t_4}^t \!\! dt_3 \! \int_{t_3}^t \!\! dt_2 \! \int_{0}^t \!\! dt_1 & \Big[ \langle A_\mu^a(h_{{\bf v}}^{{\bf L}}(t_4)) A_\nu^b(h_{{\bf v}}^{{\bf L}}(t_3)) \rangle_T \langle A_\rho^c(h_{{\bf v}}^{{\bf L}}(t_2)) A_\sigma^d(h_{{\bf v}}^{{\bf 0}}(t_1)) \rangle_T \nonumber \\
    &  + \langle A_\mu^a(h_{{\bf v}}^{{\bf L}}(t_4)) A_\rho^c(h_{{\bf v}}^{{\bf L}}(t_2)) \rangle_T \langle A_\nu^b(h_{{\bf v}}^{{\bf L}}(t_3)) A_\sigma^d(h_{{\bf v}}^{{\bf 0}}(t_1)) \rangle_T \nonumber \\
    & + \langle  A_\nu^b(h_{{\bf v}}^{{\bf L}}(t_3)) A_\rho^c(h_{{\bf v}}^{{\bf L}}(t_2)) \rangle_T \langle A_\mu^a(h_{{\bf v}}^{{\bf L}}(t_4)) A_\sigma^d(h_{{\bf v}}^{{\bf 0}}(t_1)) \rangle_T \nonumber \Big] \, .
\end{align}
It is direct to see that the two-point functions that only contain gauge fields from the Wilson line at ${\bf L}$ are exactly independent of ${\bf L}$. What remains to be checked is whether the time integrations make the result dependent on ${\bf L}$ or not.

Crucially, the integral over $t_1$ in Eq.~\eqref{eq:W4-second-line} is not constrained by the others. This means that, in the limit $t \to \infty$ (focusing on the extensive contribution), it becomes
\begin{align}
    &\int_{-\infty}^\infty dt_1 \langle A_\mu^a(h_{{\bf v}}^{{\bf L}}(t)) A_\sigma^d(h_{{\bf v}}^{{\bf 0}}(t_1)) \rangle_T \nonumber \\
    &= \delta^{ad} \int_{-\infty}^\infty dt_1 \int \frac{d^3 k}{(2\pi)^3} \frac{d\omega}{(2\pi)} \frac{e^{\omega/T} }{e^{\omega/T} - 1 } (-2\pi g_{\mu \sigma}) {\rm sgn}(\omega) \delta(\omega^2 - {\bf k}^2) e^{ i {\bf k} \cdot {\bf L} }   e^{-i\omega (t-t_1) } e^{i (t-t_1) {\bf k} \cdot {\bf v}  } \nonumber \\
    &= \delta^{ad} \int \frac{d^3 k}{(2\pi)^3} \frac{e^{{\bf k} \cdot {\bf v}/T} }{|e^{{\bf k} \cdot {\bf v}/T} - 1| } (-2\pi g_{\mu \sigma}) \delta(({\bf k} \cdot {\bf v})^2 - {\bf k}^2) e^{ i {\bf k} \cdot {\bf L} }    \, ,
\end{align}
and therefore explicitly decouples from the other integrals, as there is no remaining $t$ dependence (no matter whether it is $t_2$, $t_3$, or $t_4$). The result of this contribution is then ${\bf L}$-independent in the same way as the $2$-point function discussed in Appendix~\ref{app:calculation-structure-2}.
The argument to establish that the $4$-th line of~\eqref{eq:W4-app} does not give rise to a dependence on ${\bf L}$ is completely analogous.

Therefore, we conclude that, except for the third line in Eq.~\eqref{eq:W4-app}, we may set ${\bf L} = {\bf 0}$ without changing the result. Furthermore, it is also clear that if we also set ${\bf L} = {\bf 0}$ in the third line, the value of $W_4$ must vanish. One can check this by noting that the sum of the temporal $4$-volumes from the integration domains in each line add up to zero with the signs as given, and since all of the correlators are the same when ${\bf L} = {\bf 0}$, the cancellation is exact. One is then left with the third line in Eq.~\eqref{eq:W4-app}, which can be straightforwardly written out as~\eqref{eq:W4-simp}.

We begin by analyzing the time integrals, in the limit $t\to\infty$ from the start. There are two distinct cases,
\begin{align}
    \tilde{B}_1(\omega,\nu) &\equiv \int_{-\infty}^\infty dt_4 \int_{t_4}^\infty dt_3 \int_{-\infty}^\infty dt_2 \int_{-\infty}^{t_2} dt_1 e^{-i\omega(t_4-t_2)} e^{-\nu (t_3-t_1) } \, , \\
    \tilde{B}_2(\omega,\nu) &\equiv \int_{-\infty}^\infty dt_4 \int_{t_4}^\infty dt_3 \int_{-\infty}^\infty dt_2 \int_{-\infty}^{t_2} dt_1 e^{-i\omega(t_4-t_1)} e^{-\nu (t_3-t_2) } \, ,
\end{align}
which --- due to their obvious divergences controlled by the integration limits --- should be viewed as distributions as a function of $\omega$ and $\nu$. Because of their distributional nature, it proves convenient to analyze them in terms of their Fourier transforms. For the first one we obtain
\begin{align}
    \int_{\omega,\nu} e^{i\omega s} e^{i\nu r} \tilde{B}_1(\omega,\nu) &= \int_{-\infty}^\infty dt_4 \int_{t_4}^\infty dt_3 \int_{-\infty}^\infty dt_2 \int_{-\infty}^{t_2} dt_1 \, \delta(t_4 - t_2 - s) \delta(t_3 - t_1 - r) \nonumber \\
    &= \int_{-\infty}^\infty dt_4 \int_{t_4}^\infty dt_3 \int_{-\infty}^\infty dt_2 \, \delta(t_4 - t_2 - s) \theta(t_2 - t_3 + r) \nonumber \\
    &= \int_{-\infty}^\infty dt_4 \int_{t_4}^\infty dt_3 \, \theta(t_4 - s - t_3 + r) \nonumber \\
    &=  \int_{-\infty}^\infty dt_4 \int_{t_4}^{t_4 + r - s} dt_3 \, \theta(r-s) \nonumber \\
    &= t (r - s) \theta(r-s) \, ,
\end{align}
where in the last step we have identified $\int_{-\infty}^{\infty} dt \longrightarrow t$ as the total time interval. Similarly, the second integral yields
\begin{align}
    \int_{\omega,\nu} e^{i\omega s} e^{i\nu r} \tilde{B}_1(\omega,\nu) &= \int_{-\infty}^\infty dt_4 \int_{t_4}^\infty dt_3 \int_{-\infty}^\infty dt_2 \int_{-\infty}^{t_2} dt_1 \, \delta(t_4 - t_1 - s) \delta(t_3 - t_2 - r) \nonumber \\
    &= \int_{-\infty}^\infty dt_4 \int_{t_4}^\infty dt_3 \int_{-\infty}^\infty dt_2 \, \theta(t_2 - t_4 + s) \delta(t_3 - t_2 - r) \nonumber \\
    &= \int_{-\infty}^\infty dt_4 \int_{t_4}^\infty dt_3 \, \theta(t_3 - r + s - t_4) \nonumber \\
    &= \int_{-\infty}^\infty dt_4 \int_{t_4}^\infty dt_3 \, \left[ 1 - \theta(t_4 - s - t_3 + r) \right] \nonumber \\
    &= \frac{t^2}{2} - t (r - s) \theta(r-s) \, .
\end{align}

Returning to Eq.~\eqref{eq:W4-simp} and using $P_{\mu \nu}(\omega,{\bf k}) = g_{\mu \nu}$ in Feynman gauge, it follows that
\begin{align}   
    W_4({\bf L};{\bf v},t) &= \frac{1}{N_c} n^\mu n^\nu n^\rho n^\sigma \int_0^t \!\! dt_4 \! \int_{t_4}^t \!\! dt_3 \! \int_{0}^t \!\! dt_2 \! \int_{0}^{t_2} \!\! dt_1 \int_{\omega,{\bf k}} \int_{\Omega,{\bf q}} e^{i ( {\bf k} + {\bf q}) \cdot {\bf L}  } D^>(\omega,{\bf k}) D^>(\Omega,{\bf q}) \nonumber \\
    & \quad \times \Bigg[ {\rm Tr}_c \! \left[ T^a T^b T^a T^b \right]  e^{-i(\omega - {\bf k} \cdot {\bf v} ) (t_4-t_2) } e^{-i(\Omega - {\bf q}\cdot {\bf v}) (t_3-t_1)} P_{\mu \rho}(\omega,{\bf k}) P_{\nu \sigma}(\Omega,{\bf q}) \nonumber \\
    & \quad \quad +  {\rm Tr}_c \! \left[ T^a T^b T^b T^a \right] e^{-i(\omega - {\bf k} \cdot {\bf v} ) (t_4-t_1) } e^{-i(\Omega - {\bf q}\cdot {\bf v}) (t_3-t_2)} P_{\mu \sigma}(\omega,{\bf k}) P_{\nu \rho}(\Omega,{\bf q}) 
    \Bigg] \nonumber \\ 
    & \quad - \Big\{ ({\bf L} = {\bf 0})  \Big\} \nonumber \\
    &= \frac{(n^2)^2}{N_c} \int_{\omega,{\bf k}} \int_{\Omega,{\bf q}} e^{i ( {\bf k} + {\bf q}) \cdot {\bf L}  } D^>(\omega,{\bf k}) D^>(\Omega,{\bf q}) \int_{r,s} e^{-i(\omega - {\bf k} \cdot {\bf v} ) s} e^{-i(\Omega - {\bf q}\cdot {\bf v})r} \nonumber \\
    & \quad \times  \Bigg[ {\rm Tr}_c \! \left[ T^a [T^b, T^a] T^b \right] t (r-s) \theta(r-s) + \frac{t^2}2  {\rm Tr}_c \! \left[ T^a T^b T^b T^a \right]  
    \Bigg] \nonumber \\ 
    & \quad - \Big\{ ({\bf L} = {\bf 0})  \Big\}\, . \label{eq:W4-simp-app}
\end{align}
The color traces are straightforward to evaluate: ${\rm Tr}_c \! \left[ T^a [T^b, T^a] T^b \right] = - N_c^2 C_F / 2$ and $ {\rm Tr}_c \! \left[ T^a T^b T^b T^a \right] = N_c C_F^2$. One can also evaluate the integrals over the auxiliary time variables $r,s$:
\begin{align}
    \int_{r,s} e^{-i \omega s} e^{-i \nu r} &= (2\pi)^2 \delta(\omega) \delta(\nu) \, , \\
    \int_{r,s} e^{-i \omega s} e^{-i \nu r} (r-s) \theta(r-s) &= \int_s e^{-i(\omega+\nu) s} \int_u e^{-i \nu u}\, u \, \theta(u) \nonumber \\
    &= (2\pi) \delta(\omega + \nu) \int_0^\infty du \, u \, e^{-i\nu u - u\epsilon} \nonumber \\
    &= (2\pi) \delta(\omega + \nu) \frac{(-1)}{(\nu - i\epsilon)^2} \, ,
\end{align}
where we introduced an $i\epsilon$ to define the resulting distribution, consistent with the admissible deformations of the $\nu$ integration contour into the complex plane.

Using these results, we obtain
\begin{align}
    W_4({\bf L};{\bf v},t) &= \frac12 \left( t C_F n^2 \int_{\omega,{\bf k}} D^>(\omega,{\bf k}) e^{i {\bf k} \cdot {\bf L} \int_s e^{-i (\omega - {\bf k} \cdot {\bf v} ) s } } \right)^2 \nonumber \\
    & \quad  + t (n^2)^2 \frac{N_c C_F}{2}  \int_{\omega,{\bf k}} \int_{\Omega,{\bf q}} e^{i ( {\bf k} + {\bf q}) \cdot {\bf L}  } D^>(\omega,{\bf k}) D^>(\Omega,{\bf q}) \frac{(2\pi) \delta( \omega - {\bf k} \cdot {\bf v} + \Omega - {\bf q} \cdot {\bf v} ) }{( \Omega - {\bf q} \cdot {\bf v} - i\epsilon )^2 } \nonumber \\
    & \quad - \Big\{ ({\bf L} = {\bf 0})  \Big\} \nonumber \\
    &= \frac12 \left[W_2({\bf L};{\bf v},t)|_{\mathcal{O}(g^0)}\right]^2 + t (n^2)^2 \frac{N_c C_F}{2} \int_{k,p} D^>(k)D^>(p-k) e^{i {\bf p} \cdot {\bf L} } \frac{(2\pi) \delta(n \cdot p) }{(n \cdot k - i\epsilon)^2} \nonumber \\
    & \quad - \Big\{ ({\bf L} = {\bf 0})  \Big\} \, , \label{eq:W4-app-decomp}
\end{align}
where we made the slightly inconvenient relabelings $k = (\Omega,{\bf q})$ and $p = (\omega + \Omega,{\bf k} + {\bf q})$. Having arrived at this last expression, we have demonstrated that $W_4$ contains the square of the lower order contribution $W_2$ (which in fact vanishes) and a fully connected piece up to this order in perturbation theory. It is immediate that Eq.~\eqref{eq:W4-result} follows.

\subsection{Calculation of the perturbative contributions} \label{app:calculation-perturbative}

In this section of the Appendix we provide detailed steps of the calculation needed to arrive at the explicit integral expression~\eqref{eq:bigintegral}. We focus on the perturbative pieces, for which more intermediate steps are needed than for the HTL-resummed contributions.

\subsubsection{$2$-point contribution}
 \label{app:calculation-perturbative-2}

The gauge field spectral function is an object that appears in many QCD applications, and as such, has been evaluated many times in the literature. Here, we mostly follow the approach of Ref.~\cite{Binder:2021otw}, where the spectral function is obtained from the real part of the retarded gauge field propagator (self-energy) in Feynman gauge. Comparing with Eqs.~(3.29) and~(3.35) in~\cite{Binder:2021otw}, we see that the bosonic (gluons, ghosts, and, in $\mathcal{N}=4$ SYM, adjoint scalars) contribution to the spectral function is
\begin{align}
    n^\mu n^\nu \varrho_{\mu\nu}^{(b)}(p_0,{\bf p}) =  \int_k  2 {\rm Re} \left\{ N(p,k) [D^R(p)]^2 \left( D^{\mathcal{T}}(k) D^{\mathcal{T}}(p-k) - D^<(k) D^<(p-k)  \right) \right\} \, , \label{eq:glue-ghost-loop}
\end{align}
where the ``numerator'' $N(p,k)$ can be worked out from Table 2 in~\cite{Binder:2021otw} (except for the real adjoint scalars that are present in $\mathcal{N}=4$ SYM; we add those in the next equation). Without simplifications, it is
\begin{align}
    N(p,k) = \frac{N_c}{2} n_\lambda n_\sigma & \left\{ \left[g^\lambda_\mu (p+k)_\nu + g_{\mu \nu} (p-2k)^\lambda +g_\nu^\lambda (k-2p)_\mu \right] \right. \nonumber \\ & \quad \left. \times \left[ g^{\sigma \mu}(-p-k)^\nu + g^{\mu \nu} (2k-p)^\sigma + g^{\nu \sigma} (2p-k)^\mu \right] \right. \nonumber \\ & \quad \left. - 2  k^\lambda (p-k)^\sigma -  N_X (2k-p)^\lambda (2k-p)^{\sigma} \right\} \, ,
\end{align}
where $N_X = 6$ for $\mathcal{N}=4$ SYM, and $N_X = 0$ for QCD. (The product of square brackets comes from the gluon loop, the first term in the last line from the ghost loop, and the last term from the scalar loop.)
After some algebra, this expression reduces to
\begin{align}
    N(p,k) = \frac{N_c}{2} \bigg[ n^2 (-4p^2 - k^2 - (p-k)^2 ) + \left(4D-8  \right) (n\cdot k) (n \cdot (p-k) ) & \nonumber \\
     + (6-D) (p \cdot v)^2 - N_X (n \cdot (2k-p) )^2 & \bigg] \, .
\end{align}

More simplifications become apparent once one exploits the properties of the propagators, and the fact that the spectral function only enters Eq.~\eqref{eq:W2-app} when $p_0 = {\bf p} \cdot {\bf v}$. Indeed, if one uses this kinematic constraint, it is straightforward to check that the free retarded propagators in Eq.~\eqref{eq:glue-ghost-loop} are always purely imaginary\footnote{Except possibly at ${\bf p} = {\bf 0}$ if the integrand develops a singularity there; however, this is exactly the regime where HTL resummation takes over. Hence, this is not a concern.}. It then follows that the nontrivial action of taking the real part is to select the specific combinations of contributions from the time-ordered propagators in Eq.~\eqref{eq:glue-ghost-loop}.

The subsequent steps are mostly algebraic. We use that because $N(p,k)$ is an analytic function of $k$ (i.e., with no singularities), we have
\begin{align}
    &\int_k N(p,k) \left( D^{\mathcal{T}}(k) D^{\mathcal{T}}(p-k) - D^<(k) D^<(p-k)  \right) \nonumber \\
    &= \int_k N(p,k) \left( D^{>}(k) D^{R}(p-k) + D^R(k) D^<(p-k)  \right) \, ,
\end{align}
together with the explicit expressions
\begin{align}
    2 \, {\rm Re} \! \left\{ D^R(k) \right\} &= {\rm sgn}(k_0) (2\pi) \delta(k^2)  \, , \\
    D^>(k) &= \frac{1}{1-e^{-k_0/T}} {\rm sgn}(k_0) (2\pi) \delta(k^2) \, , \\
    D^<(k) &= \frac{1}{e^{k_0/T}-1} {\rm sgn}(k_0) (2\pi) \delta(k^2) \, ,
\end{align}
to arrive at
\begin{align}
    n^\mu n^\nu \varrho_{\mu\nu}^{(b)}(p_0,{\bf p}) = \int_k & \frac{(-1) N(p,k) }{(p^2)^2} \left[ \frac{1}{1-e^{-k_0/T}} + \frac{1}{e^{(p_0 - k_0)/T}-1} \right] \nonumber \\ &  \times {\rm sgn}(k_0) {\rm sgn}(p_0 - k_0) (2\pi)^2 \delta(k^2) \delta((p-k)^2) \, .
\end{align}

The delta functions set $k^2 = (p-k)^2 = 0$, in addition to the kinematic constraint $p\cdot n = 0$. Effectively, this means that the numerator simplifies to
\begin{equation}
    N(p,k) = 2N_c \left[ - n^2 p^2 - \left(2 + N_X  \right) (n \cdot k)^2 \right] \, ,
\end{equation}
where we have set $D = 4$. However, imposing the delta functions makes the integration domain slightly more involved. Explicitly, the integration domain is determined by the conditions
\begin{align}
    k^2 = k_0^2 - {\bf k}^2 = 0 \, , \\
    (p-k)^2 = p^2 - 2 p\cdot k + k^2 = p_0^2 - {\bf p}^2 - 2 p_0 k_0 + 2 {\bf p}\cdot {\bf k} = 0 \, , \\
    p \cdot n = p_0 - {\bf p} \cdot {\bf v} = 0 \, ,
\end{align}
where we used the first condition in the third expression of the second condition. We use the first and third conditions to eliminate $p_0$ and $|{\bf k}|$ in terms of ${\bf p} \cdot {\bf v}$ and $k_0$, respectively. To impose the second condition it is convenient to introduce the cosine of the angle between ${\bf p}$ and ${\bf k}$
\begin{equation}
    u \equiv \frac{ {\bf p} \cdot {\bf k} }{ |{\bf p}| |{\bf k}| } \, ,
\end{equation}
with which the second condition above becomes
\begin{equation}
     u = \frac{ {\bf p}^2 + 2 ({\bf p} \cdot {\bf v}) k_0 -  ({\bf p} \cdot {\bf v})^2 }{2 |{\bf p}| |k_0| } \, . \label{eq:u-in-terms-of-k0-vecp-app}
\end{equation}
By definition, $|u| \leq 1$, and so a Heaviside step function $\theta(1 - u^2)$ remains after carrying out the angular integral, with $u$ given by~\eqref{eq:u-in-terms-of-k0-vecp-app} in the resulting expression after integrating over the angle between ${\bf p}$ and ${\bf k}$. 

Finally, we also have to express $(n \cdot k)^2$ in the numerator $N(p,k)$ in terms of $k_0$ and ${\bf p}$ after the integrations described above have been carried out. Because ${\bf v}$ specifies an additional direction that is not aligned in any specific way with ${\bf p}$ or ${\bf k}$, this term is sensitive to the azimuthal angle of the ${\bf k}$ integration (where the polar direction is specified by ${\bf p}$). That being said, since the dependence on the azimuthal angle only appears through $(n \cdot k)^2$, carrying out the corresponding integral is not difficult as it can be written in terms of elementary trigonometric integrals.

When all of these steps are carried out, we obtain
\begin{align}
    &n^\mu n^\nu \varrho_{\mu\nu}^{(b)}( \pv ,{\bf p} ) \nonumber \\
    &= \int_{-\infty}^\infty \frac{dk_0}{2\pi} \frac{2N_c }{4 |{\bf p}| [{\bf p}^2 - (\pv)^2]^2 } \frac{e^{\pv/T} - 1 }{|e^{k_0/T}-1||e^{(\pv-k_0)/T}-1|} \Theta(1-u^2) \times \nonumber \\ & \left[ (1 - {\bf v}^2) [(\pv)^2 - {\bf p}^2] + (2 + N_X) \! \left( \!\! \left(k_0 - u |k_0| \frac{\pv}{|{\bf p}|} \right)^2 +  \frac{k_0^2}{2} \left( {\bf v}^2 - \frac{(\pv)^2}{{\bf p}^2} \right)(1-u^2) \! \right)  \right]
\end{align}
from the bosonic contribution to the gluon self-energy in Feynman gauge.

The self-energy also receives a fermionic contribution, which exhibits a different structure due to the statistics of such fields. Following the same logic as before (again, the vertex factors and propagator structures can be read off from~\cite{Binder:2021otw}), one obtains
\begin{align}
    n^\mu n^\nu \varrho_{\mu\nu}^{(f)}(p_0,{\bf p}) = N_f T_R \int_k \frac{(-1)n_\mu n_\nu}{(p^2)^2} 2 {\rm Re} \left\{  {\rm Tr} \left[ \gamma^\mu S^{\mathcal{T}}(k) \gamma^\nu S^{\mathcal{T}}(k-p)  - \gamma^\mu S^<(k) \gamma^\nu S^>(k-p) \right]  \right\} \, , \label{eq:fermion-loop}
\end{align}
where $S^\mathcal{T}$, $S^<$ and $S^>$ are free (massless) fermion propagators. They are determined by
\begin{align}
    S^>(k) &= \slashed{k} (2\pi) \delta(k^2) \left[ \theta(k_0) - n_F(|k_0|) \right] = \slashed{k} (2\pi) \delta(k^2) {\rm sgn}(k_0) (1 - n_F(k_0)) \, , \\
    S^<(k) &= \slashed{k} (2\pi) \delta(k^2) \left[ \theta(-k_0) - n_F(|k_0|) \right] = -\slashed{k} (2\pi) \delta(k^2) {\rm sgn}(k_0)  n_F(k_0) \, , \\
    S^R(k) &= \frac{i\slashed{k}}{k^2 + i {\rm sgn}(k_0) 0^+ } \, ,  \\
    S^{\mathcal{T}}(k) &= S^R(k) + S^<(k) \, ,
\end{align}
where $n_F(k_0) = [\exp(k_0/T)+1 ]^{-1}$, and we have also introduced the retarded correlator for convenience.

Making use of these identities, one can show that Eq.~\eqref{eq:fermion-loop} simplifies to
\begin{align}
    n^\mu n^\nu \varrho_{\mu\nu}^{(f)}(p_0,{\bf p}) = N_f T_R \int_k  & \frac{ n_\mu (p-k)_\rho n_\nu k_\sigma }{(p^2)^2} {\rm Tr} \left[\gamma^\mu \gamma^\rho \gamma^\nu \gamma^\sigma \right] \nonumber \\& \times 2 {\rm Re} \big\{  \big( - D^R(k-p) n_F(k_0) (2\pi) {\rm sgn}(k_0) \delta(k^2) \nonumber \\& \quad\quad\quad\quad - n_F(k_0-p_0) (2\pi) {\rm sgn}(k_0-p_0) \delta((k-p)^2)  D^R(k)  \nonumber \\& \quad\quad\quad\quad + n_F(k_0) (2\pi)^2 {\rm sgn}(k_0) \delta(k^2) {\rm sgn}(k_0-p_0) \delta((k-p)^2)   \big) \big\} \, , \label{eq:fermion-loop-2}
\end{align}
where $D^R$ is the scalar retarded propagator we used in the calculation of the bosonic contribution to the self-energy. The real part can be taken as before, and the Dirac trace can be read off from textbooks~\cite{Schwartz:2014sze}. After some algebraic manipulations, we obtain
\begin{align}
    n^\mu n^\nu \varrho_{\mu\nu}^{(f)}(\pv,{\bf p}) = 2N_f T_R \int_k  & \frac{ 4 (n\cdot k)^2 + n^2 p^2 }{(p^2)^2} \frac{e^{k_0/T} - e^{(k_0-\pv)/T}  }{(e^{k_0/T} + 1 )(e^{(k_0-\pv)/T} + 1 ) } \nonumber \\& \times  (2\pi)^2 {\rm sgn}(k_0( k_0- \pv) ) \delta(k^2) \delta((k-p)^2)  \, , \label{eq:fermion-loop-3}
\end{align}
where $p_\mu = (\pv,{\bf p})$ whenever $p$ appears.

The delta functions can be handled as before and the resulting phase space is exactly the same; the same considerations also apply to the term in the numerator proportional to $(n\cdot k)^2$. The explicit expression that results from here is
\begin{align}
    &n^\mu n^\nu \varrho_{\mu\nu}^{(f)}(\pv,{\bf p}) \nonumber \\
    &= \int_{-\infty}^\infty \frac{dk_0}{2\pi} \frac{2N_f T_R }{4 |{\bf p}| [{\bf p}^2 - (\pv)^2]^2 } \frac{e^{k_0/T} - e^{(k_0-\pv)/T}  }{(e^{k_0/T} + 1 )(e^{(k_0-\pv)/T} + 1 ) } {\rm sgn}(k_0( k_0- \pv) ) \Theta(1-u^2) \nonumber \\ & \quad\quad \times \left[ (1 - {\bf v}^2) [(\pv)^2 - {\bf p}^2] + 4\! \left( \!\! \left(k_0 - u |k_0| \frac{\pv}{|{\bf p}|} \right)^2 +  \frac{k_0^2}{2} \left( {\bf v}^2 - \frac{(\pv)^2}{{\bf p}^2} \right)(1-u^2) \! \right)  \right] \, ,
\end{align}
where the sign function may in fact be dropped due to the kinematic constraint $|u| \leq 1$. This follows from the fact that this constraint is explicitly given by
\begin{equation}
     \left| \frac{ {\bf p}^2 + 2 ({\bf p} \cdot {\bf v}) k_0 -  ({\bf p} \cdot {\bf v})^2 }{2 |{\bf p}| |{\bf k}| } \right| \leq 1 \, ,
\end{equation}
which, after algebraic manipulations, becomes
\begin{align}
    k_0 \leq - \frac{1}{2}(| {\bf p}| - \pv ) \, , & & {\rm or} & & k_0 \geq \frac12 (|{\bf p}| + \pv) \, .
\end{align}
Because $k_0$ satisfies this condition, then $k_0 >0$ implies that $k_0 - \pv \geq (|{\bf p}| - \pv)/2 \geq 0$, and on the other hand, $k_0 < 0$ implies that $k_0 - \pv \leq - (|{\bf p}| + \pv )/2 \leq 0$. Therefore, both factors inside the sign function have the same sign, and therefore we may simply omit it.

In conclusion, we have obtained that the gluon self-energy contributions amount to 
\begin{align}
    &n^\mu n^\nu \left( \varrho_{\mu\nu}^{(b)}(\pv,{\bf p}) + \varrho_{\mu\nu}^{(f)}(\pv,{\bf p}) \right) \nonumber \\
    &= \int_{-\infty}^\infty \frac{dk_0}{2\pi} \frac{2N_c }{4 |{\bf p}| [{\bf p}^2 - (\pv)^2]^2 } \frac{e^{\pv/T} - 1 }{|e^{k_0/T}-1||e^{(\pv-k_0)/T}-1|} \Theta(1-u^2) \times \nonumber \\ & \left[ (1 - {\bf v}^2) [(\pv)^2 - {\bf p}^2] + (2 + N_X) \! \left( \!\! \left(k_0 - u |k_0| \frac{\pv}{|{\bf p}|} \right)^2 +  \frac{k_0^2}{2} \left( {\bf v}^2 - \frac{(\pv)^2}{{\bf p}^2} \right)(1-u^2) \! \right)  \right]
    \nonumber \\ &+ \int_{-\infty}^\infty \frac{dk_0}{2\pi} \frac{2N_f T_R }{4 |{\bf p}| [{\bf p}^2 - (\pv)^2]^2 } \frac{e^{k_0/T}(1 - e^{-\pv/T})  }{(e^{k_0/T} + 1 )(e^{(k_0-\pv)/T} + 1 ) } \Theta(1-u^2) \times  \nonumber \\ & \quad\quad\quad \left[ (1 - {\bf v}^2) [(\pv)^2 - {\bf p}^2] + 4\! \left( \!\! \left(k_0 - u |k_0| \frac{\pv}{|{\bf p}|} \right)^2 +  \frac{k_0^2}{2} \left( {\bf v}^2 - \frac{(\pv)^2}{{\bf p}^2} \right)(1-u^2) \! \right)  \right] \, ,
\end{align}
which enters the Wilson loop as specified by~\eqref{eq:W2-app}. Then, all that remains is to obtain the contribution from the connected part of the gluon $4$-point correlator.

\subsubsection{$4$-point contribution}
 \label{app:calculation-perturbative-4}

From Eq.~\eqref{eq:W4-app-decomp}, we have that
\begin{equation}
    W_4({\bf L};{\bf v},t)_{\rm conn.} = t (n^2)^2 \frac{N_c C_F}{2} \int_{k,p} D^>(k)D^>(p-k) \left[ e^{i {\bf p} \cdot {\bf L} } -1 \right] \frac{(2\pi) \delta(n \cdot p) }{(n \cdot k - i\epsilon)^2} \, ,
\end{equation}
which, for convenience, can be cast in the form of the spectral function that describes the contributions to $W_2$; we therefore introduce
\begin{equation}
    n^\mu n^\nu \varrho_{\mu\nu}^{(4)}(p_0,{\bf p}) = (1 - e^{-p_0/T}) (n^2)^2 \frac{N_c}{2} \int_{k} D^>(k)D^>(p-k) \frac{1}{(n \cdot k - i\epsilon)^2} \, ,
\end{equation}
which contributes to the total result for the Wilson loop in the same way as the self-energy contributions in the $2$-point function --- i.e., the various expressions labeled by $n^\mu n^\nu \varrho_{\mu\nu}$ may simply be added up.

It follows that 
\begin{equation}
    n^\mu n^\nu \varrho_{\mu\nu}^{(4)}(p_0,{\bf p}) = (1 - e^{-p_0/T}) (1 - {\bf v}^2)^2 \frac{N_c}{2} \int_{k} \frac{(2\pi)^2 \delta(k^2) \delta((p-k)^2) }{|1-e^{-k_0/T}||1-e^{-(p_0-k_0)/T}|} \frac{1}{(n \cdot k - i\epsilon)^2} \, ,
\end{equation}
where the two delta functions restrict the integration domain in the same way as before for the self-energy contributions. The only qualitative difference with the previous integrations over $k$ we had to carry out lies in the denominator, where $n \cdot k = k_0 - {\bf v} \cdot {\bf k}$ is sensitive to the azimuthal angle in a more involved way than before as it is not an elementary trigonometric integral. That being said, while not elementary, they can be carried out analytically. Since we earlier introduced $u$ as the cosine of the angle between ${\bf k}$ and ${\bf p}$, it is convenient to introduce
\begin{align}
    v_\perp \equiv \sqrt{ {\bf v}^2 - \frac{ (\pv)^2 }{{\bf p}^2} } \, , & & {\rm and} & & v_\parallel \equiv \frac{\pv}{|{\bf p}|} \, .
\end{align}
We then have
\begin{equation}
    \frac{1}{(n \cdot k - i\epsilon)^2} = \frac{1}{(k_0 - u|k_0| v_\parallel - \cos (\phi) \sqrt{1-u^2} |k_0| v_\perp -i\epsilon )^2} \, ,
\end{equation}
where $\phi$ is the azimuthal angle defined such that $\phi = 0$ when ${\bf k}$ lies on the plane defined by ${\bf p}$ and ${\bf v}$. Note that we used the constraint $|{\bf k}| = |k_0|$ imposed by $\delta(k^2)$. Also, note that for all physical heavy quark velocities $|{\bf v}|< 1$, the $i\epsilon$ is inconsequential because $n \cdot k$ never vanishes (unless $k_0 = 0$, which is ruled out by the kinematic constraint $\Theta(1-u^2)$ for all $|{\bf p}|>0$; the fact that the sum of all contributions gives an IR finite integrand means that the limiting case $|{\bf p}| = 0$ does not contribute).

As such, we need the value of the integral
\begin{equation}
    I_\phi = \int_0^{2\pi} \frac{d\phi}{2\pi} \frac{1}{(k_0 - u|k_0| v_\parallel - \cos (\phi) \sqrt{1-u^2} |k_0| v_\perp )^2} \, ,
\end{equation}
which can be obtained by using that
\begin{equation}
    \int_0^{2\pi} \frac{d\phi}{2\pi} \frac{1}{(1 + b \cos \phi)^2} = \frac{1}{(1-b^2)^{3/2}} \quad {\rm if} \quad |b| < 1 \, . 
\end{equation}
It is immediate that
\begin{equation}
    I_\phi = \frac{1}{k_0^2} \frac{1 -  u v_\parallel \, {\rm sgn}(k_0)  }{[(1 -  u v_\parallel \, {\rm sgn}(k_0))^2 - (1-u^2) v_\perp^2 ]^{3/2}} \, ,
\end{equation}
which translates to
\begin{align}
    &n^\mu n^\nu \varrho_{\mu\nu}^{(4)}(p_0,{\bf p}) \nonumber \\ 
    &=  \int_{-\infty}^\infty \frac{dk_0}{2\pi} \frac{N_c}{8|{\bf p}| k_0^2} \frac{ (e^{\pv / T} - 1) \Theta(1-u^2) }{|e^{k_0/T}-1||e^{(\pv -k_0)/T}-1|}   \frac{(1 - {\bf v}^2)^2 [1 -  u v_\parallel \, {\rm sgn}(k_0)]   }{[(1 -  u v_\parallel \, {\rm sgn}(k_0))^2 - (1-u^2) v_\perp^2 ]^{3/2}} \, .
\end{align}

Therefore, the ``hard'' contribution that enters the kernel $S({\bf L};{\bf v})$ is finally given by
\begin{equation}
    \mathcal{I}_{\rm Hard}({\bf p},{\bf v}) = n^\mu n^\nu \left( \varrho_{\mu\nu}^{(b)}(\pv,{\bf p}) + \varrho_{\mu\nu}^{(f)}(\pv,{\bf p}) + \varrho_{\mu\nu}^{(4)}(\pv,{\bf p}) \right) \, ,
\end{equation}
fully fixed by the expressions in this Appendix.

\subsection{Explicit expressions for the HTL contributions}
\label{app:calculation-HTL}

\subsubsection{Resummed}

Here we consider the HTL-resummed self-energies, which give a partial answer containing the contributions from the soft particles in the medium. 
In terms of the expressions given in~\cite{Laine:2016hma}, one finds that
\begin{align}
    S_{HTL}^{\rm resummed}({\bf L}; {\bf v}) &= -\frac{g^2 C_F}{T} \!\! \int \! \frac{d^3 {\bf k} }{(2\pi)^3} \frac{ \left( e^{i{\bf k} \cdot {\bf L}} - 1 \right) e^{ {\bf k} \cdot {\bf v}/T } }{e^{ {\bf k} \cdot {\bf v}/T } - 1  } \times  \nonumber \\  & \quad \bigg[ \frac{ 2 \Gamma_E(v \cos \theta)   }{ \Sigma_E^2( k_0 = kv \cos \theta, k ) + \Gamma_E^2(v \cos \theta) } 
    + \frac{2 v^2 \sin^2 \theta \, \Gamma_T(v \cos \theta)  }{ \Sigma_T^2( k_0 = kv \cos \theta, k ) + \Gamma_T^2(v \cos \theta) } \bigg] \, , \label{eq:S-HTL}
\end{align}
where $\cos \theta \equiv \tfrac{{\bf v} \cdot {\bf k}}{|{\bf v}| |{\bf k}|} $ and $v = |{\bf v}|$. The functions $\Sigma_{T/E}$ and $\Gamma_{T/E}$ introduce one additional scale through the Debye mass $m_D$, which in QCD is given by $m_D^2 \equiv g^2 T^2 \left[ \tfrac{N_c}{3} + \tfrac{N_f}{6} \right]$ and in $\mathcal{N}=4$ SYM by $m_D^2 = 2 \lambda T^2$~\cite{Kim:1999sg,Andersen:2002ey,Czajka:2012gq,Du:2020odw}, where $\lambda = N_c g^2$. The general formula for the Debye mass in a gauge theory with $N_f$ Dirac fermions in representation $R$ and $N_X$ real scalars in representation $X$ is $m_D^2 \equiv g^2 T^2 \left[ \tfrac{N_c}{3} + \tfrac{N_f T_R}{3} + \frac{N_X T_X}{6} \right]$. For our purposes, since we only consider scalars in the adjoint representation, we have $T_X = N_c$.

To be explicit, the functions $\Sigma_{T/E}$ and $\Gamma_{T/E}$ are
\begin{align}
    \Sigma_T(k_0 = k \eta, k) &= k^2 (1 - \eta^2 ) + \frac{m_D^2}{2} \left[ \eta^2 + \frac{\eta (1 - \eta^2 ) }{2} \ln \left| \frac{1 + \eta}{1 -\eta} \right| \right] \, , \\
    \Sigma_E(k_0 = k \eta, k) &= k^2 + m_D^2 \left[ 1 - \frac{ \eta }{2} \ln \left| \frac{1 + \eta}{1 - \eta} \right| \right] \, , \\
    \Gamma_T(\eta) &= \frac{\pi m_D^2}{4} \eta (1 - \eta^2) \, , \\
    \Gamma_E(\eta) &= \frac{\pi m_D^2}{2} \eta \, ,
\end{align}
where $\eta$ is to be evaluated as $\eta = v \cos \theta$.

In terms of the $\mathcal{I}$ expressions used in the main text, the above results define
\begin{align}
    \mathcal{I}_{\rm Soft}({\bf p},{\bf v}) =  \frac{ 2 \Gamma_E( ({\bf p} \cdot {\bf v})/|{\bf p}|  )   }{ \Sigma_E^2( p_0 = {\bf p} \cdot {\bf v} , p ) + \Gamma_E^2( ({\bf p} \cdot {\bf v})/|{\bf p}|  ) } + \frac{2 {\bf v}^2 \left( 1 - \frac{({\bf p} \cdot {\bf v})^2}{{\bf p}^2 {\bf v}^2} \right) \, \Gamma_T( ({\bf p} \cdot {\bf v})/|{\bf p}| )  }{ \Sigma_T^2( p_0 = {\bf p} \cdot {\bf v}, p ) + \Gamma_T^2( ({\bf p} \cdot {\bf v})/|{\bf p}| ) } \, .
\end{align}

\subsubsection{Unresummed (perturbative expansion of the HTL piece)}

The unresummed HTL result --- which in the main text we refer to as the ``matching'' contribution --- is obtained by keeping the leading terms in the perturbative expansion of the HTL self-energies. This simply amounts to dropping the terms proportional to $m_D^2$ in the denominators of Eq.~\eqref{eq:S-HTL}. Concretely,
\begin{align}
    S_{HTL}^{\rm naive}({\bf L}; {\bf v}) = -\frac{g^2 C_F}{T} \!\! \int \! \frac{d^3 {\bf k} }{(2\pi)^3} & \frac{ \left( e^{i{\bf k} \cdot {\bf L}} - 1 \right) e^{ {\bf k} \cdot {\bf v}/T } }{e^{ {\bf k} \cdot {\bf v}/T } - 1  }   \nonumber \\  & \times \bigg[ \frac{ 2 \Gamma_E(v \cos \theta)   }{ k^4 } 
    + \frac{2 v^2 \sin^2 \theta \, \Gamma_T(v \cos \theta)  }{ k^4 (1 - v^2 \cos^2 \theta)^2 } \bigg] \, .
\end{align}
As expected, this is IR divergent as $k \to 0$, and diverges in exactly the same way as the hard piece.
In terms of the notation used in the main text, this expression defines
\begin{equation}
    \mathcal{I}_{\rm Match}({\bf p},{\bf v}) = \frac{ 2 \Gamma_E( ({\bf p} \cdot {\bf v})/|{\bf p}|  )   }{ |{\bf p}|^4 }  + \frac{2 {\bf v}^2 \left( 1 - \frac{({\bf p} \cdot {\bf v})^2}{{\bf p}^2 {\bf v}^2 } \right) \, \Gamma_T( ({\bf p} \cdot {\bf v})/|{\bf p}|  )  }{ |{\bf p}|^4 \left(1 -  \frac{({\bf p} \cdot {\bf v})^2}{{\bf p}^2  }  \right)^2 } \, .
\end{equation}

\section{Numerical Representation of the Fixed-Order Kernel}
\label{app:coord_transforms}

This appendix gives the representation used to numerically evaluate the longitudinal fixed-order kernel
\begin{equation}
K(x,v)\equiv S_{\rm FO}({\bf L}=-ix\hat{\bf v};{\bf v}=v\hat{\bf v}).
\end{equation}
We measure momenta in units of the temperature,
\begin{equation}
q\equiv\frac{p}{T},\qquad \omega\equiv\frac{k_0}{T},\qquad \mu\equiv\frac{{\bf p}\cdot{\bf v}}{pv},\qquad \eta\equiv\mu v,
\end{equation}
so that
\begin{equation}
\int_{\bf p}=\frac{T^3}{4\pi^2}\int_0^\infty dq\,q^2\int_{-1}^1d\mu.
\end{equation}

\subsection{Hard contribution}

Let $\widehat{\mathcal H}(\omega,q,\mu,v)$ denote the dimensionless integrand of $\mathcal I_{\rm Hard}$ in Eq.~\eqref{eq:kernel-compact},
\begin{equation}
T^2\mathcal I_{\rm Hard}(p,\mu,v)=\int_{-\infty}^{\infty}\frac{d\omega}{2\pi}\,\widehat{\mathcal H}(\omega,q,\mu,v),
\end{equation}
where the explicit bosonic and fermionic terms are those in Eq.~\eqref{eq:bigintegral}. The phase-space constraint restricts the loop energy to
\begin{equation}
\omega\geq\frac{q(1+\eta)}{2},\qquad \omega\leq-\frac{q(1-\eta)}{2}.
\end{equation}
Both branches can be mapped to $k\geq1$ by
\begin{equation}
\omega_+(k)=\frac{q}{2}(k+\eta),\qquad \omega_-(k)=-\frac{q}{2}(k-\eta),
\end{equation}
with $d\omega=(q/2)dk$. Hence
\begin{equation}
\widehat{\mathcal I}_{\rm Hard}(q,\mu,v)\equiv T^2\mathcal I_{\rm Hard}(p,\mu,v)=\int_1^\infty dk\,\frac{q}{4\pi}\left[\widehat{\mathcal H}\!\left(\omega_+(k),q,\mu,v\right)+\widehat{\mathcal H}\!\left(\omega_-(k),q,\mu,v\right)\right].
\label{eq:IHard-folded}
\end{equation}

\subsection{Matching and soft contributions}

Writing
\begin{equation}
\frac{m_D^2}{g^2T^2}=\frac{\mathcal N}{3},\qquad \chi_T(\mu,v)\equiv\frac{v^2(1-\mu^2)}{2(1-\mu^2v^2)},
\end{equation}
the unresummed matching kernel is
\begin{equation}
\widehat{\mathcal I}_{\rm Match}(q,\mu,v)\equiv\frac{T^2}{g^2}\mathcal I_{\rm Match}(p,\mu,v)=\frac{\pi \mathcal{N}\,\mu v}{3q^4}\left[1+\chi_T(\mu,v)\right].
\end{equation}
For the numerical evaluation the hard and matching contributions are combined before the
radial integration. Pairing the $\mu$ and $-\mu$ contributions, the part with $q<1$ is
\begin{align}
K_{{\rm H+M},<}(x,v)
={}&-\frac{g^4C_F}{4\pi^2}
\int_0^1dq\,q^2\int_0^1d\mu\,
\Bigg\{
\frac{e^{q\mu Tx}-1}{1-e^{-q\mu v}}\,
\widehat{\mathcal I}_{\rm Hard}(q,\mu,v)
+\frac{e^{-q\mu Tx}-1}{1-e^{q\mu v}}\,
\widehat{\mathcal I}_{\rm Hard}(q,-\mu,v)
\nonumber\\
&\hspace{31mm}
-\frac{\pi\mathcal N}{3q^3}\,
\mu^2\,Tx(Tx+v)\,
\left[1+\chi_T(\mu,v)\right]
\Bigg\}.
\label{eq:Khard-match-IR}
\end{align}
For $q>1$ only the hard contribution remains,
\begin{align}
K_{{\rm H},>}(x,v)
={}&-\frac{g^4C_F}{4\pi^2}
\int_1^\infty dq\,q^2\int_0^1d\mu\,
\Bigg[
\frac{e^{q\mu Tx}-1}{1-e^{-q\mu v}}\,
\widehat{\mathcal I}_{\rm Hard}(q,\mu,v)
+\frac{e^{-q\mu Tx}-1}{1-e^{q\mu v}}\,
\widehat{\mathcal I}_{\rm Hard}(q,-\mu,v)
\Bigg].
\label{eq:Khard-UV}
\end{align}
The cancellation between the hard and matching terms is therefore implemented directly
in the $q<1$ integrand. 

The HTL resummation enters the strict fixed-order result only through the remaining quadratic soft contribution: the soft radial integral is performed before expanding in $m_D/T$, producing the logarithmic and finite terms in Eq.~\eqref{eq:SFO-main}. For the longitudinal kernel,
\begin{align}
K_{\rm Soft}(x,v)=\frac{g^4C_F\mathcal{N}}{12\pi}\,Tx(Tx+v)\Bigg\{&\ln\!\left(\frac{m_D}{T}\right)\frac{(v^2-1)\arctanh v+v}{2v^3}\nonumber\\
&+\int_0^1d\mu\,\mu^2\left[\Phi(a_E,b_E)+\chi_T(\mu,v)\Phi(a_T,b_T)\right]\Bigg\}.
\label{eq:Ksoft-numerical}
\end{align}

The complete fixed-order kernel used numerically is therefore
\begin{equation}
K(x,v)=K_{{\rm H+M},<}(x,v)+K_{{\rm H},>}(x,v)+K_{\rm Soft}(x,v).
\label{eq:KFO-numerical-assembly}
\end{equation}
Since the matching and soft contributions are quadratic in $x$, all cumulants with
$n\geq3$ are determined entirely by the hard contribution.

\section{Cumulants and Analytic Structure}\label{app:momentasymptotics}

\subsection{$v\to0$ limit of $\kappa_L$}\label{app:v0kappa}

At $v=0$ rotational invariance implies $\kappa_L=\kappa_T\equiv\kappa$,
with $\kappa=T\mathfrak M_2(0)$. Taking two derivatives of
Eq.~\eqref{eq:KFO-numerical-assembly}, combining the two $k_0$ branches
as in Eq.~\eqref{eq:IHard-folded}, and taking $v\to0$, the $k$ and $\mu$
integrals can be performed explicitly. After evaluating the separately
finite radial integrals, this gives
\begin{align}
\mathfrak M_2(0)
={}&\frac{g^2C_Fm_D^2}{4\pi}
\Bigg\{
1-\frac{N_c+T_RN_f}{3\mathcal N}
+\frac23\left[\ln\!\left(\frac{T}{m_D}\right)-\frac12\right]
\nonumber\\
&\quad+\int_0^\infty\frac{dq}{q}
\Bigg[
\frac{4N_c(1+N_X/2)}{\pi^2\mathcal N}
\mathrm{Li}_2\!\left(e^{-q/2}\right)
\nonumber\\
&\hspace{37mm}
-\frac{8T_RN_f}{\pi^2\mathcal N}
\mathrm{Li}_2\!\left(-e^{-q/2}\right)
-\frac23\theta(1-q)
\Bigg]
\Bigg\}.
\label{eq:M2-v0-reduced}
\end{align}
Here $q=p/T$, and the term proportional to $\ln(T/m_D)-1/2$ follows
from Eq.~\eqref{eq:Ksoft-numerical}, using $\Phi(1,0)=1/2$.
The constant terms inside the square brackets cancel as $q\to0$,
making the remaining integral infrared finite.

To evaluate this integral, we insert a common factor $q^s$ with
$\operatorname{Re}s>0$. The Mellin transforms of the dilogarithms then give
\begin{align}
&\lim_{s\to0}
\left\{
\frac{4\,2^s}{\pi^2}\Gamma(s)\zeta(s+2)
\left[1+\frac{T_RN_f}{\mathcal N}\left(1-2^{-s}\right)\right]
-\frac{2}{3s}
\right\}
\nonumber\\
&\qquad=
\frac23\left[
\ln2-\gamma_{\rm E}
+\frac{T_RN_f}{\mathcal N}\ln2
+\frac{\zeta'(2)}{\zeta(2)}
\right],
\label{eq:M2-v0-mellin}
\end{align}
where the $1/s$ poles cancel before the limit is taken.
Substituting into Eq.~\eqref{eq:M2-v0-reduced} yields
\begin{equation}
\mathfrak M_2(0)=\frac{g^2C_Fm_D^2}{6\pi}
\left[
\frac12+\frac{N_cN_X}{4\mathcal N}-\gamma_{\rm E}
+\ln\!\left(\frac{2T}{m_D}\right)
+\frac{T_RN_f}{\mathcal N}\ln2
+\frac{\zeta'(2)}{\zeta(2)}
\right].
\end{equation}
Therefore, $\kappa=T\mathfrak M_2(0)$ reproduces
Eq.~\eqref{eq:kappa-v0}.

\subsection{$v\to0$ limit of higher cumulants}
\label{app:moments}

For $n\geq3$ the soft and matching terms in Eq.~\eqref{eq:SFO-main} do not contribute. At $v=0$ all odd cumulants vanish, while the even cumulants can be evaluated analytically. To make contact with the hard representation of Appendix~\ref{app:coord_transforms}, first fold the two $k_0$ branches onto $k\geq1$. After combining the branches and taking $v\to0$, the bosonic and fermionic thermal factors reduce to
\begin{equation}
\frac{\sinh^2(p\mu x/2)}{\sinh^2(pk/4T)},\qquad
\frac{\sinh^2(p\mu x/2)}{\cosh^2(pk/4T)},
\end{equation}
respectively. Consequently, for even $n$,
\begin{equation}
\left.\frac{\partial^n}{\partial x^n}\frac{\sinh^2(p\mu x/2)}{\sinh^2(pk/4T)}\right|_{x=0}
=\frac{(p\mu)^n}{2\sinh^2(pk/4T)},\qquad
\left.\frac{\partial^n}{\partial x^n}\frac{\sinh^2(p\mu x/2)}{\cosh^2(pk/4T)}\right|_{x=0}
=\frac{(p\mu)^n}{2\cosh^2(pk/4T)}.
\end{equation}
It is therefore convenient to define
\begin{align}
\mathcal J^{(\sinh)}_{n,m}(T)&\equiv\int_0^\infty dp\int_{-1}^1d\mu\int_1^\infty dk\,\frac{\tfrac12(p\mu)^n k^m}{\sinh^2(pk/4T)},\\
\mathcal J^{(\cosh)}_{n,m}(T)&\equiv\int_0^\infty dp\int_{-1}^1d\mu\int_1^\infty dk\,\frac{\tfrac12(p\mu)^n k^m}{\cosh^2(pk/4T)}.
\end{align}
Using
\begin{equation}
\frac{1}{\sinh^2z}=4\sum_{r=1}^\infty r e^{-2rz},\qquad
\frac{1}{\cosh^2z}=4\sum_{r=1}^\infty(-1)^{r+1}r e^{-2rz},
\end{equation}
one finds
\begin{align}
\mathcal J^{(\sinh)}_{n,m}(T)&=\frac{4(2T)^{n+1}}{n+1}\frac{n!}{n-m}\zeta(n),\\
\mathcal J^{(\cosh)}_{n,m}(T)&=\frac{4(2T)^{n+1}}{n+1}\frac{n!}{n-m}(1-2^{1-n})\zeta(n),
\end{align}
for the values $m=-2,0,2$ required below.

We can then express
\begin{align}
\mathfrak M_n(0)=\frac{g^2C_Fm_D^2}{4\pi T^3}\Bigg\{&\frac{3N_c}{8\pi^2\mathcal N}\left[\mathcal J^{(\sinh)}_{n,-2}-\mathcal J^{(\sinh)}_{n,0}\right]+\frac{3N_c(1+N_X/2)}{16\pi^2\mathcal N}\mathcal J^{(\sinh)}_{n,2}\nonumber\\
&+\frac{3N_fT_R}{8\pi^2\mathcal N}\left[\mathcal J^{(\cosh)}_{n,2}-\mathcal J^{(\cosh)}_{n,0}\right]\Bigg\},
\end{align}
yielding
\begin{align}
\mathfrak{M}_n(0) &=\frac{3g^2C_Fm_D^2\,2^{n-1}T^{n-2}(n-1)!\zeta(n)}{\pi^3(n+1)\mathcal N} \nonumber \\ & \quad \times \left[\frac{N_c\big(N_Xn(n+2)+2(n^2-2n+8)\big)}{8(n-2)(n+2)}+\frac{(1-2^{1-n})T_RN_f}{n-2}\right],
\end{align}
in agreement with Eq.~\eqref{eq:Mn-v0-closed}.

\subsection{Large-order longitudinal cumulants}
\label{app:large-order-arbitrary-v}

For $n\geq3$ only the hard term in Eq.~\eqref{eq:SFO-main} contributes,
\begin{equation}
\mathfrak M_n(v)=\frac{g^4C_F}{T}\int_{\bf p}\frac{(p\mu)^n}{1-e^{-p\mu v/T}}\mathcal I_{\rm Hard}({\bf p},{\bf v}),\qquad \mu\equiv\frac{{\bf p}\cdot{\bf v}}{pv}.
\label{eq:large-n-start}
\end{equation}
The two branches of the $k_0$ integration can be parameterized as
\begin{equation}
k_0=p\left(\frac{1+\mu v}{2}+s\right),\qquad
k_0=-p\left(\frac{1-\mu v}{2}+s\right),\qquad s\geq0.
\label{eq:k0-support-s}
\end{equation}
After combining the bosonic statistical factor in $\mathcal I_{\rm Hard}$ with the factor $1/(1-e^{-p\mu v/T})$ in Eq.~\eqref{eq:large-n-start}, both branches contain
\begin{equation}
\frac{\exp\!\left[-\frac{p}{T}\left(\frac{1-\mu v}{2}+s\right)\right]}
{\left[1-\exp\!\left(-\frac{p}{T}\left(\frac{1-\mu v}{2}+s\right)\right)\right]
 \left[1-\exp\!\left(-\frac{p}{T}\left(\frac{1+\mu v}{2}+s\right)\right)\right]}.
\end{equation}
At fixed $v<1$, the large-$n$ region has $p\sim nT$, so this becomes
\begin{equation}
\exp\!\left[-\frac{p}{T}\left(\frac{1-\mu v}{2}+s\right)\right]
\left[1+\mathcal O\!\left(e^{-p(1-v)/(2T)}\right)\right].
\label{eq:common-boltzmann-weight}
\end{equation}

Substituting the two branches of Eq.~\eqref{eq:k0-support-s} into Eq.~\eqref{eq:u-def} gives
\begin{align}
k_0-|k_0|\mu vu&=\pm\frac{p}{2}(1-\mu^2v^2)(1+2s),\\
k_0^2(1-u^2)&=p^2(1-\mu^2v^2)s(1+s),
\end{align}
where the upper and lower signs refer to the positive and negative $k_0$ branches, respectively. Both branches therefore give the same matrix-element factors. Using $\int_{\bf p}=(4\pi^2)^{-1}\int_0^\infty dp\,p^2\int_{-1}^1d\mu$, $|dk_0|=p\,ds$, and the leading Boltzmann weight in Eq.~\eqref{eq:common-boltzmann-weight}, their sum gives
\begin{align}
\mathfrak M_{n,B}(v)\sim{}&\frac{g^4C_FN_c}{64\pi^3T}\int_{-1}^1d\mu\,\mu^n\int_0^\infty dp\,p^n\int_0^\infty ds\,e^{-\frac{p}{T}\left(\frac{1-\mu v}{2}+s\right)}\nonumber\\
&\times\Bigg[\frac{8(1-v^2)^2(1+2s)}{\sqrt{1-\mu^2v^2}\left[1-\mu^2v^2+4(1-v^2)s(1+s)\right]^{3/2}}-\frac{8(1-v^2)}{1-\mu^2v^2}\nonumber\\
&\qquad+(4+2N_X)\left((1+2s)^2+\frac{2v^2(1-\mu^2)}{1-\mu^2v^2}s(1+s)\right)\Bigg].
\end{align}
The first term in square brackets comes from the connected Wilson-line contribution $W_4$ in Appendix~\ref{app:calculation-perturbative-4}; the remaining terms come from the bosonic self-energy in Appendix~\ref{app:calculation-perturbative-2}. The common $p^{-2}$ dependence of these matrix-element factors has canceled the $p^2$ in the radial measure.

At fixed $v<1$, the expression in square brackets is regular at $s=0$. The first two terms in its expansion are integrated using
\begin{equation}
\int_0^\infty ds\,e^{-ps/T}=\frac{T}{p},\qquad \int_0^\infty ds\,s\,e^{-ps/T}=\frac{T^2}{p^2}.
\end{equation}
Thus the leading term leaves a radial integral
\begin{equation}
\int_0^\infty dp\,p^{n-1}e^{-p(1-\mu v)/(2T)}=(n-1)!\left(\frac{2T}{1-\mu v}\right)^n,
\end{equation}
while a term linear in $s$ is suppressed relative to it by
\begin{equation}
\frac{T\int_0^\infty dp\,p^{n-2}e^{-p(1-\mu v)/(2T)}}{\int_0^\infty dp\,p^{n-1}e^{-p(1-\mu v)/(2T)}}=\frac{1-\mu v}{2(n-1)}.
\end{equation}
This justifies evaluating the matrix-element factors at $s=0$ for the leading large-$n$ result. Performing the $s$ and $p$ integrals and mapping the negative $\mu$ region to positive $\mu$ then gives
\begin{align}
\mathfrak M_{n,B}(v)\sim{}&\frac{g^4C_FN_c}{64\pi^3}(n-1)!(2T)^n\\
&\times\int_0^1d\mu\left[\frac{8(1-v^2)^2}{(1-\mu^2v^2)^2}-\frac{8(1-v^2)}{1-\mu^2v^2}+4+2N_X\right]\left[\left(\frac{\mu}{1-v\mu}\right)^n+(-1)^n\left(\frac{\mu}{1+v\mu}\right)^n\right].\nonumber
\end{align}
The factor $(-1)^n$ comes from $(p\mu)^n$ in the negative $\mu$ region.

Both angular integrals are dominated by $\mu=1$. Expanding around this endpoint and taking the $n$th power gives
\begin{align}
\left(\frac{\mu}{1\mp v\mu}\right)^n&=\frac{1}{(1\mp v)^n}\left[1-\frac{1-\mu}{1\mp v}+\mathcal O\!\left((1-\mu)^2\right)\right]^n\nonumber\\
&=\frac{1}{(1\mp v)^n}\exp\left[-\frac{n(1-\mu)}{1\mp v}+\mathcal O\!\left(n(1-\mu)^2\right)\right].
\end{align}
The relevant angular interval therefore has $1-\mu\sim(1\mp v)/n$. In this interval the leading exponent is of order unity, while the correction to it is of order $1/n$. We must therefore retain the exponential rather than expand it to linear order. The factor $(1\mp v)^{-n}$ comes outside the angular integral, and the matrix-element factors approach $8-8+4+2N_X=4+2N_X$, leaving
\begin{equation}
\int_0^1d\mu\,e^{-n(1-\mu)/(1\mp v)}=\frac{1\mp v}{n}\left(1-e^{-n/(1\mp v)}\right)\sim\frac{1\mp v}{n}.
\end{equation}
Together with $(n-1)!=n!/n$ from the radial integration, this gives the $n!/n^2$ behavior.

For the fermions, the two statistical denominators contain plus signs, but their leading Boltzmann weight is the same. Using the same kinematic identities, the sum of the fermionic matrix-element factors, after extracting $p^{-2}$, has the expansion
\begin{align}
&-\frac{2(1-v^2)}{1-\mu^2v^2}+2(1+2s)^2+\frac{4v^2(1-\mu^2)}{1-\mu^2v^2}s(1+s)\nonumber\\
&\qquad=\frac{4v^2}{1-v^2}(1-\mu)+8s+\mathcal O\!\left((1-\mu)^2,s(1-\mu),s^2\right)
\end{align}
near $\mu=1$, with $1-\mu$ replaced by $1+\mu$ near $\mu=-1$. There is no constant term at either endpoint. The term proportional to $s$ gives the additional suppression derived above, while the term proportional to $1-\mu$ gives
\begin{equation}
\frac{\int_0^1d\mu\,(1-\mu)e^{-n(1-\mu)/(1\mp v)}}{\int_0^1d\mu\,e^{-n(1-\mu)/(1\mp v)}}\sim\frac{1\mp v}{n}.
\end{equation}
The fermionic contribution is therefore suppressed by one further power of $1/n$ at each endpoint. Collecting the leading bosonic contributions gives
\begin{equation}
\mathfrak M_n(v)\sim\frac{g^4C_FN_c(N_X+2)}{32\pi^3}\frac{n!}{n^2}\left[(1-v)\left(\frac{2T}{1-v}\right)^n+(-1)^n(1+v)\left(\frac{2T}{1+v}\right)^n\right].
\label{eq:large-n-result-arbitrary-v}
\end{equation}

Defining
\begin{equation}
R_+(v)\equiv\frac{1-v}{2T},\qquad R_-(v)\equiv\frac{1+v}{2T},\qquad A_\pm(v)\equiv\frac{g^4C_FN_c(N_X+2)}{32\pi^3}(1\mp v),
\label{eq:directional-radii}
\end{equation}
we can write
\begin{equation}
\mathfrak M_n(v)\sim A_+(v)\frac{n!}{n^2}R_+^{-n}+(-1)^nA_-(v)\frac{n!}{n^2}R_-^{-n}.
\label{eq:positive-large-n-compact}
\end{equation}
Thus the expansion about $x=0$ has radius
\begin{equation}
R(v)=R_+(v)=\frac{1-v}{2T},
\end{equation}
while the singularity on the negative real axis is at $x=-R_-(v)$. Their separation,
\begin{equation}
R_-(v)-R_+(v)=\frac{v}{T},
\end{equation}
is fixed by the equilibrium symmetry~\eqref{eq:K-analytic-property}.

The $n^{-2}$ prefactor determines the form of the endpoint singularities. Up to terms less singular at $x=R_+$ and $x=-R_-$,
\begin{equation}
K_{\rm sing}(x,v)=-A_+(v)\Li_2\!\left(\frac{x}{R_+(v)}\right)-A_-(v)\Li_2\!\left(-\frac{x}{R_-(v)}\right).
\label{eq:dilog-singular-kernel}
\end{equation}
Since
\begin{equation}
\Li_2(1-\epsilon)=\frac{\pi^2}{6}+\epsilon\ln\epsilon-\epsilon+\mathcal O(\epsilon^2\ln\epsilon),
\end{equation}
$K$ remains finite at the boundary while its first derivative diverges logarithmically.

\subsection{Moment resummation and analytic continuation}
\label{app:moment-resummation}

\paragraph{Resummation for Numerical Continuation}
For $n\geq3$ we define the residual cumulants after subtracting the known leading large-order contribution,
\begin{equation}
\delta\mathfrak M_n\equiv\mathfrak M_n-A_+\frac{n!}{n^2}R_+^{-n}-(-1)^nA_-\frac{n!}{n^2}R_-^{-n}.
\end{equation}
Keeping the first two cumulants from the complete fixed-order kernel gives
\begin{align}
K_N(x,v)={}&-\mathfrak M_1x-\frac{\mathfrak M_2}{2}x^2-A_+\left[\Li_2\!\left(\frac{x}{R_+}\right)-\frac{x}{R_+}-\frac{x^2}{4R_+^2}\right]\nonumber\\
&-A_-\left[\Li_2\!\left(-\frac{x}{R_-}\right)+\frac{x}{R_-}-\frac{x^2}{4R_-^2}\right]-\sum_{n=3}^{N}\frac{\delta\mathfrak M_n}{n!}x^n.
\label{eq:resummed-K}
\end{align}

This subtraction improves numerical convergence near the endpoints, but the residual cumulant series retains the original radius: the subleading large-order terms remain singular at the same locations.

A stronger subtraction is obtained by removing the contribution of the leading Boltzmann factor in Eq.~\eqref{eq:common-boltzmann-weight} before taking the large-$n$ limit, retaining the full matrix elements rather than their threshold values. Using the variables of Eq.~\eqref{eq:k0-support-s} and writing $\alpha=[(1-\mu v)/2+s]/T$, the radial integral is
\begin{equation}
\int_0^\infty dp\,e^{-\alpha p}\left[e^{p\mu x}-1-p\mu x-\frac{(p\mu x)^2}{2}\right]=\frac{(\mu x)^3}{\alpha^3(\alpha-\mu x)}.
\label{eq:boltzmann-radial}
\end{equation}
The equality holds for $\operatorname{Re}(\alpha-\mu x)>0$; the expression on the right provides its analytic continuation. The subtracted contribution is added back in this integrated form. Removing its Taylor terms through quadratic order leaves the matched $\mathfrak M_1$ and $\mathfrak M_2$ unchanged. This implements the same subtraction-and-restoration procedure as Eq.~\eqref{eq:resummed-K}, while retaining the associated subleading terms as well.

The slowest remaining terms in the statistical factors are proportional to $e^{-p(1-\mu v+2s)/T}$ and $e^{-p(1+2s)/T}$. The integral for the remainder therefore converges in
\begin{equation}
-1<\operatorname{Re}(Tx)<1-v,
\label{eq:residual-convergence-strip}
\end{equation}
so its Taylor radius about $x=0$ is $2R_+$. The corresponding construction about $x=-v/T$ follows from Eq.~\eqref{eq:K-analytic-property}. The original branch points of $K$ have not moved: their contribution is restored analytically. Within $-R_-<\operatorname{Re}x<R_+$ this representation provides an alternative to direct evaluation of the fixed-order integral; it also permits continuation past the first branch points.

\begin{figure}[t]
\centering
\includegraphics[width=\linewidth]{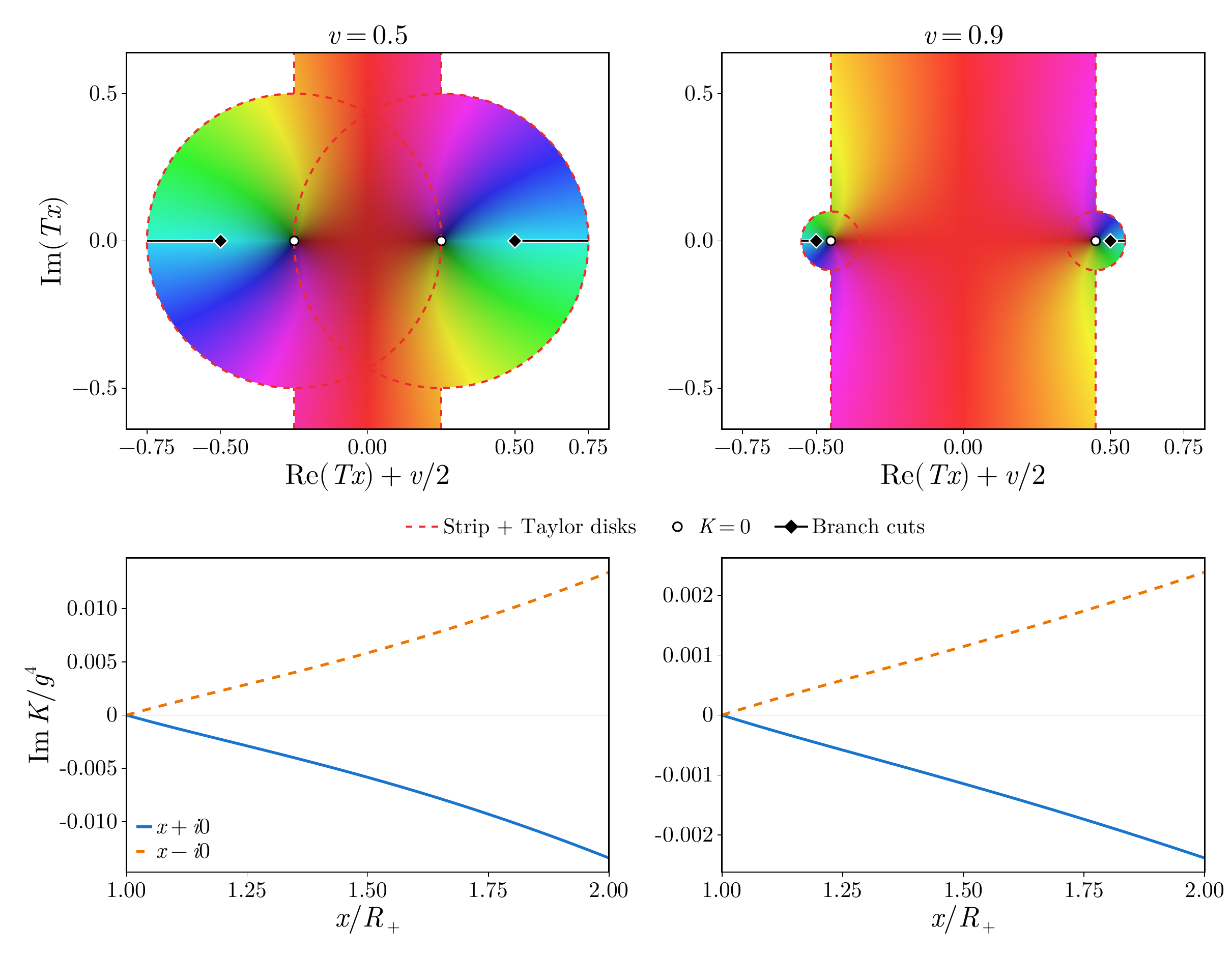}
\caption{Analytic continuation of $K(x;v)$ in QCD at $g=0.1$, for $v=0.5$ (left) and $v=0.9$ (right), with the color convention of Fig.~\ref{fig:complexKplot}. Top: the displayed region is the union of the KMS strip and the convergence disks of the remainder, $|x|<2R_+$ and $|x+v/T|<2R_+$ (red dashed boundaries). Regions outside this union are left white; this restriction does not identify the convergence boundary of the original fixed-order integral. White circles mark the zeros; black diamonds and horizontal lines mark the first branch points and cuts. Bottom: $\operatorname{Im}K/g^4$ on the two sides of the positive cut, $x+i0$ (solid blue) and $x-i0$ (dashed orange).}
\label{fig:kernel-continuation}
\end{figure}

Figure~\ref{fig:kernel-continuation} shows this continuation. For this figure the remainder is evaluated directly as an integral, without truncating its cumulant expansion. Pole contributions are subtracted and restored analytically before numerical quadrature. The values on the two sides of a real cut, $K(x+i0;v)$ and $K(x-i0;v)$, are complex conjugates, and the negative cut follows by the equilibration condition.

\subsection{Transverse radius and the full analytic domain}
\label{app:large-order-transverse-arbitrary-v}

Choose ${\bf v}=v\hat{\bf z}$ and consider one Cartesian transverse component,
\begin{equation}
K_\perp(x,v)\equiv S_{\rm FO}({\bf L}=-ix\hat{\bf x};{\bf v}=v\hat{\bf z})
=-\sum_{n=1}^\infty\frac{\mathfrak M_n^\perp(v)}{n!}x^n.
\end{equation}
Reflection symmetry implies $\mathfrak M_{2m+1}^\perp=0$. For even $n\geq4$ only the hard contribution survives,
\begin{equation}
\mathfrak M_n^\perp(v)=\frac{g^4C_F}{T}\int_{\bf p}\frac{p_x^n}{1-e^{-{\bf p}\cdot{\bf v}/T}}\mathcal I_{\rm Hard}({\bf p},{\bf v}).
\end{equation}
Using the same $k_0$ threshold variable $s$ as in the longitudinal calculation and writing
\begin{equation}
p_x=p\sqrt{1-\mu^2}\cos\phi,
\end{equation}
the $p$ and $s$ integrations again give one factor of $1/n$. Up to a smooth matrix-element factor that does not affect the large-$n$ exponential or power-law dependence, the remaining angular integral is
\begin{equation}
\mathfrak M_n^\perp(v)\propto\frac{n!T^n}{n}\int_{-1}^1d\mu\int_0^{2\pi}d\phi\left[\frac{2\sqrt{1-\mu^2}\cos\phi}{1-v\mu}\right]^n.
\end{equation}
For even $n$ this integral is dominated by
\begin{equation}
\mu=v,\qquad \phi=0,\pi,
\end{equation}
where
\begin{equation}
\frac{2\sqrt{1-\mu^2}|\cos\phi|}{1-v\mu}=2\gamma.
\end{equation}
Expanding around either maximum,
\begin{equation}
\ln\!\left[\frac{2\sqrt{1-\mu^2}|\cos\phi|}{1-v\mu}\right]
=\ln(2\gamma)-\frac{\gamma^4}{2}(\mu-v)^2-\frac12(\phi-\phi_\star)^2+\ldots,
\end{equation}
so the two angular integrations together supply a second factor of $1/n$. Therefore
\begin{equation}
\mathfrak M_n^\perp(v)\sim C_\perp(v)\,(2\gamma T)^n\frac{n!}{n^2},\qquad n\to\infty,\quad n\ {\rm even},
\label{eq:transverse-large-n-result}
\end{equation}
where $C_\perp(v)$ is independent of $n$. The transverse radius of convergence is consequently
\begin{equation}
\boxed{R_\perp(v)=\frac{1}{2\gamma T}=\frac{\sqrt{1-v^2}}{2T}}.
\label{eq:transverse-radius}
\end{equation}
The $n^{-2}$ prefactor implies logarithmic branch points at $x=\pm R_\perp$, just as in the longitudinal case.

The longitudinal and transverse radii follow directly from the full convergence domain. Setting ${\bf L}=-i{\bf x}$, the large-$p$ hard integrand contains
\begin{equation}
\exp\left\{p\left[\hat{\bf p}\cdot{\bf x}-\frac{1-{\bf v}\cdot\hat{\bf p}}{2T}\right]\right\}.
\end{equation}
The radial integral converges for every momentum direction if and only if
\begin{equation}
\boxed{|2T{\bf x}+{\bf v}|<1}.
\label{eq:full-convergence-ball}
\end{equation}
Thus the analytic domain is a ball of radius $1/(2T)$ centered at $-{\bf v}/(2T)$. Its longitudinal diameter gives $R_\pm$, while its intersection with a transverse line gives $R_\perp$. In particular,
\begin{equation}
R_\perp(v)^2=R_+(v)R_-(v).
\end{equation}

The angular maximum also clarifies the meaning of the transverse tail. A large value of one transverse component is dominated by momentum-transfer directions
\begin{equation}
\hat{\bf p}_\star^{\,\pm}=v\hat{\bf v}\pm\sqrt{1-v^2}\,\hat{\bf e}_\perp,
\end{equation}
for which
\begin{equation}
\frac{p_\parallel}{|p_\perp|}=\gamma v.
\end{equation}
Hence $R_\perp$ characterizes the marginal distribution of a transverse component, with the unobserved longitudinal transfer integrated over. It is not the tail exponent of the slice constrained to $k_\parallel=0$, whose leading exponential scale is instead $1/(2T)$.

\bibliographystyle{jhep.bst}
\bibliography{main.bib}

\end{document}